\documentclass[a4paper,11pt]{article}
\pdfoutput=1 

\usepackage{jheppub} 

\usepackage[T1]{fontenc} 
\usepackage[colorlinks=true]{hyperref}

\title{\boldmath Modular Linear Differential Equations for Hecke and Fricke Groups}

\usepackage{mathrsfs}
\usepackage{xcolor}
\usepackage{colortbl}
\usepackage{bbold}

\usepackage{scalerel,stackengine}
\stackMath
\newcommand\reallywidehat[1]{%
\savestack{\tmpbox}{\stretchto{%
  \scaleto{%
    \scalerel*[\widthof{\ensuremath{#1}}]{\kern-.6pt\bigwedge\kern-.6pt}%
    {\rule[-\textheight/2]{1ex}{\textheight}}
  }{\textheight}%
}{0.5ex}}%
\stackon[1pt]{#1}{\tmpbox}%
}
\usepackage{tcolorbox}

\usepackage{relsize}

\newtheorem{theorem}{Theorem}[section]

\newtheorem{lemma}[theorem]{Lemma}

\newmuskip\pFqmuskip
\newcommand*\pFq[6][8]{%
  \begingroup 
  \pFqmuskip=#1mu\relax
  \mathcode`\,=\string"8000
  \begingroup\lccode`\~=`\,
  \lowercase{\endgroup\let~}\pFqcomma
  {}_{#2}F_{#3}{\left[\genfrac..{0pt}{}{#4}{#5};#6\right]}%
  \endgroup
}
\newcommand{\pFqcomma}{\mskip\pFqmuskip}

\usepackage{tikz}
\tikzset{Witten diagram/.style={execute at begin picture={%
\draw[blue,fill=blue!20] circle[radius=\pgfkeysvalueof{/tikz/Witten/radius}];
\path node (X){\phantom{X}};
},baseline={(X.base)}},vertex/.style={circle,fill,inner sep=1.5pt,node
contents={}},
Witten/.cd,radius/.initial=1.5cm}

\usepackage{tikz-cd}
\usetikzlibrary{cd}

\author{Naveen Balaji Umasankar}

\affiliation{University of Amsterdam,\\ Institute for Theoretical Physics (ITFA),\\
Science Park 904, 1098 XH Amsterdam.}

\emailAdd{naveenandmetallica@gmail.com}

\abstract{
Modular linear differential equations (MLDE) play a significant role in the classification of two-dimensional CFTs, where the modular forms in the equations belonged to the space of $\text{SL}(2,\mathbb{Z})$. A systematic study of the differential equations and their solutions for the Hecke groups $\Gamma_{0}(N)$ and Fricke groups $\Gamma_{0}^{+}(N)$ would better our understanding of CFT classification as there has not been significant work on the MLDE analysis for subgroups of $\text{SL}(2,\mathbb{Z})$. In this paper, we set up and solve MLDEs for Hecke and Fricke groups at levels $N\leq 12$ and report on admissible character-like solutions obtained in each group. We find that only the first four genus zero groups $\Gamma_{0}^{+}(p)$ where $p$ is a prime divisor of the Monster group $\mathbb{M}$ possess admissible single character solutions and we argue that the solutions for $\Gamma_{0}^{+}(11)$ are rendered inadmissible due to its Hauptmodul while those for $\Gamma_{0}^{+}(13)$ are rendered inadmissible due to the nature of the basis decomposition of the space of modular forms. We present a new quasi-character solution at the single character level for the Hecke groups $\Gamma_{0}(2)$, $\Gamma_{0}(7)$, and the subsequent group in its modular tower, $\Gamma_{0}(49)$. We also extend all of the results for single character solutions of Fricke groups to all prime divisor levels of $\mathbb{M}$ and remark on favorable properties in each group that could play a role in obtaining admissible solutions. Finally, we find the $\Theta$-series associated with levels $p = 2,3,5,7$ and the corresponding lattice data of Kissing numbers and lattice radii for each case. We find that the Fricke $\Theta$-series of level $p = 2$ has distinctive ties to the odd Leech lattice in $24$-dimensions.}

\begin{document} 
\maketitle
\flushbottom

\section{Introduction}
The idea of using MLDEs to look for character theories of rational conformal field theories (RCFT) was first put forward in \cite{Sen-Mukhi}, with a focus on modular forms from the group $\text{SL}(2,\mathbb{Z})$. Since then, progress has been made in both physics \cite{Phys1,Phys2,Ramesh_Mukhi,Hampapura-Mukhi} and mathematics \cite{Math1,Math2,Math3,Math4}. A clear overview of the current status of the RCFT classification utilizing the MLDE approach can be found in \cite{Mukhi_status_report}. The use of MLDE with modular forms belonging to congruence subgroups has not been thoroughly investigated. While Bae et al. \cite{Bae_Hecke_2} and Cheng et al. \cite{Cheng_Hecke_2} have investigated this using level-two congruence subgroups of $\text{SL}(2,\mathbb{Z})$, there has been no advancement in the categorization of admissible solutions at other levels of Hecke and Fricke groups. Due to their close relationship to even, self-dual lattices, and automorphic forms, single character or meromorphic theories are mathematically significant. These theories have a central charge with a $c = 8n$, or $n\in \mathbb{Z}$, multiple of eight. In order to set up and categorize single character solutions, we turn to the groups $\Gamma_{0}(N)$ and $\Gamma_{0}^{+}(N)$ and examine the theory of modular forms in these groups. 
\noindent
\subsection*{Why care about MLDEs?}
MLDEs show up in and play important roles in two-dimensional conformal field theory and number theory. RCFTs are a special subclass of 
CFTs whose central charges and conformal dimensions are rational numbers. Characters of an RCFT were found to possess transformation properties under the modular group \cite{Sen-Mukhi, RCFT_Moore}. Thanks to periodic boundary conditions on a torus, we can write the torus partition function as a trace, i.e $Z(\tau_{1},\tau_{2}) = \text{Tr}_{\mathscr{H}}e^{2\pi i\tau_{1}P}e^{2\pi i\tau_{2}H}$, where the trace is taken over all the physical states in the Hilbert space $\mathscr{H}$. Expressing the operators $H$ and $P$ in terms of $L_{0}$ and $\overline{L}_{0}$ respectively, we can write
\begin{align}
    Z(\tau,\overline{\tau}) = \text{Tr}_{\mathscr{H}}q^{L_{0} - \tfrac{c}{24}}\overline{q}^{\overline{L}_{0} - \tfrac{c}{24}},
\end{align}
where $q = e^{2\pi i\tau}$. Since the partition function depends on the torus' moduli, it is clearly invariant under modular transformations. Decomposing the Hilbert space
into a sum of tensor products of Verma modules, the torus partition function can be expressed as
follows
\begin{align}
    Z(\tau,\overline{\tau}) = \sum\limits_{i,j}M_{ij}\chi_{i}(\tau)\overline{\chi}_{j}(\overline{\tau}),\ \chi_{\Delta_{i}}(\tau) = q^{-\tfrac{c}{24} + \Delta_{i}}\sum\limits_{n=0}^{\infty}a(n)^{n},
\end{align}
where each $a(n)$ is the degeneracy of the Verma module at level $n$ that is bounded by $P(n)$, the partition of integers, $M_{ij}$ are non-negative integers which represent the multiplicity of the $(ij)^{\text{th}}$ representation, and each character $\chi_{i}(\tau)$ belongs to a particular representation that encodes data about a primary of weight $\Delta_{i}$. The characters possess well-defined transformations under the $S$ and $T$ matrices. These modular transformation properties were derived from first principles by Zhu \cite{Zhu} who suggested that the MLDE is a consequence of a null-vector relation in the vacuum Verma module. This was later proven by Gaberdiel et al. \cite{Gaberdiel}. Although an infinite number of RCFTs are known to exist, there exists no known classification of them. The chiral and fusion algebras of RCFTs are two significant algebraic structures that define them. The conventional approach to classification is based on chiral algebras. A minimal series of CFTs may be written down for every given chiral algebra thanks to the structure of its null vectors. In the simplest instance, the chiral algebra is merely the Virasoro algebra, and this process produces the Virasoro minimal models. Kac-Moody algebras have been key in the classification of potential chiral algebras. One may build coset models by using Wess-Zumino-Witten models, which are the minimal series of a given Kac-Moody algebra. This construction helps generate numerous RCFTs, including the various minimal series already obtained using null vector approaches.  This technique, however, is not exhaustive, and there are other RCFTs that do not fit inside this classification method. Mathur, Mukhi, and Sen \cite{Sen-Mukhi} classified RCFTs (called the MMS classification), whose different characters of the finitely many irreducible highest weight representations, satisfied a common second-order MLDE. The MMS classification was an alternative method that categorized the theories using fusion algebras and is based on the fact that the modular invariant partition function $Z(\tau,\overline{\tau})$ is not holomorphic and that the holomorphic characters $\chi_{i}(\tau)$ are not modular invariant. However, these characters may be expressed as solutions to a differential equation that is both holomorphic and modular invariant. Together, these two characteristics are quite restricting and may be used to group MLDEs that provide the characters of RCFTs. The RCFT is then identified by the pair $(n,\ell)$, where $n$ denotes the number of characters and $\ell$ denotes the Wronskian index which is an integral (non-negative) number associated with the structure of zeroes of the Wronskian that can be expressed solely in terms of the RCFT data, $\{c,n,\Delta_{i}\}$.\\\\
\noindent
The simplest way to classify RCFTs would be to pick those fusion algebras that possess trivial fusion. This is observed only when the primary is the vacuum state and all other states are the vacuum's descendants. The torus partition function in this case is given by the character itself, $Z(\tau,\overline{\tau}) = \vert\chi(\tau)\vert^{2}$. The vacuum character $\chi(\tau)$ satisfies a specific MLDE of order one and we can find the central charge using the Riemann-Roch theorem. Schellekens classified all $c = 24$ single character theories and found that there are a total of $71$ CFTs, all of which possess a partition function of the form $Z(\tau) = J(\tau) + 12m$, where $m\in\left\{0, 2, 3, 4, 5, 6, 7, 8, 9, 10, 12, 13, 14, 16, 18, 20, 22, 24, 25, 26, 28, 30, 32, 34, 38, 46, 52, 62, 94\right\}$ \cite{Schellekens}. However, a particular subtlety to note is that not all the $71$ CFTs correspond to unique admissible character-like solutions since some characters describe more than one CFT.  A notable example in this list is the CFT that corresponds to the minimal value $m = 0$ and has Klein's $J$-function as the partition function. This is the much celebrated Monster CFT constructed by Frenkel, Lepowsky, and Meurman (FLM) as a $\mathbb{Z}_{2}$ orbifold of free bosons on the Leech lattice $\Lambda_{24}$. From the $q$-series expansion of the $J$-function, $J(\tau) = q^{-1} + 196884q + \ldots$, FLM observed that the CFT has an identity operator $1$ that corresponds to the unique vacuum state and $196884 = 1 + 196883$ dimension-$2$ operators among which one is the stress tensor, a second-order operator of the identity, and the other $196883$ being primaries that transform under the monster group representations. It is also interesting to note that from the perspective of number theory, Kaneko and Zagier used MLDEs to study supersingular $j$-invariants of elliptic curves \cite{Kaneko-Zagier}. It was later shown in \cite{Kaneko-Zagier-MMS} that the Kaneko-Zagier equation is equivalent to the MMS classification. Finding (admissible) solutions to MLDEs helps us better understand the nature and behaviour of the characters which we could then relate to an existing RCFT. The primary focus of this paper will be on single character solutions which we will obtain by directly solving the MLDE. Although we find admissible solutions, we haven't related them to known theories in all cases. This work reports on the classification of admissible character-like solutions in the Hecke and Fricke groups of level $N\leq 12$. The relation of these solutions and solutions of higher-order MLDEs to known theories is reserved for future work. 

\noindent

\subsection*{Monster and friends}
The monstrous moonshine refers to the connection between the representation of the Monster group $\mathbb{M}$ (also known as the Fischer-Griess group $F_{1}$) and modular forms. It was first observed by John McKay and the term was first coined by John Conway \cite{Conway}. The Monster group is the largest of the sporadic groups and is of the order $\approx 8\times 10^{53}$ which can be expressed in terms of prime divisors of the Monster group $\mathbb{M}$ as follows
\begin{align}\label{prime_divisors_Monster}
    \begin{split}
        \vert\mathbb{M}\vert
        = 2^{46}\cdot3^{20}\cdot5^{9}\cdot7^{6}\cdot11^{2}\cdot13^{3}\cdot17\cdot19\cdot23\cdot29\cdot31\cdot41\cdot59\cdot71
    \end{split}
\end{align}
The $J$-function constructed from the Hauptmodul of $\Gamma = \text{PSL}(2,\mathbb{Z})$ reads
\begin{align}
        J(\tau) = j(\tau) -744 =q^{-1} +& 196884 q + 21493760 q^2 + 864299970 q^3 + 20245856256 q^4 + \ldots
\end{align}
McKay made the following observation: $ 196884 = 1 + 196883$, where the summands are exactly equal to the dimensions of the first two irreducible representations of the conjugacy class $1A$ of $\mathbb{M}$. These dimensions of irreps are nothing but the degrees of the irreducible characters of the group. The next few coefficients also possess a decomposition in terms of the characters. This unexpected relation between representation theory and modular forms culminated in the conjecture of the existence of an infinite-dimensional $\mathbb{Z}$-graded Monster module, $V = \bigoplus\limits_{n = 1}^{\infty}V_{n}$, such that the graded dimension of the module is equal to the coefficients of the $J$-function, i.e. 
\begin{equation}
    \text{dim}\ V_{n} = c_{n},\ \forall n\geq -1
\end{equation}
Hence, we can express the $J$-function as follows
\begin{equation}
    J(\tau) = \sum\limits_{n=-1}^{\infty}\text{dim}\ V_{n}q^{n},
\end{equation}
with $\text{dim}\ V_{0} = 0$ corresponding to coefficient $c_{0} = 0$. Now, this definition does not prevent us from stating that, for example, $a_{2} = 21493760 = 21493760\chi_{1}$, i.e. we can always compose each $V_{n}$ with $n$ copies of the trivial representation of $\mathbb{M}$ to reproduce the $J$-function. To make the conjecture more concrete, Thompson proposed to look at graded characters of $V$. We define the McKay-Thompson series as follows
\begin{equation}
    T_{g}(\tau) \equiv \sum\limits_{n=-1}^{\infty}\text{Tr}_{V_{n}}(g)q^{n},
\end{equation}
with $g\in\mathbb{M}$ and $T_{e} = J$, where $e$ is the identity element. The constant term of this series is null-valued, i.e. $\text{Tr}_{V_{0}} = 0$. With $G$ denoting the group the representation $(V,\rho)$, where $\rho_{n}:G\to \text{GL}(V)$ is a group homomorphism, the character $\chi_{\rho}:G\to \mathbb{C}$ is defined by the trace of the representation matrices, i.e. $\chi_{\rho} \equiv \text{Tr}(\rho(G)) = \text{Tr}_{V}(g)$ with $g\in G$. The character possesses the following property
\begin{align}
     \chi_{\rho}(g) =  \chi_{\rho}(hgh^{-1}) \ \Leftrightarrow\ T_{g} = T_{hgh^{-1}},\ \forall g,h\in G.
\end{align}
This defines a conjugacy and there are $194$ such conjugacy classes in $\mathbb{M}$. The monstrous moonshine conjecture states that for each $g\in\mathbb{M}$ there exists a genus zero subgroup $\Gamma_{g}\subset\Gamma$ such that the McKay-Thompson series $T_{g}(\tau)$ is a Hauptmodul on $\Gamma_{g}$, i.e. $T_{g}(\tau) = J_{\Gamma_{g}}(\tau)$. 
A Hecke group of level $N$, $\Gamma_{0}(N)$, is a discrete subgroup of the modular group and an index $k\in\mathbb{Z}$ subgroup of the Fricke group of level $N$, $\Gamma_{0}^{+}(N)$, which is a slightly larger group. The index $k$ is easy to determine using the multiplicative property. A clear description of the definition and the properties of these groups is discussed in appendix \ref{appendix:A}. Groups $\Gamma_{0}(p)$ where $p\in\mathbb{P}$ are of genus-zero when $(p-1)\vert 24$ and the $J$-function is equal to the McKay-Thompson series corresponding to type $B$ conjugacy classes of $\mathbb{M}$ \cite{Toshiki}, 
\begin{align}\label{J_function_McKay}
    J_{p}(\tau) = T_{pB}(\tau) = \left(\frac{\eta(\tau)}{\eta(p\tau)}\right)^{\frac{24}{p-1}} + \frac{24}{p-1}.
\end{align}
The groups $\Gamma_{0}^{+}(p)$ are of genus zero for values $2\leq p\leq 31$ or when $p \in\left\{41,47,59,71\right\}$ (which cover all the prime divisors of $\mathbb{M}$ as shown in \ref{prime_divisors_Monster}) and the $J$-function is equal to the McKay-Thompson series corresponding to type $A$ conjugacy classes of $\mathbb{M}$ \cite{Toshiki},
\begin{equation}\label{J_function_McKay_Fricke}
    J_{p^{+}}(\tau) = T_{pA}(\tau) = \left(\frac{\eta(\tau)}{\eta(p\tau)}\right)^{\frac{24}{p-1}} + \frac{24}{p-1} + p^{\frac{12}{p-1}}\left(\frac{\eta(p\tau)}{\eta(\tau)}\right)^{\frac{24}{p-1}}.
\end{equation}

\subsection*{Hecke and Fricke groups in physics}
Bae et al. have found a large class of Fermionic RCFTs using order one and two MLDEs for level-two congruence subgroups \cite{Bae_Hecke_2}, and the analysis was extended to explore the landscape of three-character Fermionic RCFTs in \cite{Bae_Hecke_3}. Hecke groups play an important role in the computation of one-loop modular integrals in string theory involving the Narain partition function \cite{Hecke_unfolding_1, Hecke_unfolding_2}. The Hauptmodules of Hecke groups are used to obtain an exact non-perturbative relation between each effective and the bare couplings in $\mathcal{N} = 2$ SQCD \cite{SQCD}. The theory of Hecke groups is related to dessin d'enfants that show up in $\mathcal{N} = 2$ supersymmetric gauge theories and topological strings among others. A review of this connection can be found in \cite{dessins}. In string theory, Fricke dualities are found to play an important role in the dualities of CHL models \cite{CHL_1, CHL_2}.\\

\noindent
The main results of this paper are the following:
\begin{itemize}
    \item Analysis of MLDE solutions for groups $\Gamma_{0}(N)$ and $\Gamma_{0}^{+}(N)$ at levels $N\leq 12$.
    \item Admissible single character MLDE solutions for modular forms in the Fricke groups $\Gamma_{0}^{+}(p)$ for $p = 2,3,5,7$.
    \item New type of quasi-character solutions for groups $\Gamma_{0}(2)$ , $\Gamma_{0}(7)$, $\Gamma_{0}(49)$, and $\Gamma_{0}^{+}(49)$.
    \item Explicit constructions and $q$-series expansions of $\Theta$-series and kissing numbers obtained from single character solutions.
    \item Relation between $\Theta$-series of the odd Leech lattice and the one obtained via the single character solution in $\Gamma_{0}^{+}(2)$.
\end{itemize}

\subsection*{Outline}
In the next section, we introduce the idea of the MLDE and the general approach to obtaining admissible solutions. In section \ref{sec:Gamma_0_2+}, we study the theory of modular forms in $\Gamma_{0}^{+}(2)$ and set up differential equations to find single character solutions. In sections \ref{sec:Gamma_0_3+} and \ref{sec:Gamma_0_5+}, we study the theory of modular forms in $\Gamma_{0}^{+}(3)$ and $\Gamma_{0}^{+}(5)$ and set up differential equations to find single character solutions. In section \ref{sec:level_7}, we turn to the groups at level $N = 7$, study the theory of modular forms, and discuss the applicability of finding solutions at this level. Setting up differential equations, we find a new type of quasi-character in $\Gamma_{0}(7)$ and single character solutions in $\Gamma_{0}^{+}(7)$. We conclude this section with a study of admissible solutions in the groups associated with the modular towers, i.e. $\Gamma_{0}(49)$ and $\Gamma_{0}^{+}(49)$. In sections \ref{sec:Gamma_0_11+} and \ref{sec:Gamma_0_13+}, we study the groups $\Gamma_{0}^{+}(11)$ and $\Gamma_{0}^{+}(13)$ and discuss why there are no admissible single character solutions. In section \ref{sec:other_groups}, we highlight the issues associated with finding single character solutions in Hecke groups by taking $\Gamma_{0}(3)$ as an example, present the quasi-characters found in $\Gamma_{0}(2)$, and briefly discuss the idea of an MLDE set up for the groups $\Gamma_{1}(N)$. Section \ref{sec:Kissing} deals with finding explicit $q$-series expansions for the associated $\Theta$-series of the single character solutions of Fricke groups at prime levels. Following a quick discussion on kissing numbers of lattices and known lower and upper bounds, we find connections between the coefficients of the $\Theta$-series associated with an MLDE solution of $\Gamma_{0}^{+}(2)$ to the odd Leech lattice and $T_{23A}(\tau)$, and the $\Theta$-series and kissing numbers of lattices obtained from single character solutions in groups $\Gamma_{0}^{+}(3)$, $\Gamma_{0}^{+}(5)$, and $\Gamma_{0}^{+}(7)$ are presented. In section \ref{sec:conclusion}, we finish with conclusions and future prospects. All fundamental domains of the Hecke and Fricke groups were obtained using the Mathematica program in \cite{Mathematica_Fundamental_domain}. Several appendices complement the theory and results of the main sections. We provide the mathematical foundations necessary to understand the paper in appendix \ref{appendix:A}. This comprises definitions and theorems that we heavily rely on throughout the main sections. Finally, we explicitly find the $\Theta$-series in appendix \ref{appendix:C}, together with the corresponding lattice data, using single character solutions to the Fricke groups of levels $5 $ and $7$. 

\section{Modular linear differential equation}\label{sec:MLDE}
Consider an $(n+1)$-dimensional square matrix made up of $\chi_{0},\chi_{1},\ldots,\chi_{n-1},f$ and their derivatives. The most general ODE that is invariant under $\text{SL}(2,\mathbb{Z})$ reads
\begin{align}\label{MLDE_first_version}
    \left[\omega_{2\ell}(\tau)\mathcal{D}^{n} + \omega_{2\ell + 2}(\tau)\mathcal{D}^{n-1} + \ldots + \omega_{2(n-1+\ell)}(\tau)\mathcal{D} + \omega_{2(n+\ell)}(\tau)\right]f(\tau) = 0,
\end{align}
where the functions $\omega_{k}$ are holomorphic modular forms of weight $k$ and the derivatives here are the Ramanujan-Serre covariant derivatives given by
\begin{align}
    \begin{split}
        \mathcal{D} \equiv& \frac{1}{2\pi i}\frac{d}{d\tau} - \frac{r}{12}E_{2}(\tau),\\
        \mathcal{D}^{n} =& \mathcal{D}_{r+2n-2}\circ\ldots\circ\mathcal{D}_{r+2}\circ\mathcal{D}_{r},
    \end{split}
\end{align}
where $r$ is the weight of the modular form on which the derivative acts on. The covariant derivative transforms a weight $r$ modular form to a weight $r+2$ modular form. Dividing \ref{MLDE_first_version} throughout by the first coefficient $\omega_{2\ell}(\tau)$, we obtain the monic form of the MLDE
\begin{align}
    \left[\mathcal{D}^{n} + \phi_{n-1}(\tau)\mathcal{D}^{n-1} + \ldots + \phi_{1}(\tau)\mathcal{D} + \phi_{0}(\tau)\right]f(\tau) = 0,
\end{align}
where the functions $\phi_{k}$ are now meromorphic modular forms. If the function $f$ is a linear combination of $n$ characters then the determinant of the $(n+1)$-dimensional matrix vanishes and we have
\begin{align}\label{prior_ODE}
    \sum\limits_{p=0}^{n}(-1)^{p}W_{p}\mathcal{D}^{p}f(\tau) = 0,
\end{align}
where the Wronskian determinant $W_{p}$ is given by
\begin{align}
    W_{p} = \text{det}\begin{pmatrix}
    \chi_{0} & \ldots & \chi_{n-1}\\
    \mathcal{D}\chi_{0}& \ldots & \mathcal{D}\chi_{n-1}\\
    \vdots & & \vdots\\
    \mathcal{D}^{p-1}\chi_{0} & \ldots & \mathcal{D}^{p-1}\chi_{n-1}\\
    \mathcal{D}^{p+1}\chi_{0} & \ldots & \mathcal{D}^{p+1}\chi_{n-1}\\
    \vdots & & \vdots\\
    \mathcal{D}^{n}\chi_{0} & \ldots & \mathcal{D}^{n}\chi_{n-1}
    \end{pmatrix}.
\end{align}
Writing $\phi_{p}(\tau) = (-1)^{n-p}\tfrac{W_{p}}{W_{n}}$ to be automorphic forms of weight $2(n-p)$ for $\text{SL}(2,\mathbb{Z})$, we can rewrite \ref{prior_ODE} as the following MLDE
\begin{align}
    \left[\mathcal{D}^{n} + \sum\limits_{p=0}^{n-1}\phi_{p}(\tau)\mathcal{D}^{p}\right]f(\tau) = 0.
\end{align}
In general, the automorphic functions $\phi_{p}$ need not be holomorphic, they can be meromorphic even though
the resulting characters are holomorphic. Functions $\phi_{p}(\tau)$ become singular at the zeros of the Wronskian, i.e. they could have poles at the zeros of $W_{n}(\tau)$. The classification of RCFTs is based on numbers $n$ and $\ell$ that denote the number of characters (or the order of the MLDE) and the total number of zeros of the Wronskian $W_{n}$. The number of zeros $\ell$ can be expressed in terms of the central charge and conformal weights $\Delta_{i}$ of the primaries associated with the $n$ characters. The number of zeros of $W_{n}$ in the fundamental domain is equal to the number of zeros of $\omega_{2\ell}(\tau)$  which can be found to be 
\begin{align}
    \# = 2\ell\cdot \frac{\mu}{12},
\end{align}
where $\mu = \mu_{0}$ and $\mu_{0}^{+}$ denote the group index of the Hecke and the Fricke groups respectively. At the cusp, $\tau = i\infty$, the character $\chi_{i}$ associated with a primary of weight $\Delta_{i}$ behaves as follows
\begin{align}\label{character_behaviour}
   \begin{split}
        \chi_{i} =& q^{-\frac{c}{24} + \Delta_{i}}\left[\prod\limits_{m=2}^{\infty} \frac{1}{1-q^{m}} + \mathcal{O}(q^{n-1})\right],\\
        \sim& q^{-\frac{c}{24} + \Delta_{i}}\left(1 + \mathcal{O}(q)\right),
   \end{split}
\end{align}
and hence, the Wronskian has the following cusp behaviour
\begin{align}\label{W_n_asymptotic}
    W_{n}(\tau) \sim q^{-\frac{nc}{24} + \sum\limits_{i}\Delta_{i}}\left(1 + \mathcal{O}(q)\right).
\end{align}
There are $\tfrac{1}{2}n(n-1)$ derivatives in the construction of the Wronskian $W_{n}$ and it transforms like a modular form of weight $k = n(n-1)$ (up to a phase). In the following sections, we shall consider the following Frobenius ansatz for the characters in $\Gamma_{0}(N)$ and $\Gamma_{0}^{+}(N)$
\begin{align}
    \chi_{i} = q^{\alpha_{i}}\sum\limits_{n=0}^{\infty}a_{n}^{(i)}q^{n},
\end{align}
where the exponents $\alpha_{i}$, the order $n$ of the differential equation, and the integer $\ell$ are related by an expression that follows from the Riemann-Roch theorem that is derived using the valence formulae of $\Gamma_{0}(N)$ and $\Gamma_{0}^{+}(N)$. The algorithm we follow to find admissible character-like solutions to first-order MLDEs is the following:
\begin{enumerate}
    \item Set up the theory of modular forms for the Hecke/Fricke group of interest.
    \item From the group data and knowledge of the valence formula, find the number of zeros in the fundamental domain.
    \item Set up the Riemann-Roch theorem and hence, find an expression for the central charge in terms of the Wronskian index $\ell$.
    \item Postulate MLDE and solve with Frobenius ansatz for all $\ell\leq 6$ to find character-like solutions that possess positive integral coefficients, $a_{n}>0$.
\end{enumerate}
\noindent
It is shown in \cite{Ramesh_Mukhi} that $\text{SL}(2,\mathbb{Z})$ characters can be expressed as follows
\begin{align}
    \chi(\tau) = j^{w_{\rho}}(j - 1728)^{w_{i}}\mathfrak{P}_{w_{\tau}}(j),
\end{align}
where $j(\tau)$ is Klein's invariant which is the only modular invariant function in $\text{SL}(2,\mathbb{Z})$, $w_{\rho} \in\left\{0,\tfrac{1}{3},\tfrac{2}{3}\right\}$, $w_{i}\in\left\{0,\tfrac{1}{2}\right\}$, $w_{i}\in\mathbb{Z}$, and $\mathfrak{P}_{w_{\tau}}(j)$ is a monic polynomial of degree $w_{\tau}$ in $j(\tau)$. This was constructed by looking at the characters at $\ell = 1$, $\ell = 2$, and $\ell = 3$ which were found to be $1$, $j^{\tfrac{1}{3}}(\tau)$, and $\left(j(\tau) - 1728\right)^{\tfrac{1}{2}}$ respectively. With this, the Riemann-Roch equation for $\text{SL}(2,\mathbb{Z})$ was found to yield
\begin{align}
    c = 24\left(w_{\rho} + w_{i} + w_{\tau}\right) = 4\ell.
\end{align}
For the values of $\ell$ in the range $\ell \leq 6$, there is a finite number of admissible solutions at even integers $\ell = 0,2,4$ and none at odd integers $\ell = 1,3,5$. But to show the typical issues faced that make the solutions inadmissible, we take into account the odd cases and explicitly solve the MLDE. The characters for cases $\ell\geq 6$ turn out to depend on arbitrary integers and are admissible over infinite ranges of these numbers. In this work, MLDE analysis is carried out in the range $0\leq \ell\leq 6$. Despite the fact that there are many free parameters involved in the $\ell = 6$ MLDE, we find that disregarding the parameters related to cusp forms yields an admissible outcome in the case of $\Gamma_{0}^{+}(2)$.

\subsection*{Quasi-characters}
Quasi-characters are defined as those solutions to the MLDE that possess all integer coefficients that need not necessarily be positive. These characters come in the following two types in literature:
\begin{enumerate}
    \item Type I: Some of the coefficients $m_{1},m_{2},\ldots, m_{n}\leq0$, while all of the $m_{k}>0$ for $k>n$.
    \item Type II: Some of the coefficients $m_{1},m_{2},\ldots,m_{n}\geq0$, while all of the $m_{k}<0$ for $k>n$.
\end{enumerate}

\noindent
\section{\texorpdfstring{$\mathbf{\Gamma_{0}^{+}(2)}$}{Γ0(2)+}}\label{sec:Gamma_0_2+}
The Fricke group of level $2$ is generated by the $T$-matrix and the Atkin-Lehner involution of level $2$, $W_{2} = \left(\begin{smallmatrix}0 & -\tfrac{1}{\sqrt{2}}\\ \sqrt{2} & 0\end{smallmatrix}\right)$, i.e. $\Gamma_{0}^{+}(2) = \langle\left(\begin{smallmatrix}1 & 1\\ 0 & 1\end{smallmatrix}\right), W_{2}\rangle$. A non-zero $f\in\mathcal{M}_{k}^{!}(\Gamma_{0}^{+}(2))$ satisfies the following valence formula
\begin{align}
    \nu_{\infty}(f) + \frac{1}{2}\nu_{\tfrac{i}{\sqrt{2}}}(f) + \frac{1}{4}\nu_{\rho_{2}}(f) + \sum\limits_{\substack{p\in\Gamma_{0}^{+}(2)\backslash\mathbb{H}^{2}\\ p\neq \tfrac{i}{\sqrt{2}},\rho_{2}}}\nu_{p}(f) =   \frac{k}{8},
\end{align}
where $\tfrac{i}{\sqrt{2}}$ and $\rho_{2} = \tfrac{-1+i}{2}$ are elliptic points and this group has a cusp at $\tau = \infty$ (see \cite{Junichi} for more details). For $k\geq2$ and $k\in2\mathbb{Z}$, the dimension of the space of modular forms reads
\begin{align}
    \begin{split}
        \text{dim}\ \mathcal{S}_{k}(\Gamma_{0}^{+}(2)) = \begin{cases}\left\lfloor\left.\frac{k}{8}\right\rfloor\right. - 1,\ &k \equiv 2\ (\text{mod}\ 8)\\
        \left\lfloor\left.\frac{k}{8}\right\rfloor\right.,\ &k\not\equiv2\ (\text{mod}\ 8).
        \end{cases},\\
        \text{dim}\ \mathcal{M}_{k}(\Gamma_{0}^{+}(2)) = 1 + \mathcal{S}_{k}(\Gamma_{0}^{+}(2)).
    \end{split}
\end{align}
Using theorem \ref{thm:genus_Fricke_index} with $(N,h,m)=(2,1,2)$, we have
\begin{align}
    \chi(\Gamma_{0}^{+}(2)) = \frac{2}{6}\prod\limits_{p\vert 2}\frac{p+1}{2p} = \frac{1}{4},\ \ 
    \left[\Gamma_{0}^{+}(2):\Gamma_{0}(2)\right] = 2,
\end{align}
i.e. $\Gamma_{0}(2)$ is an index-$2$ subgroup of $\Gamma_{0}^{+}(2)$. With this, we find the index of $\Gamma_{0}^{+}(2)$ in $\text{SL}(2,\mathbb{Z})$ to be
\begin{align}\label{fractional_ghost_index}
    \left[\text{SL}(2,\mathbb{Z}):\Gamma_{0}^{+}(2)\right] = \frac{\left[\text{SL}(2,\mathbb{Z}):\Gamma_{0}(2)\right]}{\left[\Gamma^{+}_{0}(2):\Gamma_{0}(2)\right]} = \frac{3}{2}.
\end{align}
Two comments are in order here. Firstly, the Fricke group, unlike the Hecke group, is not a subgroup of $\text{SL}(2,\mathbb{Z})$ but is an extension of the Hecke group (see appendix \ref{appendix:A}). Secondly, the word ``index" here is abused since
we notice that the above computation gives us a fraction which implies a fractional number of cosets of $\Gamma_{0}^{+}(2)$ in $\text{SL}(2,\mathbb{Z})$. The multiplicative property of the index is used to obtain that is necessary to obtain the left-hand side of the Riemann-Roch relation required to establish a relation between the Wronskian index $\ell$ and the central charge. Using \ref{fractional_ghost_index} we find the number of zeros in the fundamental domain $\mathcal{F}_{2^{+}}$ to be $\# = 2\ell\cdot \tfrac{\mu}{12} = \tfrac{1}{4}\ell$. The Riemann-Roch theorem takes the following form
\begin{align}\label{Riemann_Roch_Gamma_0_2+}
    \begin{split}
        \sum\limits_{i=0}^{n-1}\alpha_{i} =& -\frac{nc}{24} + \sum\limits_{i}\Delta_{i}\\
        =&\frac{1}{8}n(n-1) - \frac{1}{4}\ell.
    \end{split}
\end{align}
When $n = 1$, we find the following relation between the number of zeros in the fundamental domain and the central charge
\begin{align}
    c = 6\ell.
\end{align}
Let $k$ be an even integer such that $k\geq 4$, then the space of modular forms of $\Gamma_{0}^{+}(2)$ has the following basis decomposition
\begin{align}\label{basis_decomposition_Gamma_0_2+}
    \begin{split}
    \mathcal{M}_{k}(\Gamma_{0}^{+}(2)) =& E_{k - 8n}^{(2^{+})}\left[\mathbb{C}\left(\left(E_{4}^{(2^{+})}\right)^{2}\right)^{n} \oplus \mathbb{C}\left(\left(E_{4}^{(2^{+})}\right)^{2}\right)^{n-1}\Delta_{2}\oplus \ldots\oplus \mathbb{C}\left(\Delta_{2}\right)^{n}\right],\\
    n =& \text{dim}\ \mathcal{M}_{k}(\Gamma_{0}^{+}(2)) - 1,
    \end{split}
\end{align}
where $k - 8n = 0,4,6,$ or $10$. $E_{k}^{(2^{+})}(\tau)$ and $\Delta_{2}(\tau)$ are the Eisenstein series and the cusp form of $\Gamma_{0}^{+}(2)$ respectively and are defined as follows
\begin{align}
    \begin{split}
    E_{k}^{(2^{+})}(\tau) \equiv& \frac{2^{\tfrac{k}{2}}E_{k}(2\tau) + E_{k}(\tau)}{2^{\tfrac{k}{2}} + 1},\\
    \Delta_{2}(\tau) \equiv& \left(\eta(\tau)\eta(2\tau)\right)^{8}\in\mathcal{S}_{8}(\Gamma_{0}^{+}(2)).
    \end{split}
\end{align}
One might be tempted to consider the $E_{2}^{(2^{+})}(\tau)$ as a weight $2$ modular form of $\Gamma_{0}^{+}(2)$ but the quasi-modular $E_{2}(p\tau)$ Eisenstein series do not conspire to make this candidate a modular form. This is rather easy to see,
\begin{align}
    \begin{split}
        E_{k}(p\gamma(\tau)) =& (c\tau + d)^{2}E_{k}(p\tau) + \tfrac{12c}{2\pi ip}(c\tau + d),\ \gamma\in\Gamma_{0}^{+}(2),\\
        E_{2}^{(2^{+})}(\gamma(\tau)) =& (c\tau + d)^{2}E_{2}^{(2^{+})}(\tau) + \frac{8c}{2\pi i}(c\tau + d),
    \end{split}
\end{align}
with the action $\gamma(\tau) = \tfrac{a\tau + b}{c\tau + d}$. If we were to tweak the definition by a change of sign we obtain $E_{2,2^{'}}(\tau) = 2E_{2}(2\tau) - E_{2}(\tau)$, which is a modular form of weight $2$ in the Hecke group $\Gamma_{0}(2)$. Lastly, the Hauptmodul of this group is defined as follows
\begin{align}\label{Hauptmodul_Gamma_0_2+}
    \begin{split}
        j_{2^{+}}(\tau) =& \left(\frac{\left(E_{4}^{(2^{+})}\right)^{2}}{\Delta_{2}}\right)(\tau) = \left(\left(\frac{\eta(\tau)}{\eta(2\tau)}\right)^{12} + 2^{6}\left(\frac{\eta(2\tau)}{\eta(\tau)}\right)^{12}\right)^{2}\\
        =& q^{-1} + 104 + 4372q + 96256q^{2} + 1240002q^{3} + \ldots
    \end{split}
\end{align}

\subsection{Single character solutions}
\subsection*{\texorpdfstring{$\ell = 0$}{Lg}}
Let us consider the simple case with no poles $\ell = 0$ and a single character. In this case, the only primary is the vacuum state and all other states are descendants of the vacuum. The single vacuum character behaves as $f_{0}(\tau) \sim q^{-\tfrac{c}{24}}\left(1 + \mathcal{O}(q)\right)$ and satisfies the following MLDE
\begin{align}
    \left[\omega_{2\ell}(\tau)\mathcal{D} + \omega_{2\ell + 2}(\tau)\right]f(\tau) = 0.
\end{align}
For $\ell = 0$, we have to make choices for the functions $\omega_{0}(\tau)$ and $\omega_{2}(\tau)$. Since the spaces $\mathcal{M}_{0}(\Gamma_{0}^{+}(2))$ and $\mathcal{M}_{2}(\Gamma_{0}^{+}(2))$ are one-dimensional, we choose $\omega_{0}(\tau) = 1 = \omega_{2}(\tau)$. Plugging in the behaviour of $f_{0}(\tau)$, we find $\mu = \tfrac{c}{24} = 0$ since $c =0$. Hence, the only trivial solution is $f_{0}(\tau) = 1$. 

\subsection*{\texorpdfstring{$\ell = 1$}{Lg}}
If an admissible theory exists for $\ell = 1$, then it would have a central charge $c = 32$. The MLDE in this case takes the following form
\begin{align}
    \left[\omega_{2}(\tau)\mathcal{D} + \omega_{4}(\tau)\right]f(\tau) = 0.
\end{align}
Since the space $\mathcal{M}_{4}(\Gamma_{0}^{+}(2))$ is one-dimensional, we make the following choice for $\omega_{4}(\tau)$
\begin{align}
    \begin{split}
        \omega_{4}(\tau) =& \mu E_{4}^{(2^{+})}(\tau)\\
        =& \mu\left(1 + 48 q + 624 q^2 + 1344 q^3 + 5232 q^4 + \ldots\right)
    \end{split}
\end{align}
Now, to build $\tfrac{\omega_{4}(\tau)}{\omega_{2}(\tau)}$, which is a modular form of weight two, we consider ratios of higher weight modular forms. We find the following unique choice
\begin{align}
    \frac{\omega_{4}(\tau)}{\omega_{2}(\tau)} = \mu_{1}\left(\frac{\left(E_{4}^{(2^{+})}\right)^{2}}{E_{2}^{(2^{+})}}\right)(\tau) + \mu_{2}\left(\frac{\Delta_{2}}{E_{6}^{(2^{+})}}\right)(\tau).
\end{align}
The MLDE with this basis reads
\begin{align}
    \left[\Tilde{\partial} + \mu_{1}\left(\frac{E_{6}^{(2^{+})}}{E_{4}^{(2^{+})}}\right)(\tau) + \mu_{2}\left(\frac{\Delta_{2}}{E_{6}^{(2^{+})}}\right)(\tau)\right]f(\tau) = 0,
\end{align}
where $\widetilde{\partial} = \tfrac{1}{2\pi i}\tfrac{d}{d\tau}$. Since we know the $q$-series expansions in the equation, we can solve this ODE order by order in $q$ which yields a character that is modular invariant (up to a phase) for every value of the free parameters. The character $f(\tau) = \chi(\mu,\tau)$ will possesses the following $q$-series expansion
\begin{align}\label{correct_ansatz}
    \chi(\mu,\tau) = q^{\alpha}\left(1 + m_{1}(\mu)q + m_{2}(\mu)q^{2} + \ldots\right),
\end{align}
where the coefficient $m_{0} = 1$ since the corresponding state is the vacuum state of the theory, which is to be unique. With $f(\tau) = q^{-\tfrac{c}{24}}\sum_{n=0}^{\infty}a_{n}q^{n}$ where $m_{0}(\mu) = 1$, we find for $\mu_{1} = \tfrac{c}{24} = \tfrac{1}{4}$ and $\mu_{2} = -256$, the following admissible solution
\begin{align}
    \begin{split}
        \chi^{(\ell = 1)}(\tau) =& j_{2^{+}}^{\tfrac{1}{4}}(\tau)\\
        =& q^{-\tfrac{1}{4}}\left(1 + 26 q + 79 q^2 + 326 q^3 + 755 q^4 + \ldots\right).
    \end{split}
\end{align}

\subsection*{\texorpdfstring{$\ell = 2$}{Lg}}
If an admissible theory exists for $\ell = 2$, then it would have a central charge $c = 12$.
The MLDE for this case takes the following form
\begin{align}
    \left[\omega_{4}(\tau) \mathcal{D} + \omega_{6}(\tau)\right]f(\tau) = 0. 
\end{align}
The space $\mathcal{M}_{6}(\Gamma_{0}^{+}(2))$ is one-dimensional and using basis decomposition \ref{basis_decomposition_Gamma_0_2+}, we fix $\omega_{6}(\tau)$ to be the following
\begin{align}
    \begin{split}
        \omega_{6}(\tau) =& \mu_{1} E_{6}^{2^{+}}(\tau)\\
        =& \mu_{1}\left(1 - 56 q - 2296 q^2 - 13664 q^3 - 73976 q^4 + \ldots\right)
    \end{split}
\end{align}
The MLDE now takes the following form
\begin{align}
    \left[\Tilde{\partial} + \mu\left(\frac{E_{6}^{(2^{+})}}{E_{4}^{(2^{+})}}\right)(\tau)\right]f(\tau) = 0,
\end{align}
where $\mu = \tfrac{1}{2}$. The first few coefficients of the solution read $m_{1}(\mu) = 52, m_{2}(\mu) = 834, m_{3}(\mu) = 4760, \ldots$, and so on. The solution now has
all integral coefficients that are non-negative and can be expressed as a power of the Hauptmodul $j_{2^{+}}(\tau)$ defined in \ref{Hauptmodul_Gamma_0_2+} as follows
\begin{align}
    \begin{split}
    \chi^{(\ell = 2)}(\tau) =& j_{2^{+}}^{\tfrac{1}{2}}(\tau)\\
    =& q^{-\tfrac{1}{2}}\left(1 + 52 q + 834 q^2 + 4760 q^3 + 24703 q^4 + \ldots\right).
    \end{split}
\end{align}

\subsection*{\texorpdfstring{$\ell = 3$}{Lg}}
The central charge of an admissible theory would now be $c = 18$ with the following MLDE
\begin{align}
    \left[\omega_{6}(\tau)\mathcal{D} + \omega_{8}(\tau)\right]f(\tau) = 0.
\end{align}
The space $\mathcal{M}_{k}(\Gamma_{0}^{+}(2))$ is two-dimensional and using \ref{basis_decomposition_Gamma_0_2+}, we have the following decomposition
\begin{align}
    \mathcal{M}_{8}(\Gamma_{0}^{+}(2)) = \mathbb{C}\left(E_{4}^{(2^{+})}\right)^{2}\oplus \mathbb{C}\Delta_{2}.
\end{align}
This gives us two choices for $\omega_{8}(\tau)$ and the $q$-series expansions of $\tfrac{\omega_{8}(\tau)}{\omega_{6}(\tau)}$ in these two cases read
\begin{align}\label{omega_8/omega_6}
    \begin{split}
        \frac{\omega_{8}(\tau)}{\omega_{6}(\tau)} =& \mu\left(\frac{\left(E_{4}^{(2^{+})}\right)^{2}}{E_{6}^{(2^{+})}}\right)(\tau)\\
        =& \mu\left(1 + 152 q + 14360 q^2 + 1229408 q^3 + 104497176 q^4 + \ldots\right),\\
        \frac{\omega_{8}(\tau)}{\omega_{6}(\tau)} =& \mu\left(\frac{\Delta_{2}}{E_{6}^{(2^{+})}}\right)(\tau)\\
        =& \mu\left(q + 48 q^2 + 4996 q^3 + 403712 q^4 + \ldots\right)
    \end{split}
\end{align}
The MLDE can be rewritten as follows
\begin{align}
    \left[\mathcal{D} + \mu_{1}\left(\frac{\left(E_{4}^{(2^{+})}\right)^{2}}{E_{6}^{(2^{+})}}\right)(\tau) + \mu_{2}\left(\frac{\Delta_{2}}{E_{6}^{(2^{+})}}\right)(\tau)
    \right]f(\tau) = 0.
\end{align}
For $\mu = \tfrac{3}{4}$, we find $m_{1}(\mu) = -114, -3$ for the first and the second choice and higher coefficients continue to be negative. Prior to abandoning the case of $\ell = 3$, let us try to carefully choose the coefficients $\mu_{i}$ so that all the expansion coefficients $m_{i}(\mu_{i})$ turn out to be positive integers. It turns out that all the coefficients $m_{i}(\mu_{i})$, $i= 1,2,\ldots$ fail to be positive integers for positive values of the coefficient $\mu_{2}$. Although this suggests that the solution at $\ell = 3$ proves to be inadmissible, we note that when we take a closer look at  the higher-order coefficients pertaining to the first choice, we observe the following behaviour
\begin{align}
    \begin{split}
        \chi^{(\ell = 3)}(\tau) = q^{-\tfrac{3}{4}}(1 -& 114q + 1113q^{2} + 59614q^{3} + 1989624q^{4} + 75221748q^{5}\\
        +& 3483999169q^{6} + 183921024594q^{7} + 10536468170103q^{8}\\ 
        +& 638465956243752q^{9} + 40328828586484503q^{10} + \ldots),
    \end{split}
\end{align}
\\
i.e. all coefficients $m_{i}(\mu)$ where $i\neq 1$ are positive integers. This qualifies as a type I quasi-character. Now, checking for negative values of $\mu_{2}$, we find that at $\mu_{2} = -192$, we obtain an admissible solution which reads
\begin{align}
    \begin{split}
        \chi^{(\ell = 3)}(\tau) =& j_{2^{+}}^{\tfrac{3}{4}}(\tau)\\
        =& q^{-\tfrac{3}{4}}\left(1 +78 q + 2265 q^2 + 30878 q^3 + 232056 q^4 + \ldots\right).
    \end{split}
\end{align}

\subsection*{\texorpdfstring{$\ell = 4$}{Lg}}
An admissible theory with $c = 24$ possesses the following MLDE
\begin{align}
    \left[\omega_{8}(\tau)\mathcal{D} + \omega_{10}(\tau)\right]f(\tau) = 0.
\end{align}
The space $\mathcal{M}_{10}(\Gamma_{0}^{+}(2))$ is one-dimensional we fix $\omega_{10}(\tau)$ to be the following
\begin{align}
    \begin{split}
    \omega_{10}(\tau) =& \mu E_{10}^{(2^{+})}(\tau) = \frac{\mu}{33}\left(32 E_{6}(2\tau)E_{4}(2\tau) + E_{6}(\tau)E_{4}(\tau)\right)\\
    =& \mu\left(1 - 8 q - 4360 q^2 - 157472 q^3 - 2232584 q^4 + \ldots\right).
    \end{split}
\end{align}
The final form of the MLDE reads
\begin{align}\label{eqn_l=4}
    \left[\mathcal{D} + \mu_{1} \left(\frac{E_{10}^{(2^{+})}}{\left(E_{4}^{(2^{+})}\right)^{2} + \mu_{2}\Delta_{2}}\right)(\tau)\right]f(\tau) = 0,
\end{align}
We shall now consider the cases when the coefficient $\mu_{2}$ is turned off and then turned on. With $\mu_{2} = 0$, the solution now has all integral coefficients that are non-negative that is surprisingly equal to the Hauptmodul $j_{2^{+}}(\tau)$, defined in \ref{Hauptmodul_Gamma_0_2+}, as shown below
\begin{align}
    \begin{split}
    \chi^{(\ell = 4)}(\tau) =& j_{2^{+}}(\tau)\\
    =& q^{-1}\left(1 + 104q + 4372q^{2} + 96256q^{3} + 1240002q^{4} + \ldots\right).
    \end{split}
\end{align}
This has a curious connection to the baby Monster group $2\cdot \mathbb{B}$. The coefficients of the character $\chi^{(\ell = 3)}(\tau)$, or more precisely, the $J$-function of $\Gamma_{0}^{+}(2)$, $J_{2^{+}}(\tau) = j_{2^{+}}(\tau) - 104$, can be expressed as a sum of character degrees of the $2A$ conjugacy class of $2\cdot\mathbb{B}$ as shown below (see table \ref{tab:Baby_Monster_character_degrees})
\begin{align}
    \begin{split}
        &4372 = \chi_{1} + \chi_{2},\ 96256 = \chi_{1} + \chi_{3},\ 124002 = 2\chi_{1} + \chi_{2} + \chi_{3} + \chi_{4},\\
        &10698752 = 2\chi_{1} + \chi_{2} + \chi_{3} + \chi_{4} + \chi_{5},\ldots
    \end{split}
\end{align}
\begin{table}[htb!]
    \centering
    \begin{tabular}{||c|c|c|c|c|c|c|c||}
    \hline
    n & 1 & 2 & 3 & 4 & 5 & 6 & 7\\ [1ex]
    \hline\hline
    $2A$ &  1 & 4371 & 96255 & 1139374 & 9458750 & 9550635 & 63532485\\
    $2C$ & 1 & 275 & 2047 & 8878 & 37950 & 35627 & 134597\\ [1ex]
    \hline
    \end{tabular}
    \caption{Partial character table of the $2A$ and $2C$ conjugacy classes of the baby Monster group $2\cdot \mathbb{B}$.}
    \label{tab:Baby_Monster_character_degrees}
\end{table}
\\
It was found in \cite{Hampapura-Mukhi} that a $(n,\ell) = (3,0)$  RCFT with data $(c,\Delta_{1},\Delta_{2},\Delta_{3}) = (\tfrac{47}{2},0,\tfrac{31}{16},\tfrac{3}{2})$ was related to the baby Monster. These characters have the following $q$-series expansions
\begin{align}\label{Mukhi_characters}
    \begin{split}
        \chi_{0}(\tau) =& q^{-\tfrac{47}{48}}\left(1 + 96256q^{2}  + 9646891q^{3} +  366845011q^{4} + \ldots\right),\\
        \chi_{1}(\tau) =& q^{\tfrac{25}{48}}\left(4371 + 1143745q + 64680601q^{2} + 1829005611q^{3} + \ldots\right),\\
        \chi_{2}(\tau) =& q^{\tfrac{23}{24}}\left(96256 + 110602496q + 420831232q^{2} + 96854952512q^{3} + \ldots\right),
    \end{split}
\end{align}
where the coefficients can be expressed in terms of the character degrees of the $2A$ conjugacy class of $2\cdot\mathbb{B}$. Linear combinations of these characters were found to be equal to the characters of the baby Monster vertex algebra $\chi_{V\mathbb{B}^{\sharp}}(\tau)$. Bae et al. \cite{Bae_Monster_anatomy} showed that a bilinear relation of the twinned characters of \ref{Mukhi_characters} with the $2C$ conjugacy class of $2\cdot \mathbb{B}$ and the Ising characters precisely yields the Hauptmodul of $\Gamma_{0}^{+}(2)$. The twinning operation can be understood via an elementary example. Consider the character $\chi_{1}(\tau)$ in \ref{Mukhi_characters}, let the $2C$ twin be $\Tilde{\chi}^{2C}_{1}(\tau)$ which is constructed by mimicking the way the coefficients of $\chi_{1}(\tau)$ are formed out of the character degrees, i.e. (see table \ref{tab:Baby_Monster_character_degrees})
\begin{align}
    \begin{split}
        \chi_{1}(\tau) =& q^{\tfrac{25}{48}}\left(\underbrace{4371}_{\chi^{2A}_{2}} + \underbrace{1143745}_{\chi^{2A}_{2} + \chi^{2A}_{4}}\cdot q + \underbrace{64680601}_{2\chi^{2A}_{1} + \chi^{2A}_{4} + \chi^{2A}_{7}}\cdot q^{2} + \ldots\right)\\
        \Tilde{\chi}^{2C}_{1}(\tau) =& q^{\tfrac{25}{48}}\left(\underbrace{275}_{\chi^{2C}_{2}} + \underbrace{9153}_{\chi^{2C}_{2} + \chi^{2C}_{4}}\cdot q + \underbrace{144025}_{2\chi^{2C}_{1} + \chi^{2C}_{4} + \chi^{2C}_{7}}\cdot q^{2} + \ldots\right).
    \end{split}
\end{align}
The bilinear combination that yields $j_{2^{+}}(\tau)$ reads
\begin{align}\label{bilinear_2A}
    \chi^{(\ell = 4)}(\tau) = j_{2^{+}}(\tau) =  \chi_{0}(\tau)\cdot\Tilde{\chi}^{2C}_{0}(\tau) + \chi_{\tfrac{1}{2}}(\tau)\cdot \Tilde{\chi}^{2C}_{1} + \chi_{\tfrac{1}{16}}(\tau)\cdot \Tilde{\chi}^{2C}_{2}(\tau).
\end{align}
Now, with $\mu_{2} = x$, we find the order $\mathcal{O}(1)$ coefficient $m_{1} = 104 + x$ where the number can be traced from the pole in \ref{eqn_l=4} at
\begin{align}
    \mu_{2} = -\left(\frac{\left(E_{4}^{(2^{+})}\right)^{2}}{\Delta_{2}}\right)(\tau) = -j_{2^{+}}(\tau) = -q^{-1} - 104 - \ldots
\end{align}
Hence, this tells us that for any $\mu_{2}\geq-104$, we have an admissible solution.

\subsection*{\texorpdfstring{$\ell = 5$}{Lg}}
An admissible theory with $c = 30$ possesses the following MLDE
\begin{align}
    \left[\omega_{10}(\tau)\mathcal{D} + \omega_{12}(\tau)\right]f(\tau) = 0.
\end{align}
The space $\mathcal{M}_{12}(\Gamma_{0}^{+}(2))$ is two-dimensional with the decomposition
\begin{align}
    \mathcal{M}_{12}(\Gamma_{0}^{+}(2)) = E^{(2^{+})}_{4}\left(\mathbb{C}\left(E^{(2^{+})}_{4}\right)^{2}\oplus \mathbb{C}\Delta_{2}\right).
\end{align}
With this we have two choices for the ratio $\tfrac{\omega_{12}(\tau)}{\omega_{10}(\tau)}$ a shown below
\begin{align}
    \begin{split}
    \frac{\omega_{12}(\tau)}{\omega_{10}(\tau)} =& \mu \left(\frac{\left(E_{4}^{(2^{+})}\right)^{3}}{E_{10}^{(2^{+})}}\right)(\tau),\\
    \frac{\omega_{12}(\tau)}{\omega_{10}(\tau)} =& \mu\left(\frac{E_{4}^{(2^{+})}\Delta_{2}}{E_{10}^{(2^{+})}}\right)(\tau).
    \end{split}
\end{align}
These precisely correspond to the choices for the ratio $\tfrac{\omega_{8}(\tau)}{\omega_{6}(\tau)}$ mentioned in \ref{omega_8/omega_6}. The MLDE in this case reads
\begin{align}
    \left[\mathcal{D} + \mu_{1}\left(\frac{\left(E_{4}^{(2^{+})}\right)^{3}}{E_{10}^{(2^{+})}}\right)(\tau) + \mu_{2}\left(\frac{E_{4}^{(2^{+})}\Delta_{2}}{E_{10}^{(2^{+})}}\right)(\tau)\right]f(\tau) = 0.
\end{align}
From Riemann-Roch, we fix $\mu_{1} = \tfrac{5}{4}$. Now, to fix $\mu_{2}$, let us assume the ansatz $\chi(\tau) = j_{2^{+}}^{\tfrac{5}{4}}(\tau)$ that follows from the trend of admissible solutions we have observed thus far. From the $q$-series expansion, we have $m_{1} = 130$, and we find the following equation at order $\mathcal{O}(1)$,
\begin{align}
    190 + \mu_{2} + \left(\frac{9}{4} + \mu_{1}\right)m_{1} = 0.
\end{align}
Solving this helps us fix $\mu_{2} = -320$. Thus, the admissible solution at $\ell = 5$ reads
\begin{align}
    \begin{split}
        \chi^{(\ell = 5)}(\tau) =& j_{2^{+}}^{\tfrac{5}{4}}(\tau)\\
        =& q^{-\tfrac{5}{4}}\left(1 + 130 q + 7155 q^2 + 218470 q^3 + 4122705 q^4 + \ldots\right).
    \end{split}
\end{align}

\subsection*{\texorpdfstring{$\ell = 6$}{Lg}}
An admissible theory with $c = 36$ possesses the following MLDE
\begin{align}
    \left[\omega_{12}(\tau)\mathcal{D} + \omega_{14}(\tau)\right]f(\tau) = 0.
\end{align}
The space $\mathcal{M}_{14}(\Gamma_{0}^{+}(2))$ is two-dimensional with the decomposition
\begin{align}
    \mathcal{M}_{14}(\Gamma_{0}^{+}(2)) = E^{(2^{+})}_{6}\left(\mathbb{C}\left(E^{(2^{+})}_{4}\right)^{2}\oplus \mathbb{C}\Delta_{2}\right).
\end{align}
With this we have three unique choices for the ratio $\tfrac{\omega_{14}(\tau)}{\omega_{12}(\tau)}$ a shown below
\begin{align}\label{choices}
    \begin{split}
        \frac{\omega_{14}(\tau)}{\omega_{12}(\tau)} =& \mu\left(\frac{E_{6}^{(2^{+})}}{E_{4}^{(2^{+})}}\right)(\tau),\\
        =& \mu\left(1 - 104 q + 2072 q^2 - 49568 q^3 + 1146904 q^4 + \ldots\right),\\
        \frac{\omega_{14}(\tau)}{\omega_{12}(\tau)} =& \mu\left(\frac{E_{6}^{(2^{+})}\Delta_{2}}{\left(E_{4}^{(2^{+})}\right)^{3}}\right)(\tau)\\
        =& \mu\left(q - 208 q^2 + 19332 q^3 - 1246976 q^4 + \ldots\right),\\
        \frac{\omega_{14}(\tau)}{\omega_{12}(\tau)} =& \mu\left(\frac{E_{6}^{(2^{+})}\left(E_{4}^{(2^{+})}\right)^{2}}{\Delta_{2}}\right)(\tau)\\
        =& \mu\left(q^{-1} + 48 - 3748 q - 401024 q^2 - 15683478 q^3 - 347352704 q^4 + \ldots\right).
    \end{split}
\end{align}
Also, we note that we can consider the choice with $\omega_{12} = \omega_{6}^{2} = E_{6}^{2}$ in the first and second expressions above. Including all of these in the MLDE and performing a series expansion with the ansatz $\chi(\tau) = j_{2^{+}}^{\tfrac{3}{2}}(\tau)$, we first see that the solution takes the form, $\chi(\tau) = q^{-\tfrac{3}{2}}\left(\tfrac{3}{2}\mu_{3}q^{-1} + 1 + \ldots\right)$,  where $\mu_{3}$ is the coefficient corresponding to the third choice in \ref{choices}. This isn't the correct form of the character ansatz (see \ref{correct_ansatz}). Hence, we set $\mu_{3} = 0$. Next, if we were to include the choice $\mu E_{6}^{(2^{+})}\Delta_{2}/\left(E_{6}^{(2^{+})}\right)^{2}$, we find that the solution reads, $\chi(\tau) = q^{-\tfrac{3}{2}}\left(1 +\tfrac{3}{2}\mu + \ldots\right)$, which tells us that we have to set $\mu=0$ in order to have a unique vacuum. The MLDE now takes the following form
\begin{align}
    \left[\mathcal{D} + \mu_{1}\left(\frac{E_{6}^{(2^{+})}}{E_{4}^{(2^{+})}}\right)(\tau) + \mu_{2}\left(\frac{E_{6}^{(2^{+})}\Delta_{2}}{\left(E_{4}^{(2^{+})}\right)^{3}}\right)(\tau) + \mu_{3}\left(\frac{E_{6}^{(2^{+})}E_{4}^{(2^{+})}}{\left(E_{6}^{(2^{+})}\right)^{2}}\right)(\tau)\right]f(\tau) = 0.
\end{align}
Performing a series expansion gives us the following equation at order $\mathcal{O}(q^{-\tfrac{1}{2}})$,
\begin{align}
    156 + \frac{3}{2}\left(\mu_{2} + \mu_{3} - 104\right) = 0.
\end{align}
This gives us the relation $\mu_{2} = -\mu_{3}$. Using this at the next order, we find $\mu_{2} =0$. Hence, with $\mu_{2} = \mu_{3} = 0$ and with $\mu_{1} = \tfrac{3}{2}$, we obtain a solution that is equal to the Hauptmodul raised to the power $\tfrac{3}{2}$, i.e.
\begin{align}
    \begin{split}
    \chi^{(\ell = 6)}(\tau) =& j_{2^{+}}^{\tfrac{3}{2}}(\tau)\\
    =& q^{-\tfrac{3}{2}}\left(1 + 156q + 10614q^{2} + 415096q^{3} + 10411305q^{4} + \ldots\right).
    \end{split}
\end{align}
Here we note that the parameters we have disregarded, namely $\mu_{2}$ and $\mu_{3}$, are associated with the cusp form $\Delta_{2}(\tau)$. It turns out that including these forms in the MLDE renders the solutions inadmissible. This is a feature we will see repeating throughout the paper. We can now express the $\Gamma_{0}^{+}(2)$ characters for $\ell\leq 6$ as follows 
\begin{align}
    \begin{split}
    &\chi(\tau) = j_{2^{+}}^{w_{\rho}}(\tau) + x\delta_{\ell,4},\\
    &c = 24w_{\rho} = 6\ell,
    \end{split}
\end{align}
where $w_{\rho}\in\left\{0,\tfrac{1}{4},\tfrac{1}{2}, \tfrac{3}{4}, 1,\tfrac{5}{4},\tfrac{3}{2}\right\}$ corresponding to the characters at $\ell \in\{0,1,2,3,4,5,6\}$ respectively, and $x\geq -104$.

\section{\texorpdfstring{$\mathbf{\Gamma_{0}^{+}(3)}$}{Γ0(3)+}}\label{sec:Gamma_0_3+}
The Fricke group of level $3$ is generated by the $T$-matrix and the Atkin-Lehner involution of level $3$, $W_{3} = \left(\begin{smallmatrix}0 & -\tfrac{1}{\sqrt{3}}\\ \sqrt{3} & 0\end{smallmatrix}\right)$, i.e. $\Gamma_{0}^{+}(3) = \langle\left(\begin{smallmatrix}1 & 1\\ 0 & 1\end{smallmatrix}\right), W_{3}\rangle$. A non-zero $f\in\mathcal{M}_{k}^{!}(\Gamma_{0}^{+}(3))$ satisfies the following valence formula
\begin{align}
    \nu_{\infty}(f) + \frac{1}{2}\nu_{\tfrac{i}{\sqrt{3}}}(f) + \frac{1}{4}\nu_{\rho_{3}}(f) + \sum\limits_{\substack{p\in\Gamma_{0}^{+}(3)\backslash\mathbb{H}^{2}\\ p\neq \tfrac{i}{\sqrt{3}},\rho_{3}}}\nu_{p}(f) =   \frac{k}{6},
\end{align}
where $\tfrac{i}{\sqrt{3}}$ and $\rho_{3} = -\tfrac{1}{2}+\tfrac{i}{2\sqrt{3}}$ are elliptic points and this group has a cusp at $\tau = \infty$ (see \cite{Junichi} for more details). For $k>2$ and $k\in2\mathbb{Z}$, the dimension of the space of modular forms reads
\begin{align}
    \begin{split}
        \text{dim}\ \mathcal{S}_{k}(\Gamma_{0}^{+}(3)) = \begin{cases}\left\lfloor\left.\frac{k}{6}\right\rfloor\right. - 1,\ &k \equiv 2,6\ (\text{mod}\ 12)\\
        \left\lfloor\left.\frac{k}{6}\right\rfloor\right.,\ &k\not\equiv2,6\ (\text{mod}\ 12).
        \end{cases},\\
        \text{dim}\ \mathcal{M}_{k}(\Gamma_{0}^{+}(3)) = 1 + \mathcal{S}_{k}(\Gamma_{0}^{+}(3)).
    \end{split}
\end{align}
Using theorem \ref{thm:genus_Fricke_index},with $(N,h,m)=(3,1,3)$, we have
\begin{align}
    \chi(\Gamma_{0}^{+}(3)) = \frac{3}{6}\prod\limits_{p\vert 3}\frac{p+1}{2p} = \frac{1}{3},\ \ 
    \left[\Gamma_{0}^{+}(3):\Gamma_{0}(3)\right] = 2,
\end{align}
i.e. $\Gamma_{0}(3)$ is an index-$2$ subgroup of $\Gamma_{0}^{+}(3)$. With this, we find the index of $\Gamma_{0}^{+}(3)$ in $\text{SL}(2,\mathbb{Z})$ to be
\begin{align}
    \left[\text{SL}(2,\mathbb{Z}):\Gamma_{0}^{+}(3)\right] = \frac{\left[\text{SL}(2,\mathbb{Z}):\Gamma_{0}(3)\right]}{\left[\Gamma^{+}_{0}(3):\Gamma_{0}(3)\right]} = \frac{4}{2} = 2.
\end{align}
Hence, we find the number of zeros in the fundamental domain $\mathcal{F}_{3^{+}}$ to be $\# = 2\ell\cdot \tfrac{\mu}{12} = \tfrac{1}{3}\ell$. The Riemann-Roch theorem takes the following form
\begin{align}
    \begin{split}
        \sum\limits_{i=0}^{n-1}\alpha_{i} =& -\frac{nc}{24} + \sum\limits_{i}\Delta_{i}\\
        =& \frac{1}{6}n(n-1) - \frac{1}{3}\ell.
    \end{split}
\end{align}
From this for $n = 1$, we find the following relation between the number of zeros in the fundamental domain and the central charge
\begin{align}
    c = 8\ell.
\end{align}
The space of modular forms of $\Gamma_{0}^{+}(3)$ has the following basis decomposition
\begin{align}\label{basis_decomposition_Gamma_0_3+}
    \begin{split}
    \mathcal{M}_{k}(\Gamma_{0}^{+}(3)) =& E_{\overline{k},3^{'}}\left[\mathbb{C}\left(\left(E_{2,3^{'}}\right)^{3}\right)^{n} \oplus \mathbb{C}\left(\left(E_{2,3^{'}}\right)^{3}\right)^{n-1}\Delta_{3}\oplus \ldots\oplus \mathbb{C}\left(\Delta_{3}\right)^{n}\right],\\
    n =& \text{dim}\ \mathcal{M}_{k}(\Gamma_{0}^{+}(3)) - 1,
    \end{split}
\end{align}
where 
\begin{align}
    \begin{split}
        E_{\overline{k},3^{'}}\equiv \begin{cases}
        1,\ &k\equiv 0\ (\text{mod}\ 12),\\
        \left(E_{2,3^{'}}\right)^{2}E_{4,3^{'}},\ &k\equiv 2\ (\text{mod}\ 12),\\
        \left(E_{2,3^{'}}\right)^{2},\ &k\equiv 4\ (\text{mod}\ 12),\\
        E_{2,3^{'}}E_{4,3^{'}},\ &k\equiv 6\ (\text{mod}\ 12),\\
        E_{2,3^{'}},\ &k\equiv 8\ (\text{mod}\ 12),\\
        E_{4,3^{'}},\ &k\equiv 10\ (\text{mod}\ 12).,
        \end{cases}
    \end{split}
\end{align}
and
\begin{align}
   \begin{split}
       E_{2,3^{'}}(\tau)\equiv& \frac{3E_{2}(3\tau) - E_{2}(\tau)}{2}\in\mathcal{M}_{2}(\Gamma_{0}(3)),\\
       E_{k}^{(3^{+})}(\tau)\equiv& \frac{3^{\tfrac{k}{2}}E_{k}(3\tau) + E_{k}(\tau)}{3^{\tfrac{k}{2}} + 1},\ k\geq 4,\\
       E_{4,3^{'}}(\tau) \equiv& \left(\frac{E_{6}^{(3^{+})}}{E_{2,3^{'}}}\right)(\tau),\\
       \Delta_{3}(\tau)\equiv& \left(\eta(\tau)\eta(3\tau)\right)^{6}.
   \end{split} 
\end{align}
The forms $E_{4,3^{'}}(\tau)$ and $\Delta_{3}(\tau)$ are weight $4$ and weight $6$ semi-modular forms of $\Gamma_{0}^{+}(3)$ respectively. The Hauptmodul of $\Gamma_{0}^{+}(3)$ is defined as follows
\begin{align}
    \begin{split}
        j_{3^{+}}(\tau) \equiv& \left(\frac{\left(E_{2,3^{'}}\right)^{3}}{\Delta_{3}}\right)(\tau) = \left(\left(\frac{\eta(\tau)}{\eta(3\tau)}\right)^{6} + 3^{3}\left(\frac{\eta(3\tau)}{\eta(\tau)}\right)^{6}\right)^{2}\\
        =& q^{-1} + 42 + 783q + 8672q^{2} + 65367q^{3} + \ldots
    \end{split}
\end{align}

\subsection{Single character solutions}
\subsection*{\texorpdfstring{$\ell = 0$}{Lg}}
Let us consider the simple case with no poles $\ell = 0$ and a single character. Since the spaces $\mathcal{M}_{0}(\Gamma_{0}^{+}(3))$ and $\mathcal{M}_{2}(\Gamma_{0}^{+}(3))$ are one-dimensional, we choose $\omega_{0}(\tau) = 1 = \omega_{2}(\tau)$. Plugging in the behaviour of $f_{0}(\tau)$, we find $\mu = \tfrac{c}{24} = 0$. Hence, the only trivial solution is $f_{0}(\tau) = 1$. 

\subsection*{\texorpdfstring{$\ell = 1$}{Lg}}
An admissible theory would have a central charge $c = 8$. Since the space $\mathcal{M}_{4}(\Gamma_{0}^{+}(3))$ is one-dimensional using \ref{basis_decomposition_Gamma_0_3+} we make the following choice for $\omega_{4}(\tau)$,
\begin{align}
    \begin{split}
        \omega_{4}(\tau) =& \mu\left(E_{2,3^{'}}\right)^{2}(\tau)\\
        =& \mu\left(1 + 24 q + 216 q^2 + 888 q^3 + 1752 q^4 + \ldots\right).
    \end{split}
\end{align}
Now, to build $\tfrac{\omega_{4}(\tau)}{\omega_{2}(\tau)}$, which is a modular form of weight two, we consider ratios of higher weight modular forms. We find the following unique choice
\begin{align}
    \begin{split}
        \frac{\omega_{4}(\tau)}{\omega_{2}(\tau)} =& \mu\left(\frac{E_{4,3^{'}}}{E_{2,3^{'}}}\right)(\tau)\\
        =& \mu\left(1 - 42 q + 198 q^2 - 1446 q^3 + 8454 q^4 + \ldots\right)(\tau),
    \end{split}
\end{align}
where $\mu = \tfrac{1}{3}$. The final form of the MLDE reads
\begin{align}
    \left[\Tilde{\partial} + \mu\left(\frac{E_{4,3^{'}}}{E_{2,3^{'}}}\right)(\tau)\right]f(\tau) = 0.
\end{align}
The solution can be expressed in terms of Hauptmodul $j_{3^{+}}(\tau)$ as follows
\begin{align}
    \begin{split}
    \chi^{(\ell = 1)}(\tau) =& j_{3^{+}}^{\tfrac{1}{3}}(\tau)\\
    =& q^{-\tfrac{1}{3}}\left(1  +14 q + 65 q^2 + 156 q^3 + 456 q^4 + \ldots\right).
    \end{split}
\end{align}

\subsection*{\texorpdfstring{$\ell = 2, 4, 5$}{Lg}}
Redoing the calculation at $\ell = 2, 4, 5$, we find that the unique ratio we obtain from the space of modular forms in each case is $E_{4,3^{'}}/E_{2,3^{'}}$, and this yields the following admissible solutions corresponding to central charges $c = 16, 32, 40$ respectively,
\begin{align}
    \begin{split}
    \chi^{(\ell =2)}(\tau) =& j_{3^{+}}^{\tfrac{2}{3}}(\tau)\\
    =& q^{-\tfrac{2}{3}}\left(1 + 28 q + 326 q^2 + 2132 q^3 + 9505 q^4 + \ldots\right),\\
    \chi^{(\ell = 4)}(\tau) =& j_{3^{+}}^{\tfrac{4}{3}}(\tau)\\
    =& q^{-\tfrac{4}{3}}\left(1 + 56 q + 1436 q^2 + 22520 q^3 + 244678 q^4 + \ldots\right),\\
    \chi^{(\ell =5)}(\tau) =& j_{3^{+}}^{\tfrac{5}{3}}(\tau)\\
    =& q^{-\tfrac{5}{3}}\left(1 + 70 q + 2285 q^2 + 46420 q^3 + 662490 q^4 + \ldots\right).
    \end{split}
\end{align}

\subsection*{\texorpdfstring{$\ell = 3$}{Lg}}
An admissible theory with central charge $c = 24$ and since the spaces $\mathcal{M}_{6}(\Gamma_{0}^{+}(3))$ and $\mathcal{M}_{8}(\Gamma_{0}^{+}(3))$ are one- and two-dimensional respectively, we obtain the following choices 
\begin{align}
    \begin{split}
        \frac{\omega_{8}(\tau)}{\omega_{6}(\tau)} =& \mu_{1}\left(\frac{\left(E_{2,3^{'}}\right)^{3}}{E_{4,3^{'}}}\right)(\tau),\\
        \frac{\omega_{8}(\tau)}{\omega_{6}(\tau)} =& \mu_{2}\left(\frac{\Delta_{3}}{E_{2,3^{'}}E_{4,3^{'}}}\right)(\tau),
    \end{split}
\end{align}
where $\mu_{1} = 1$. Assuming the ansatz $\chi(\tau) = j_{3^{+}}(\tau)$, a $q$-series expansion yields the equation $-66+\mu_{2}+m_{1}=0$. Now, from the series expansion of $j_{3^{+}}(\tau)$, we see that $m_{1} = 42$ and this sets $\mu_{2} = -108$. Thus, for any $\mu_{2}\geq -108$, we have an admissible solution.

\subsection*{\texorpdfstring{$\ell = 6$}{Lg}}
An admissible theory would have a central charge $c = 48$. Since the spaces $\mathcal{M}_{12}(\Gamma_{0}^{+}(3))$ and $\mathcal{M}_{14}(\Gamma_{0}^{+}(3))$ are both two-dimensional, we obtain four unique choices for the ratio $\tfrac{\omega_{14}(\tau)}{\omega_{12}(\tau)}$ and performing an analysis done previously for the case $\ell = 6$ for $\Gamma_{0}^{+}(2)$, we find that the only ratio that yields non-negative integer coefficients is
\begin{align}
    \begin{split}
        \frac{\omega_{14}(\tau)}{\omega_{12}(\tau)} =& \mu\left(\frac{E_{4,3^{'}}}{E_{2,3^{'}}}\right)(\tau),
    \end{split}
\end{align}
The solution to the MLDE is equal to the square of the Hauptmodul $j_{3^{+}}(\tau)$, i.e.
\begin{align}
    \begin{split}
        \chi^{(\ell = 4)}(\tau) =& j_{3^{+}}^{2}(\tau)\\
        =& q^{-2}\left(1 + 84 q + 3330 q^2 + 83116 q^3 + 1472271 q^4 + \ldots\right).
    \end{split}
\end{align}
We can now express the $\Gamma_{0}^{+}(3)$ characters for $\ell\leq 6$ as follows
\begin{align}
    \begin{split}
        &\chi(\tau) = j^{w_{\rho}}_{3^{+}}(\tau) + x\delta_{\ell,3},\\
        &c = 24w_{\rho} = 8\ell,
    \end{split}
\end{align}
where $w_{\rho}\in\left\{0,\tfrac{1}{3}, \tfrac{2}{3}, 1, \tfrac{4}{3},\tfrac{5}{3}, 2\right\}$ corresponding to the characters at $\ell \in\{0,1,2,3,4,5,6\}$ respectively, and $x\geq -108$.

\section{\texorpdfstring{$\mathbf{\Gamma_{0}^{+}(5)}$}{Γ0(5)+}}\label{sec:Gamma_0_5+}
The Fricke group of level $5$ is generated by the $T$-matrix, the Atkin-Lehner involution of level $5$, $W_{5} = \left(\begin{smallmatrix}0 & -\tfrac{1}{\sqrt{5}}\\ \sqrt{5} & 0\end{smallmatrix}\right)$, and $S^{-1}T^{-2}S^{-1}T^{-3}S$ i.e. $\Gamma_{0}^{+}(5) = \langle\left(\begin{smallmatrix}1 & 1\\ 0 & 1\end{smallmatrix}\right), W_{5}, \left(\begin{smallmatrix} & 1\\ 5 & 2\end{smallmatrix}\right)\rangle$. A non-zero $f\in\mathcal{M}_{k}^{!}(\Gamma_{0}^{+}(5))$ satisfies the following valence formula
\begin{align}
    \nu_{\infty}(f) + \frac{1}{2}\nu_{\tfrac{i}{\sqrt{5}}}(f) + \frac{1}{2}\nu_{\rho_{5,1}}(f) + \frac{1}{2}\nu_{\rho_{5,2}}(f) + \sum\limits_{\substack{p\in\Gamma_{0}^{+}(5)\backslash\mathbb{H}^{2}\\ p\neq \tfrac{i}{\sqrt{5}},\rho_{5,1}, \rho_{5,2}}}\nu_{p}(f) =   \frac{k}{4},
\end{align}
where $\tfrac{i}{\sqrt{5}}$, $\rho_{5,1} = -\tfrac{1}{2}+\tfrac{i}{2\sqrt{5}}$, and $\rho_{5,2} = \tfrac{-2 + i}{5}$ are elliptic points and this group has a cusp at $\tau = \infty$ (see \cite{Junichi} for more details). 
Using theorem \ref{thm:genus_Fricke_index}, with $(N,h,m)=(5,1,5)$, we have
\begin{align}
    \chi(\Gamma_{0}^{+}(5)) = \frac{5}{6}\prod\limits_{p\vert 5}\frac{p+1}{2p} = \frac{1}{2},\ \ 
    \left[\Gamma_{0}^{+}(5):\Gamma_{0}(5)\right] = 2,
\end{align}
i.e. $\Gamma_{0}(5)$ is an index-$2$ subgroup of $\Gamma_{0}^{+}(5)$. With this, we find the index of $\Gamma_{0}^{+}(5)$ in $\text{SL}(2,\mathbb{Z})$ to be
\begin{align}
    \left[\text{SL}(2,\mathbb{Z}):\Gamma_{0}^{+}(5)\right] = \frac{\left[\text{SL}(2,\mathbb{Z}):\Gamma_{0}(5)\right]}{\left[\Gamma^{+}_{0}(5):\Gamma_{0}(5)\right]} = \frac{6}{2} = 3.
\end{align}
Hence, we find the number of zeros in the fundamental domain $\mathcal{F}_{5^{+}}$ to be $\# = 2\ell\cdot \tfrac{\mu}{12} = \tfrac{1}{2}\ell$. The Riemann-Roch theorem takes the following form
\begin{align}
    \begin{split}
        \sum\limits_{i=0}^{n-1}\alpha_{i} =& -\frac{nc}{24} + \sum\limits_{i}\Delta_{i}\\
        =& \frac{1}{4}n(n-1) - \frac{1}{2}\ell.
    \end{split}
\end{align}
From this for $n = 1$, we find the following relation between the number of zeros in the fundamental domain and the central charge
\begin{align}
    c = 12\ell.
\end{align}
The space of modular forms of $\Gamma_{0}^{+}(5)$ has the following basis decomposition
\begin{align}\label{basis_decomposition_Gamma_0_5+}
    \begin{split}
    \mathcal{M}_{4n}(\Gamma_{0}^{+}(5)) =& \mathbb{C}\left(\left(E_{2,5^{'}}\right)^{2}\right)^{n} \oplus \mathbb{C}\left(\left(E_{2,5^{'}}\right)^{2}\right)^{n-1}\Delta_{5}\oplus \ldots\oplus \mathbb{C}\left(\Delta_{5}\right)^{n},\\
    \mathcal{M}_{4n + 6}(\Gamma_{0}^{+}(5)) =& E_{6,5^{+}}\mathcal{M}_{4n}(\Gamma_{0}^{+}(5)),
    \end{split}
\end{align}
where 
\begin{align}
    \begin{split}
        E_{2,5^{'}}(\tau) =& \frac{5E_{2,5^{'}}(5\tau) - E_{2}(\tau)}{4}\in\mathcal{M}_{2}(\Gamma_{0}(5))\\
        =& 1 + 6 q + 18 q^2 + 24 q^3 + 42 q^4 + \ldots\\
        \Delta_{5}(\tau) =& \left(\eta(\tau)\eta(5\tau)\right)^{4}\in\mathcal{S}_{4}(\Gamma_{0}^{+}(5))\\
        =& q - 4 q^2 + 2 q^3 + 8 q^4 + \ldots\\
        E_{6,5^{+}}(\tau) =& \frac{5^{\tfrac{6}{2}}E_{6}(5\tau) + E_{6}(\tau)}{5^{\tfrac{6}{2}} + 1}\in\mathcal{M}_{6}(\Gamma_{0}^{+}(5))\\
        =& 1 - 4 q - 132 q^2 - 976 q^3 - 4228 q^4 + \ldots
    \end{split}
\end{align}
The Hauptmodul of $\Gamma_{0}^{+}(5)$ is defined as follows
\begin{align}
    \begin{split}
        j_{5^{+}}(\tau) =& \left(\tfrac{\left(E_{2,5^{'}}\right)^{2}}{\Delta_{5}}\right)(\tau)\\
        =& q^{-1} + 16 + 134 q + 760 q^2 + 3345 q^3 + 12256 q^4 + \ldots
    \end{split}
\end{align}

\subsection{Single character solutions}
\subsection*{\texorpdfstring{$\ell = 0$}{Lg}}
Consider the trivial case of $\ell =0$. Since there are no modular forms of weight $0$ and $2$ on $\Gamma_{0}^{+}(5)$, we choose $\omega_{0}(\tau) = 1 = \omega_{2}(\tau)$. Plugging in the behaviour of $f_{0}(\tau)$, we find $\mu = \tfrac{c}{24} = 0$. Hence, the only trivial solution is $f_{0}(\tau) = 1$.

\subsection*{\texorpdfstring{$\ell = 2$}{Lg}}
An admissible theory would have a central charge $c = 24$. We make the following choice for the ratio $\tfrac{\omega_{6}(\tau)}{\omega_{4}(\tau)}$,
\begin{align}
   \begin{split}
       \frac{\omega_{6}(\tau)}{\omega_{4}(\tau)} =& \mu_{1}\left(\frac{E_{6,5^{+}}}{\left(E_{2,5^{'}}\right)^{2}}\right)(\tau) + \mu_{2}\left(\frac{\Delta_{5}}{\left(E_{2,5^{'}}\right)^{2}}\right)(\tau,
   \end{split} 
\end{align}
where $\mu_{1} = 1$. With the ansatz, $\chi^{(\ell = 2)}(\tau) = j_{5^{+}}(\tau)$ and matching series expansions, we set $\mu_{2} = 0$. Hence, we have 
\begin{align}
        \chi^{(\ell = 4)}(\tau) = j_{5^{+}}(\tau).
\end{align}
Similar to how we expressed the Hauptmodul $j_{2^{+}}(\tau)$ as a bilinear relation in \ref{bilinear_2A}, we can also express $j_{5^{+}}(\tau)$ as a bilinear relation of the Ising characters and the twinned characters of $2\cdot\mathbb{B}$ and the $5A$ conjugacy class of $\mathbb{M}$ \cite{Bae_Monster_anatomy} as follows
\begin{align}
    \chi^{(\ell = 2)}(\tau) = j_{5^{+}}(\tau) = \chi_{0}(\tau)\cdot\Tilde{\chi}_{0}^{5A} + \chi_{\tfrac{1}{16}}(\tau)\cdot\Tilde{\chi}_{1}^{5A}(\tau) + \chi_{\tfrac{1}{2}}(\tau)\cdot\Tilde{\chi}^{5A}_{2}.
\end{align}

\subsection*{\texorpdfstring{$\ell = 4$}{Lg}}
An admissible theory would have a central charge $c = 48$. We make the following choice for the ratio $\tfrac{\omega_{10}(\tau)}{\omega_{8}(\tau)}$,
\begin{align}
   \begin{split}
       \frac{\omega_{10}(\tau)}{\omega_{8}(\tau)} =& \mu\left(\frac{E_{6,5^{+}}}{\left(E_{2,5^{'}}\right)^{2}}\right)(\tau) = \frac{\omega_{6}(\tau)}{\omega_{4}(\tau)}
   \end{split} 
\end{align}
where $\mu = 2$. The solution to the MLDE is equal to the square of the Hauptmodul $j_{5^{+}}(\tau)$, i.e.
\begin{align}
    \begin{split}
        \chi^{(\ell = 4)}(\tau) =& j_{5^{+}}^{2}(\tau)\\
        =& q^{-2}\left(1 + 32 q + 524 q^2 + 5808 q^3 + 48966 q^4 + \ldots\right).
    \end{split}
\end{align}
We note that there are no admissible solutions at $\ell = 1,3,6$. We can now express the $\Gamma_{0}^{+}(5)$ characters for $\ell\leq 6$ as follows
\begin{align}
    \begin{split}
        &\chi(\tau) = j_{5^{+}}^{w_{\rho}}(\tau),\\
        &c = 24w_{\rho} = 12\ell,
    \end{split}
\end{align}
where $w_{\rho}\in\left\{0,1,2\right\}$ corresponding to the three characters at $\ell\in\{0,2,4\}$ respectively.
\section{Level-7 groups}\label{sec:level_7}
We will be using the $\Theta$-function of lattices to describe CFTs. The $q$-series expansion of a $\Theta$-function of a $d$-dimensional lattice $\Lambda$ is a weight $\tfrac{d}{2}$ modular form given by \cite{Monster_extension_24k}
\begin{align}
    \Theta_{\Lambda}(\tau) = \sum\limits_{x\in\Lambda}N(m)q^{m},
\end{align}
where the sum runs over all the vectors $x$ in the lattice $\Lambda$ with length $m=x\cdot x$, $N(m)$ is the number of vectors of norm $m$, and $q \equiv e^{\pi i\tau}$. The partition function of the lattice $\mathcal{Z}$ is defined as follows
\begin{align}\label{partition_theta_lattice}
    \mathcal{Z}(\tau) \equiv \frac{\Theta_{\Lambda}(\tau)}{\eta^{d}(\tau)}.
\end{align}
For an even, self-dual lattice of integer dimension $d$, we can always define a holomorphic modular-covariant CFT with central charge $c$ with the partition function of the theory given by \ref{partition_theta_lattice}. It was shown by Dixon et al. \cite{Dixon} that when the central charge is a multiple of $24$, the partition function $\mathcal{Z}(\tau)$ will be modular invariant, and when the central charge is $c = 8\ (\text{mod}\ 24)$ or $c = 16\ (\text{mod}\ 24)$, $\mathcal{Z}(\tau)$ will be a  modular-covariant function, i.e. a weight $0$ modular form with a non-trivial multiplier which means that a phase of $e^{2\pi i\tfrac{c}{24}}$ is picked up under the modular transformation $\tau\mapsto \tau + 1$. It was also proved that the partition function of an arbitrary holomorphic CFT with central charge $c = 24$ based on the torus $\mathbb{R}^{24}/\Lambda$, where $\Lambda$ could be any of the $24$ even self-dual Niemeier lattices in $24$ dimensions, can be written as follows
\begin{align}
    \mathcal{Z}(\tau) = \frac{\Theta_{\Lambda}(\tau)}{\eta^{24}(\tau)} = J(\tau) + 24(h^{\vee} + 1),
\end{align}
where $J(\tau) = j(\tau) - 744$ and $h^{\vee}$ is the Coxeter number associated to lattice $\Lambda$. In $24$ dimensions, there exist $24$ such self-dual lattices: the Leech lattice $\Lambda_{24}$ which has no roots, and $23$ other Niemeier lattices with their own root system. In the case of the Leech lattice, the Coxeter number is $h^{\vee} = 0$, and hence, the partition function of the Leech CFT is $\mathcal{Z}(\tau) = J(\tau) + 24$ which matches the case with symmetry algebra $U(1)^{12}$ in table \ref{tab:central_candidates}. The monster CFT is constructed from a $\mathbb{Z}_{2}$ orbifold of the Leech CFT.
\\
\begin{table}[htb!]
    \centering
    \begin{tabular}{||c|c|c||}
    \hline
    Central charge $c$ & Partition function $\mathcal{Z}(\tau)$ & Lattice $\Lambda$\\[1ex]
    \hline\hline
    4 & $j^{\tfrac{1}{3}}(\tau)$ & The $\text{E}_{8}$ lattice\\[0.5ex]
    12 & $\substack{\mathcal{Z}_{U}(\tau) = j(\tau) - 720\\ \mathcal{Z}_{V}(\tau) = j(\tau) - 744}$ & Leech lattice $\Lambda_{24}$\\[1ex]
    \hline
    \end{tabular}
    \caption{This table shows candidate partition functions and the lattice that correspond to certain values of central charge. The subscripts $U,V$ in the $c = 12$ case correspond to the symmetry algebras $U(1)^{12}\times U(1)^{12}$ and $\text{Vir}_{12}\times\text{Vir}_{12}$, $j(\tau)$ is the usual $\text{PSL}(2,\mathbb{Z})$ $j$-function. See \cite{Hartman-Rastelli-Mazac} for more details.}
    \label{tab:central_candidates}
\end{table}

\subsection{\texorpdfstring{$\Gamma_{0}(7)$}{Γ0(7)}}
\begin{figure}[htb!]
    \centering
    \includegraphics[width = 15cm]{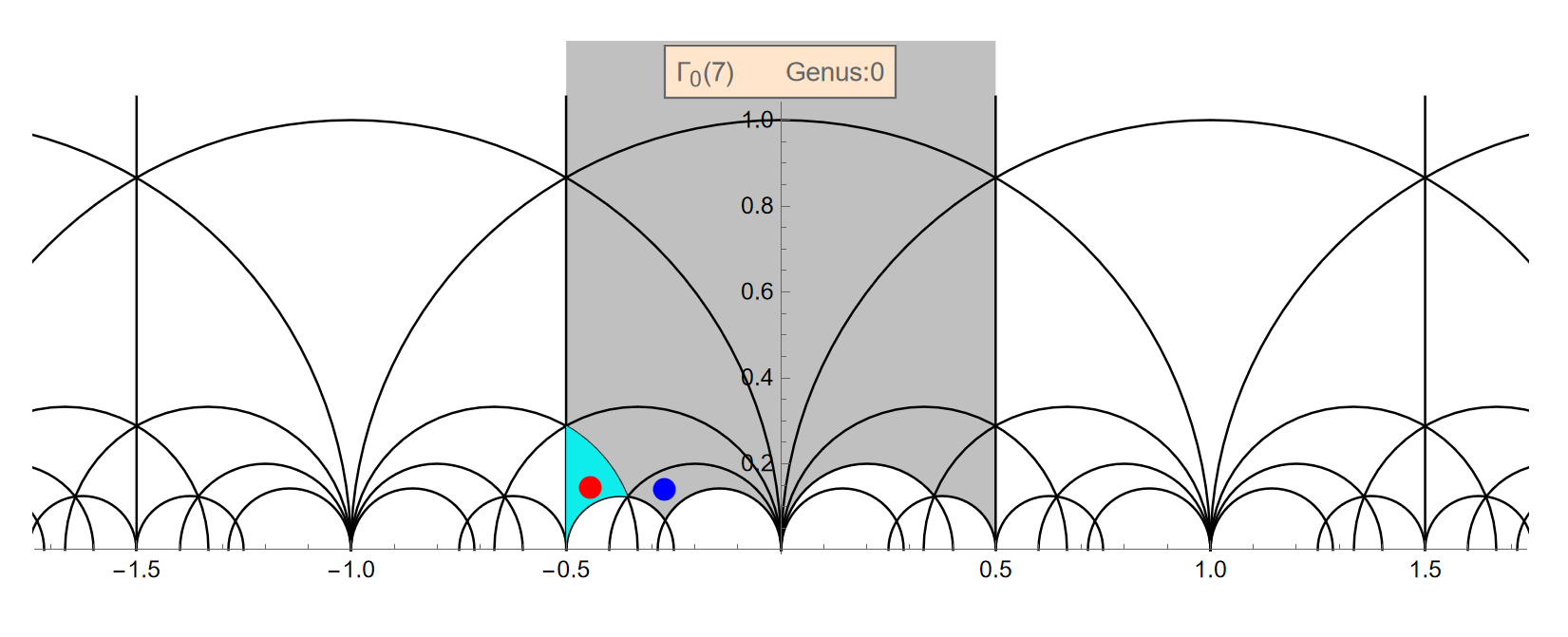}
    \caption{The matrix corresponding to the part of the fundamental domain marked with a blue dot is $ST^{3} = \left(\begin{smallmatrix}0 & -1\\ 1 & 3\end{smallmatrix}\right)$ and the equivalent matrix corresponding to the triangle shaded in turquoise is $S^{-1}T^{2}S^{-1} = \left(\begin{smallmatrix}-1 & 0\\ 2 & -1\end{smallmatrix}\right)$. The two matrices are related by $\left(ST^{3}\right)A_{1} = S^{-1}T^{2} S^{-1}$, where $A_{1} = T^{-1}S^{-3} = \left(\begin{smallmatrix}-1 & -1 \\ 1 & 0\end{smallmatrix}\right)$. A similar identification has been done for the part of the fundamental domain on to the right of the symmetric axis: $\left(ST^{-3}\right)A_{2} = S^{-1}T^{-2}S^{-1}$, where $A_{2} = TS^{-3} = \left(\begin{smallmatrix}1 & -1 \\ 1 & 0\end{smallmatrix}\right)$.}
    \label{fig:coset_rep}
\end{figure}

\begin{figure}
    \centering
    \includegraphics[width = 15 cm]{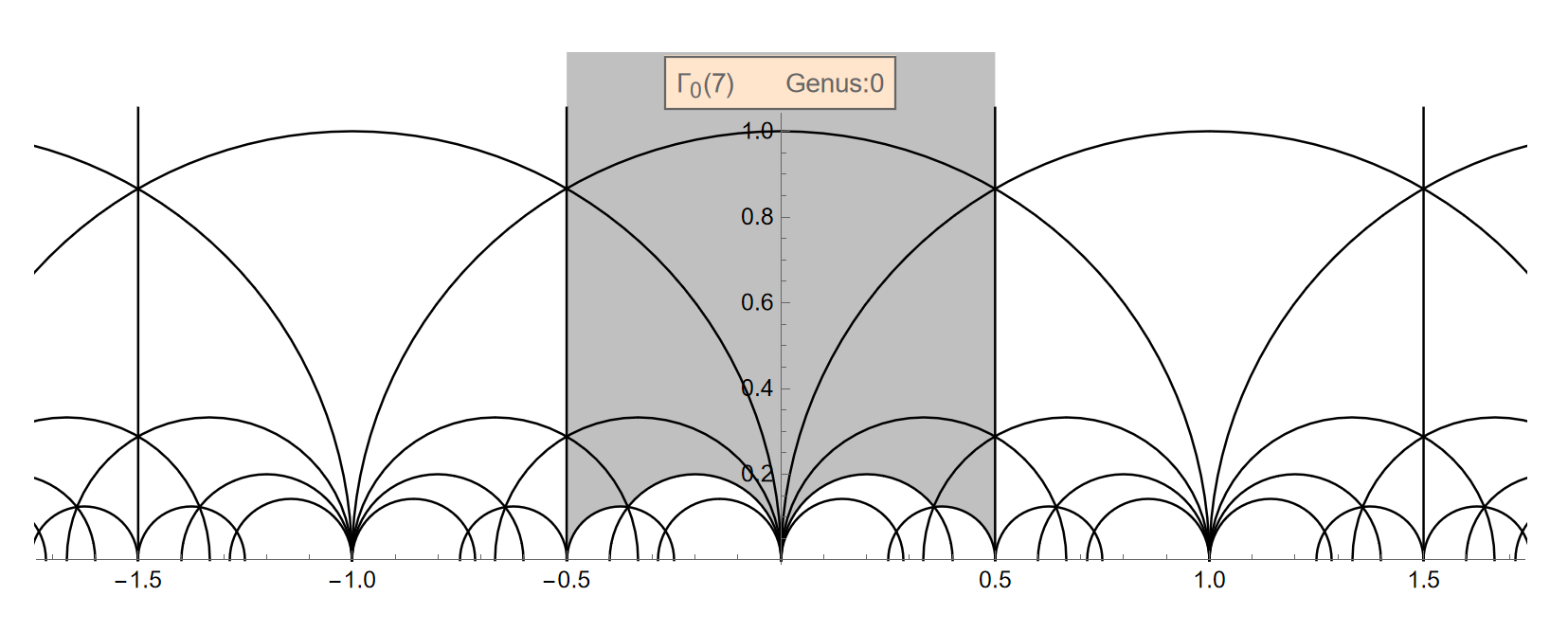}
    \caption{The equivalent fundamental domain of $\Gamma_{0}(7)$ bounded by lines $\text{Re}(\tau) = -\tfrac{1}{2}$, $\text{Re}(\tau) = \tfrac{1}{2}$, and with cusps at $\tau = 0, i\infty$.}
    \label{fig: fundamental_domain_level_7_equiv}
\end{figure}

\begin{figure}[htb!]
    \centering
    \includegraphics[width = 14cm]{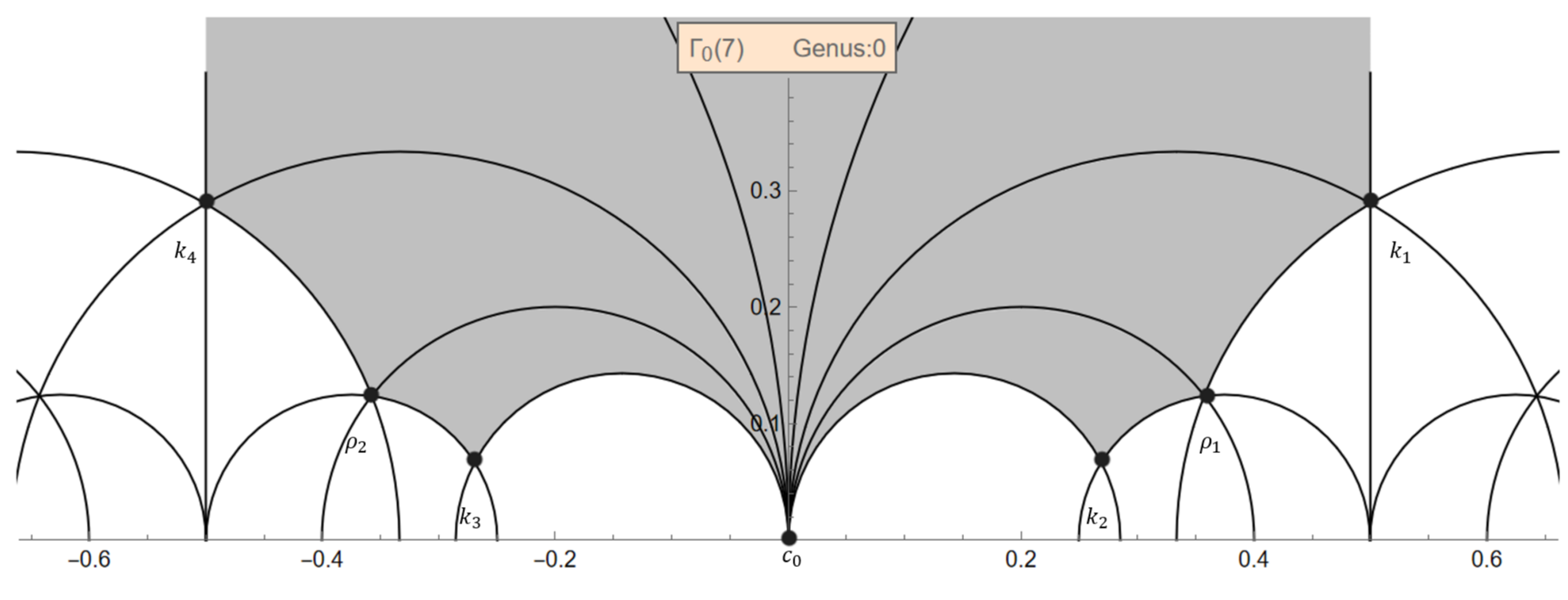}
    \caption{The points and their location on the fundamental domain are (from right to left): $k_{1} = \tfrac{7+i\sqrt{7}}{14}$, $\rho_{1} = \tfrac{5 + i\sqrt{3}}{14}$, $k_{2} = \tfrac{1}{4} + \tfrac{i}{4\sqrt{7}}$, $c_{0} = 0$, $k_{3} = -\tfrac{1}{4} + \tfrac{i}{4\sqrt{7}}$, $\rho_{2} = \tfrac{-5 + i\sqrt{3}}{14}$, $k_{4} = \tfrac{-7 + i\sqrt{7}}{14}$. In the equivalent fundamental domain, the points $k_{1}$ and $k_{4}$ are mapped to locations $\tfrac{1}{2}$ and $-\tfrac{1}{2}$ respectively.}
    \label{fig:fundamental_domain_level_7_points}
\end{figure}

We now consider the rational curve $X_{0}(7) = \Gamma_{0}(7)\backslash\mathbb{H}^{2*}$. The fundamental region of a group $\Gamma_{0}(p)$ with $p\in\mathbb{P}$ is given by \cite{Apostol}
 \begin{align}
      \mathcal{F}_{p} = \mathcal{F}\cup \bigcup\limits_{k=0}^{p-1}ST^{k}(\mathcal{F}).
\end{align}
For $p=7$, we find the fundamental domain of the Hecke subgroup $\Gamma_{0}(7)$ to be
\begin{equation}
    \mathcal{F}_{7} = \mathcal{F}\cup S\cup ST\cup ST^{2} \cup ST^{3} \cup ST^{4} \cup ST^{5} \cup ST^{6}.
\end{equation}
\begin{figure}[htb!]
    \centering
    \includegraphics[width = 14cm]{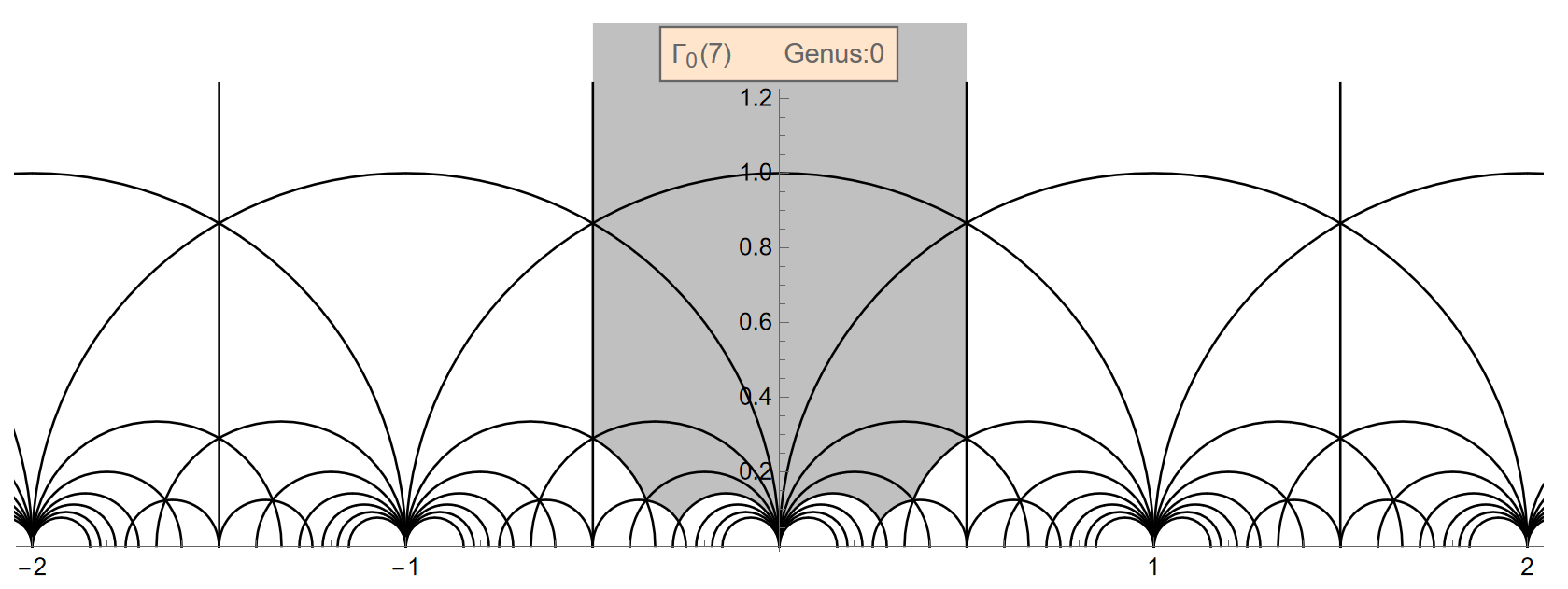}
    \label{fig:fundamental_domain_level_7_full}
    \caption{The fundamental domain of $\Gamma_{0}(7)$.}
\end{figure}
\noindent
For an index $n$ subgroup $\Gamma$ in $\text{SL}(2,\mathbb{Z})$, the set of coset representatives is defined as follows 
\begin{equation}\label{coset_rep_def} 
      \Gamma\backslash\text{SL}(2,\mathbb{Z}) = \left\{\gamma_{1},\ldots,\gamma_{n}\right\},
\end{equation}
The set of coset representatives of the curve $\Gamma_{0}(7)\backslash\text{SL}(2,\mathbb{Z})$ consists of the following eight elements
\begin{align}
    \left\{\tau\mapsto \tau, \tau\mapsto-\frac{1}{\tau},\tau\mapsto- \frac{1}{\tau + j}\right\},
\end{align}
where $j = 1,\ldots,6$. There exists an equivalent fundamental domain that is obtained by identifying coset representatives. For $A\in\text{SL}(2,\mathbb{Z})$, the right coset of a congruence subgroup $\Gamma$ in $\text{SL}(2,\mathbb{Z})$ is defined as the set $\Gamma A = \{BA:B\in\Gamma\}$. A congruence subgroup $\Gamma$ of $\text{SL}(2,\mathbb{Z})$ is a disjunctive conjunction of the right cosets of $\Gamma$ in $\text{SL}(2,\mathbb{Z})$ \cite{Hungerford}. This tells us that there exists a list of pairwise right coset representatives $A_{i}\in\text{SL}(2,\mathbb{Z})$, $i\in\{1,\ldots,n\}$, such that
\begin{equation}
    \text{SL}(2,\mathbb{Z}) = \Gamma A_{1}\cup \Gamma A_{2}\cup \ldots\cup \Gamma A_{n}.
\end{equation}
Thus, for a given part of the fundamental domain described by a matrix, we can identify an equivalent matrix with respect to a set of coset representatives. This is shown in figure \ref{fig:coset_rep}.
\\
\\
The number of cusps of $X_{0}(N)$ and the cusp width are given by \cite{Ogg}
\begin{align}
    \begin{split}
        \sigma_{\infty}(N) \equiv&  \sum\limits_{d\vert N}\phi\left(\left(d,\tfrac{N}{d}\right)\right),\\
        e_{d,N}\equiv&  \frac{N}{d\left(d,\tfrac{N}{d}\right)},
    \end{split}
\end{align}
where $\phi$ is the Euler totient function and $(\cdot,\cdot)$ stands for the greatest common divisor. The rational cusps $\tau = \tfrac{a}{d}$ are those for which $\phi\left(\left(d,\tfrac{N}{d}\right)\right) = 1$, i.e. the ones with $\left(d,\tfrac{N}{d}\right) = 1$ or $2$. When $N=7$, we have $\sigma_{\infty} = 2$ that corresponds to the two cusps at $0$ and $\infty$ of the curve $X_{0}(7)$. The cusp data for $X_{0}(7)$ is shown in table \ref{tab:cusp_X_0(7)}.
\\
\begin{table}[htb!]
    \centering
    \begin{tabular}{||c|c|c||}
    \hline
        Cusp &  $\tau = \tfrac{a}{d}$ & Cusp width $e_{d,N}$\\ [0.5ex]
        \hline\hline
        0 & $\tfrac{1}{1}$ & $e_{0,7} = 7$\\
        $\infty$ & $\tfrac{1}{7}$ & $e_{\infty,7} = 1$\\
        \hline
    \end{tabular}
    \caption{The cusp data of the curve $X_{0}(7)$.}
    \label{tab:cusp_X_0(7)}
\end{table}
\\
The new equivalent fundamental domain (with the cusps identified) of $\Gamma_{0}(7)$ is shown in figure \ref{fig: fundamental_domain_level_7_equiv}.
\\
\\
The point $\tau$ is called an elliptic point for a congruence subgroup $\Gamma$ if there exists a matrix $I \neq \gamma\in\Gamma$ such that $\gamma\tau  = \tau$. If $\tau$ is an elliptic point of $\Gamma_{0}(p)$ for $p\in\mathbb{P}$, then there exists a matrix $\gamma = \left(\begin{smallmatrix}a & b\\ c & d\end{smallmatrix}\right)$, $p\vert c$, such that $\gamma\tau = \tau$. Rearranging this and using the relation $ad-bc = 1$ yields
\begin{equation}
    \tau = \frac{a-d + \sqrt{(a+d)^{2} - 4)}}{2c}.
\end{equation}
For this to be imaginary, we require $(a+d)^{2}<4$ and hence, we see that $a+d\in\{-1,0,1\}$. This gives us two cases
\begin{enumerate}
    \item A point $\tau$ is called a $\rho$-elliptic for $\Gamma_{0}(p)$ if it is a shift of $\rho = \tfrac{1 + \sqrt{-3}}{2}$, i.e. $\tau = \gamma\rho$ for some $\gamma\in\text{SL}(2,\mathbb{Z})$. These elliptic points are all of the form $\tau = \tfrac{q + \sqrt{-3}}{2p}$ for some $q\in\mathbb{Z}$.
    \item A point $\tau$ is called a $i$-elliptic for $\Gamma_{0}(p)$ if it is a shift of $i$, i.e. $\tau = \gamma i$ for some $\gamma\in\text{SL}(2,\mathbb{Z})$. These elliptic points are all of the form $\tau = \tfrac{q + i}{p}$ for some $q\in\mathbb{Z}$.
\end{enumerate}
\noindent
The number of $\rho$- and $i$-elliptic points for $\Gamma_{0}(p)$ are given by 
\begin{align}\label{number_of_elliptic_points}
    \begin{split}
        \varepsilon_{\rho} =& 1 + \left(-\frac{3}{p}\right),\\
        \varepsilon_{i} =& 1 + \left(-\frac{1}{p}\right),
    \end{split}
\end{align}
where $\left(\cdot\right)$ is the Jacobi symbol. For $\Gamma_{0}(7)$, we see that we have $\varepsilon_{\rho} = 2$ and $\varepsilon_{i} = 0$ with the $\rho$-elliptic points located at
\begin{equation}
    \rho_{1} = \frac{5 + i\sqrt{3}}{14},\ \rho_{2} = \frac{-5 + i\sqrt{3}}{14}.
\end{equation}
The fundamental domain of $\Gamma_{0}(7)$ with labelled points is shown in figure \ref{fig:fundamental_domain_level_7_points}.
\\
\\
The curve $X_{0}(7)$ which also goes by the name Hecke modular curve of level $7$ possesses the following rational coordinate or Hauptmodul \cite{Ramanujan},
\begin{align}\label{Hauptmodul_7}
    \begin{split}
        j_{7}(\tau) = \left(\frac{\eta(\tau)}{\eta(7\tau)}\right)^{4} =& \frac{1}{q}\prod\limits_{n=1}^{\infty}\left(\frac{(1-q^{n})}{(1-q^{7n})}\right)^{4}\\
        =&q^{-1} - 4 + 2q + 8q^{2} - 5q^{3} - 4q^{4} +\ldots 
    \end{split}
\end{align}
This directly follows from proposition $4.1$ in \cite{Farkas} stated below:
\\
\noindent
For each prime $k$,
  \begin{enumerate}
      \item $f_{k}(\tau) = f^{\frac{24}{k-1}}_{k,1}(\tau) = \left(\frac{\eta(k\tau)}{\eta(\tau)}\right)^{\frac{24}{k-1}}$ is a $\Gamma_{0}(k)$-multiplicative function,
      \item for all $N\in\mathbb{Z}$. $f_{k}^{N}$ is $\Gamma_{0}(k)$-invariant if $(k-1)\vert 12N$; in particular
      \item (for $N=1$) $f_{k}$ is $\Gamma_{0}(k)$-invariant if and only if $k=2,3,5,7$ or $13$.
  \end{enumerate}
\noindent
It follows from this proposition that the case of $N = -1$ and $k=7\in\mathbb{P}$ implies we have $6\vert 12$ and hence, the Hauptmodul $j_{7}(\tau)$ in \ref{Hauptmodul_7} is a $\Gamma_{0}(7)$-multiplicative function that is related to the $f_{7}(\tau)$ as $j_{7}(\tau) = f^{-1}_{7}(\tau)$. More generally, from \ref{J_function_McKay}, since the Hauptmodul of a genus zero subgroup $\Gamma_{0}(p)$ for $p\in\{2,3,5,7,13\}$ reads (also see \cite{Ahlgren})
\begin{equation}
    j_{p}(\tau) = \left(\frac{\eta(\tau)}{\eta(p\tau)}\right)^{\frac{24}{p-1}}
\end{equation}
Clearly, this matches our expression in \ref{Hauptmodul_7} for the case of $p=7$. For $p=\{2,3,5,7,13\}$, the Hauptmodul $j_{p}(\tau)$ has a pole of order $1$ at $i\infty$ and a zero of order $\tfrac{1}{p}$ at $\tau = 0$ \cite{Apostol}. It follows from this that the Hauptmodul $j_{7}(\tau)$ has a pole of order $1$ at $i\infty$ and a zero of order $\tfrac{1}{7}$ at $\tau = 0$. The  $\Gamma_{0}(7)$ has cusps at $i\infty$ and at $0$ of which the latter can be observed from its fundamental domain in figure \ref{fig:fundamental_domain_level_7_full}. The $J$-function is given by
\begin{align}
    \begin{split}
        J_{7}(\tau) =& j_{7}(\tau) + 4\\
        =& q^{-1} + 2q+8q^{2} + \ldots
    \end{split}
\end{align}
The congruence subgroup $\Gamma(7)$ is a normal subgroup of $\Gamma_{0}(7)$, and $\Gamma_{0}(7)$ is generated by $\Gamma(7)$ and the group generated by the transformations $T$ and $H = ST^{-3}STST^{-1}ST^{3} = \left(\begin{smallmatrix}2 & 7\\ 7 & 25 \end{smallmatrix}\right)$. The quotient $\Gamma_{0}(7)/\Gamma(7)$ is of order $21$ and is isomorphic to the subgroup $\mathbf{g}\subset G$ of upper triangular matrices. The natural projection defines a Galois covering \cite{Ramanujan}
\begin{equation}
    X(7)\to X_{0}(7),
\end{equation}
with Galois group $\mathbf{g}$. We can define the $j$-invariant in terms of the Hauptmodul. The function $j(\tau)$ is holomorphic in $\mathbb{H}^{2}$, is invariant under the action of the modular group $\text{PSL}(2,\mathbb{Z})$, and has a simple pole at infinity; it takes the following form
\begin{align}\label{j_7_to_J}
    \begin{split}
        j(\tau) =& \frac{1}{j_{7}(\tau)^{7}}\left(j_{7}(\tau)^{2} + 13j_{7}(\tau) + 49\right)\left(j_{7}(\tau)^{2} + 245j_{7}(\tau) + 2401\right)^{3}\\
        =& q^{-1} + 744 + 196884 q + 21493760 q^{2} + 864299970 q^{3} +20245856256 q^{4} + \ldots
    \end{split}
\end{align}
with $J(\tau) = j(\tau)-744$ being the $J$-function on $\mathbb{H}^{2}$. We now define
\begin{align}\label{k_and_t}
    \begin{split}
    \mathbf{k}(\tau) \equiv& \frac{\eta(\tau)^{7}}{\eta(7\tau)}\\
    =& 1 - 7 q + 14 q^2 + 7 q^3 - 49 q^4 + \ldots,\\
    \mathbf{t}(\tau) \equiv& \left(\frac{\eta(7\tau)}{\eta(\tau)}\right)^{4} = \frac{1}{j_{7}(z)}\\
    =&q + 4 q^2 + 14 q^3 + 40 q^4 + \ldots
    \end{split}
\end{align}
From theorem \ref{thm:space_dim_G_0}, we find the dimension of the space of modular forms of the group $\Gamma_{0}(7)$ to be
\begin{equation}\label{dim_space_level_7}
    \text{dim}\mathcal{M}_{k}(\Gamma_{0}(7)) =1 +  2\left\lfloor\left. \frac{k}{3}\right\rfloor\right. \ \text{if}\ k\geq 1.
\end{equation}
The space $\mathcal{M}_{1}(\Gamma_{0}(7))$ is one-dimensional and the following series belongs to this space
\begin{equation}
    \theta_{7}(\tau) = \sum\limits_{\xi\in\mathbb{Z}[\alpha]}q^{\xi\overline{\xi}},
\end{equation}
where $\alpha = \tfrac{-1+\sqrt{7}}{2}$. We can express this as the following sum
\begin{align}
    \begin{split}
        \theta_{7}(\tau) = 1 + \sum\limits_{n=1}^{\infty}R(n)q^{n} =& 1 + 2\sum\limits_{n=1}^{\infty}\chi_{7}(n)\frac{q^{n}}{1-q^{n}}\\
        =& 1 + 2\left[1\cdot\frac{q}{1-q} + 1\cdot\frac{q^{2}}{1-q^{2}} + (-1)\cdot\frac{q^{3}}{1-q^{3}} +  \ldots\right]\\
        =& 1 + 2q+4q^{2} + 6q^{4}+2q^{7}+8q^{8}+2q^{9}+4q^{11}+\ldots,
    \end{split}
\end{align}
where
\begin{align}
    \begin{split}
    \chi \equiv \chi_{7}(n) =& \left(\frac{n}{7}\right)\\
    =& \begin{cases}
    1,\ \ \ \text{if}\ n\ \text{is a quadratic residue mod $7$ and}\ n\not\equiv 0(\text{mod}\ 7)\\
    -1,\ \text{if}\ n\ \text{is a non-quadratic residue mod $7$}\\
    0,\ \ \ \text{if}\ n\equiv 0(\text{mod}\ 7)
    \end{cases},
    \end{split}
\end{align}
and we have replaced the Legendre symbol with its values and $R(n)$ is defined as follows
\begin{equation}
    R(n) = \left\{(x,y)\in\mathbb{Z}^{2}\vert x^{2}+xy + 2y^{2} = n\right\} = 2\sum\limits_{d\vert n}\chi_{7}(d).
\end{equation}
he square of the level $7$ theta function, $\theta_{7}^{2}(\tau)$, and the level $7$ Eisenstein series of weight $2$, $E_{2}^{(7)}$, belong to this space with the later defined as follows
\begin{align}\label{level_7_eisenstein_2}
    \begin{split}
    E_{2}^{(7)}(\tau) \equiv& 1 + 4\sum\limits_{n=1}^{\infty}\sigma_{1}(n)\left(q^{n} - 7q^{7n}\right)\\
    =& 1 + 4 q + 12 q^2 + 16 q^3 + 28 q^4 + 24 q^5\\
    =& \theta_{7}^{2}(\tau),
    \end{split}
\end{align}
where we have used the fact that $\sum_{n=1}^{\infty}\sigma_{a}(n)q^{n} = \sum_{n=1}^{\infty}\tfrac{n^{a}q^{n}}{1-q^{n}}$ to obtain the $q$-expansion in the second line. Lastly, we make the following observation
\begin{align}
    \begin{split}
        \frac{1}{6}\left(E_{2}(\tau) - 7E_{2}(7\tau)\right)=&  \frac{1}{6}\left[\left(1-24\sum\limits_{n=1}^{\infty}\frac{nq^{n}}{1-q^{n}}\right) - 7\left(1 - 24\sum\limits_{n=1}^{\infty}\frac{nq^{7n}}{1-q^{7n}}\right)\right]\\
        =& -1 - 4 q - 12 q^2 - 16 q^3 - 28 q^4 - 24 q^5 + \ldots\\
        =& -E^{(7)}_{2}(\tau) = -\theta^{2}_{7}(\tau).
    \end{split}
\end{align}

\noindent

\subsection{Single character solutions}

The valence formula for $\Gamma_{0}(7)$ is given by 
\begin{align}\label{valence_formula_Hecke}
    \nu_{\infty}(f) + \nu_{0}(f) + \frac{1}{3}\nu_{\rho_{1}}(f) + \frac{1}{3}\nu_{\rho_{2}}(f) + \sum\limits_{\substack{p\in X_{0}(7)\\ p\neq \rho_{1},\rho_{2}}} \nu_{p}(f) = \frac{2}{3}k,
\end{align}
where $\nu_{\ell}(f) = \text{ord}_{\ell}f$ and $\rho_{1} = \tfrac{5 + i\sqrt{3}}{14}$ $\rho_{2} = \tfrac{-5 + i\sqrt{3}}{14}$
are $\rho$-elliptic points identified in figure \ref{fig:fundamental_domain_level_7_points}. From \ref{W_n_asymptotic}, we have
\begin{align}
    \nu_{i\infty}(W_{n}) = -\frac{nc}{24} + \sum\limits_{i}\Delta_{i}.
\end{align}
The number of zeros inside the fundamental domain $\mathcal{F}_{7}$ is given by 
\begin{align}
    \# =  2\ell\cdot\frac{8}{12} = \frac{4}{3}\ell.
\end{align}
With this the valence formula \ref{valence_formula_Hecke} gives us
\begin{align}\label{central_charge_valence_Hecke}
    -\frac{nc}{24} + \sum\limits_{i}\Delta_{i} + \frac{4}{3}\ell = \frac{2}{3}n(n-1),
\end{align}
where the contribution of the zeros read
\begin{align}
    \frac{4}{3}\ell = \nu_{0}(W_{n}) + \frac{1}{3}\nu_{\rho_{1}}(W_{n}) + \frac{1}{3}\nu_{\rho_{2}}(W_{n}) + \sum\limits_{\substack{p\in X_{0}(7)\\ p\neq \rho_{1},\rho_{2}}} \nu_{p}(W_{n}).
\end{align}
Since $\theta_{7}^{2}(\tau)$ is a modular form of weight $2$ in $\Gamma_{0}(7)$, $\ell$ is a non-negative integer that is either zero or greater than or equal to one, $\ell = 0,1,2,\ldots$

\subsection*{\texorpdfstring{$\ell = 0$}{Lg}}
Let us consider the simple case with no poles $\ell = 0$ and a single character. In this case, the only primary is the vacuum state and all other states are descendants of the vacuum. The single vacuum character behaves as $f_{0}(\tau) \sim q^{-\tfrac{c}{24}}\left(1 + \mathcal{O}(q)\right)$ and satisfies the following MLDE
\begin{align}
    \left[\omega_{2\ell}(\tau)\mathcal{D} + \omega_{2\ell + 2}(\tau)\right]f(\tau) = 0.
\end{align}
Since $n=1$ and $\Delta_{i} = 0$ for the vacuum state, using \ref{central_charge_valence_Hecke} we find
\begin{align}\label{c=32l}
    c = 32\ell.
\end{align}
For $\ell = 0$, we have to make choices for the functions $\omega_{0}(\tau)$ and $\omega_{2}(\tau)$. Since $\mathcal{M}_{0}(\Gamma_{0}(7))$ is one-dimensional, $\omega_{0}(\tau) = 1$ and since the space $\mathcal{M}_{2}(\Gamma_{0}(7))$ is one-dimensional with the modular form $\theta_{7}^{2}(\tau) = E_{2}^{(7)}(\tau)$, we choose
\begin{align}
    \begin{split}
        \omega_{2}(\tau) =& \mu E_{2}^{(7)}(\tau)\\
        =& \mu\left(1 + 4q + 12q^{2} + 16q^{3} + 28q^{4} + \ldots\right).
    \end{split}
\end{align}
Plugging in the behaviour of $f_{0}(\tau)$, we find $\mu = \tfrac{c}{24} = 0$ since $c =0$. Hence, the only trivial solution is $f_{0}(\tau) = 1$. 

\subsection*{\texorpdfstring{$\ell = 1$}{Lg}}
If an admissible theory exists for $\ell = 1$, then it would have a central charge $c = 32$. The MLDE in this case takes the following form
\begin{align}
    \left[\omega_{2}(\tau)\mathcal{D} + \omega_{4}(\tau)\right]f(\tau) = 0.
\end{align}
We make the same choice for $\omega_{2}(\tau)$ as before. The space $\mathcal{M}_{4}(\Gamma_{0}(7))$ is three-dimensional. The space of modular forms of $\Gamma_{0}(7)$ has the following basis decomposition
\begin{align}
    \begin{split}
     \mathcal{M}_{k}(\Gamma_{0}(7)) =& E^{(7)}_{k-\tfrac{3}{2}n}(\tau)\left(\mathbb{C}\left(\Delta_{7}^{\infty}\right)^{n}\oplus\mathbb{C}\left(\Delta_{7}^{\infty}\right)^{n-1}\Delta_{7}^{0}\oplus\mathbb{C}\left(\Delta_{7}^{\infty}\right)^{n-2}\left(\Delta_{7}^{0}\right)^{2}\oplus\ldots\oplus\mathbb{C}\left(\Delta_{7}^{0}\right)^{n}\right),\\
    \Delta_{7}^{0}(\tau) \equiv& \sqrt{\frac{\eta^{7}(\tau)}{\eta(7\tau)}} = \mathbf{k}^{\frac{1}{2}}(\tau),\ \ 
    \Delta_{7}^{\infty}(\tau) \equiv \sqrt{\frac{\eta^{7}(7\tau)}{\eta(7\tau)}} = \left(\mathbf{k}(\tau)\mathbf{t}(\tau)\right)^{\frac{1}{2}},
    \end{split}
\end{align}
where $n = \text{dim}\mathcal{M}_{k}(\Gamma_{0}(7)) - 1$, $E^{(7)}_{1}(\tau) = \sqrt{E^{(7)}_{2}(\tau)} = \theta_{7}(\tau)$, and modular forms $\mathbf{k}(\tau)$ and $\mathbf{t}(\tau)$ are defined in \ref{k_and_t}. For the case of $k=4$, we have $n=2$ and the following decomposition
\begin{align}
    \mathcal{M}_{4}(\Gamma_{0}(7)) = \theta_{7}(\tau)\left(\mathbb{C}\mathbf{k}\oplus\mathbb{C}\mathbf{k}\mathbf{t}\oplus\mathbb{C}\mathbf{k}\mathbf{t}^{2}\right),
\end{align}
where we have used $E^{(7)}_{1}(\tau) = \theta_{7}(\tau)$. This gives us three choices for $\omega_{4}(\tau)$. The $q$-series expansions of $\tfrac{\omega_{4}(\tau)}{\omega_{2}(\tau)}$ in these three cases read
\begin{align}
    \begin{split}
        \frac{\omega_{4}(\tau)}{\omega_{2}(\tau)} =& \mu\frac{\mathbf{k}(\tau)}{\theta_{7}(\tau)}\\
        =& \mu\left(1 - 9 q + 28 q^2 - 13 q^3 - 141 q^4 + \ldots\right),\\
        \frac{\omega_{4}(\tau)}{\omega_{2}(\tau)} =& \mu\frac{\mathbf{k}(\tau)\mathbf{t}(\tau)}{\theta_{7}(\tau)}\\
        =&\mu\left(q - 5 q^2 + 6 q^3 + 13 q^4 + \ldots\right),\\
        \frac{\omega_{4}(\tau)}{\omega_{2}(\tau)} =& \mu\frac{\mathbf{k}(\tau)\mathbf{t}^{2}(\tau)}{\theta_{7}(\tau)}\\
        =&\mu\left(q^2 - q^3 + 7 q^5 +  \ldots\right).
    \end{split}
\end{align}
Since we know the $q$-series expansions in the equation, we can solve this ODE order by order in $q$ which yields a character that is modular invariant (up to a phase) for every value of the free parameters. We find that with $\mu = \tfrac{c}{24} = \tfrac{4}{3}$, $m_{1}(\mu) = -12,\tfrac{4}{3}$ for the first and second choices respectively and $0$ for the third choice, and $m_{2}(\mu) = \tfrac{112}{3},-\tfrac{20}{3}, \tfrac{4}{3}$ for the first, second, and third choices respectively. These character coefficients are either negative or non-integers or both. Hence, it seems that at $\ell = 1$, the solution fails to be admissible and there exists no single character CFT with $c = 32$. Before abandoning this case, we consider the following general expression for $\omega_{4}(\tau)$
\begin{align}
    \omega_{4}(\tau) = \theta_{7}(\tau)\left(\mu_{1}\mathbf{k}(\tau) + \mu_{2}\mathbf{k}\mathbf{t}(\tau) + \mu_{3}\mathbf{k}(\tau)\mathbf{t}^{2}(\tau)\right).
\end{align}
For this basis, the most general equation is the following
\begin{align}
    \left[\Tilde{\partial} + \mu_{1}\frac{\mathbf{k}(\tau)}{\theta_{7}(\tau)} + \mu_{2}\frac{\mathbf{k}\mathbf{t}(\tau) }{\theta_{7}(\tau)} + \mu_{3}\frac{\mathbf{k}\mathbf{t}^{2}(\tau)}{\theta_{7}(\tau)}\right]f(\tau) =0.
\end{align}
Our aim now is to look at the $q$-series expansion of this sum of the ratios of modular forms and carefully choose the coefficients $\mu_{i}$ so that all the expansion coefficients $m_{i}(\mu_{i})$ turn out to be positive integers. We can repeat the approach taken in the case of Fricke groups done in the previous section, but we quickly run into an issue. If we were to consider the ansatz, $\chi(\tau) = j_{7}^{\tfrac{4}{3}}(\tau)$, the solution ceases to be admissible since the $q$-series expansion contains non-negative fractions. Thus, the Hauptmodul itself raised to a power determined by $\mu_{1} = \tfrac{4}{3}$ fails to be admissible.

\subsection*{\texorpdfstring{$\ell = 2$}{Lg}}
If an admissible theory exists for $\ell = 2$, then it would have a central charge $c = 64$.
The MLDE for this case takes the following form
\begin{align}
    \left[\omega_{4}(\tau) \mathcal{D} + \omega_{6}(\tau)\right]f(\tau) = 0. 
\end{align}
For fixing $\omega_{6}(\tau)$ , we look at the space $\mathcal{M}_{6}(\Gamma_{0}(7))$ which is five-dimensional and possesses the following basis decomposition
\begin{align}
    \mathcal{M}_{6}(\Gamma_{0}(7)) = \mathbb{C}\mathbf{k}^{2}\oplus\mathbb{C}\mathbf{k}^{2}\mathbf{t}\oplus\mathbb{C}\mathbf{k}^{2}\mathbf{t}^{2}\oplus\mathbb{C}\mathbf{k}^{2}\mathbf{t}^{3}\oplus\mathbb{C}\mathbf{k}^{2}\mathbf{t}^{4},
\end{align}
Consider the choice $\omega_{4}(\tau) = \mu_{1}\textbf{k}(\tau) + \mu_{2}\left(\textbf{k}\textbf{t}\right)(\tau)$ and $\omega_{6}(\tau) = \mu_{3}\textbf{k}^{2}(\tau)$ with $\mu_{2} = -4\mu_{1}$ and $\mu\equiv \tfrac{\mu_{3}}{\mu_{1}} = \tfrac{8}{3}$. This yields integer coefficients, although some are negative, that quickly turn fractional at higher orders of $q$. The same issue arises if we were to include higher powers of $\textbf{t}(\tau)$ in $\omega_{6}(\tau)$. Thus, we conclude that there exist no admissible solutions at $\ell = 2$. 

\subsection*{\texorpdfstring{$\ell = 3$}{Lg}}
If an admissible theory exists for $\ell = 3$, then it would have a central charge $c = 96$.
The MLDE takes the following form
\begin{align}
    \left[\omega_{6}(\tau)\mathcal{D} + \omega_{8}(\tau)\right)f(\tau) = 0.
\end{align}
For fixing $\omega_{8}(\tau)$ , we look at the space $\mathcal{M}_{8}(\Gamma_{0}(7))$ which is five-dimensional and possesses the following basis decomposition
\begin{align}
    \mathcal{M}_{8}(\Gamma_{0}(7)) = \theta^{2}_{7}(\tau)\left(\mathbb{C}\mathbf{k}^{2}\oplus\mathbb{C}\mathbf{k}^{2}\mathbf{t}\oplus\mathbb{C}\mathbf{k}^{2}\mathbf{t}^{2}\oplus\mathbb{C}\mathbf{k}^{2}\mathbf{t}^{3}\oplus\mathbb{C}\mathbf{k}^{2}\mathbf{t}^{4}\right),
\end{align}
We can expect to build an admissible solution here since $j_{7}^{\tfrac{96}{24}}(\tau)$ possesses integral coefficients. We find that the polynomial ansatz $\chi(\tau) = j_{7}^{4}\left(j_{7} - 4\right)^{-3}(\tau)$ yields
non-negative integral coefficients. Hence, we can now try to fix the unknown free parameters based on this ansatz by matching series expansions. We motivate this approach by first considering a special case to see how it requires us to choose a polynomial ansatz. Consider the choice $\omega_{6}(\tau) = \mu_{1}\textbf{k}(\tau) + \mu_{2}\left(\textbf{k}^{2}\textbf{t}\right)(\tau)$ and $\omega_{8}(\tau) = \mu_{3}\left(\theta_{7}\textbf{k}\right)^{2}(\tau)$, set $\mu_{2} = 5\mu_{1}$, and define $\mu\equiv \tfrac{\mu_{3}}{\mu_{1}}$ to obtain the following MLDE
\begin{align}
    \left[\mathcal{D} + \mu\left(\frac{\theta_{7}^{2}\textbf{k}^{2}}{\textbf{k}^{2}+ 5\textbf{k}^{2}\textbf{t}}\right)(\tau)\right]f(\tau) = 0.
\end{align}
Matching gives us $\mu = \tfrac{c}{24} = 4$. The first few coefficients are $m_{1}(\mu) = 4$, $m_{2}(\mu) = 14$, $m_{3}(\mu) = 60$, $\ldots$, and the solution to this MLDE, $f(\tau) = \chi^{(\ell = 3)}(\tau)$, has the following $q$-series expansion
\begin{align}\label{admissible_char_l=3}
    \chi^{(\ell = 3)}(\tau) = q^{-4}\left(1 + 4 q + 14 q^2 + 60 q^3 + 125 q^4 + 268 q^5  + 588 q^6 + 420 q^7 + 314 q^8 +  \ldots\right),
\end{align}
although the coefficients $m_{i}(\mu)$ seem to be non-negative integers, performing a computation to higher orders in $q$ reveals a couple of negative integer coefficients. The $q$-series expansion to order $\mathcal{O}(q^{21})$ reads
\begin{align}\label{char}
    \begin{split}
        \chi^{(\ell = 3)}(\tau) = q^{-4}(1 &+ 4 q + 14 q^2 + 60 q^3 + 125 q^4 + 268 q^5  + 588 q^6 + 420 q^7 + 314 q^8\\ &-416 q^9 - 3358q^{10} - 3500q^{11} - 2496 q^{12} + 9668q^{13} + 9596q^{14}\\ &+ 26232q^{15} + 26043q^{16} 
        - 95000q^{17} -65450q^{18} -116728q^{19} + 165684q^{20} +  \ldots)
    \end{split}
\end{align}
Although the presence of negative coefficients makes this an inadmissible solution, the coefficients still remain integers. Such characters are called quasi-characters and can be used to construct admissible solutions. For the case of $\Gamma_{0}(7)$, the only modular invariant function is the Hauptmodul $j_{7}(\tau)$. Hence, it should be possible to build the character corresponding to $\ell = 3$ as a polynomial in $j_{7}$. It turns out that with a certain choice of coefficients, the $q$-series expansion with a monic polynomial of $j_{7}$ of degree $4$ exactly matches the character found in \ref{admissible_char_l=3},
\begin{align}
    \begin{split}
    \chi^{(\ell = 3)}(\tau) =& \left(j_{7}^{4} + 20j_{7}^{3} + 150j_{7}^{2} + 500j_{7} + 500\right)(\tau)\\
    =& q^{-4}\left(1 + 4 q + 14 q^2 + 60 q^3 + 125 q^4 + 268 q^5  + 588 q^6 + 420 q^7 + 314 q^8 +  \ldots\right).
    \end{split}
\end{align}
Since the only admissible character prior to $\ell = 3$ was the trivial one corresponding to $\ell = 1$, we conclude that the generic form of a character of the Hecke group of level $7$, for all permissible values of $\ell$, takes the form
\begin{align}
    \chi(\tau) = \mathfrak{P}_{w_{\tau}}(j_{7}),
\end{align}
and the degree can be found using the Riemann-Roch theorem as follows
\begin{align}
    c = 24w_{\tau} = 32\ell.
\end{align}
This exercise tells us that certain choices of the free parameters yield a quasi-character that can be defined as a polynomial in $j_{7}(\tau)$. Thus, a polynomial that contains all non-negative integral coefficients should be possible to build by fixing the free parameters via $q$-series expansion matching. The admissible solution, in this case, should then be of the following form
\begin{align}
    \begin{split}
        \chi^{(\ell = 3)}(\tau) =& j_{7}^{4}\left(j_{7} - 4\right)^{-3}(\tau)\\
        =& q^{-1}\left(1 + 8 q + 98 q^2 + 1032 q^3 + 10299 q^4 + \ldots\right).
    \end{split}
\end{align}
Lastly, it is explicitly stated in \cite{Ramesh_Mukhi} that no quasi-characters were found outside the above classification with alternating positive and negative signs. Going back to the $q$-series expansion of the $\Gamma_{0}(7)$ quasi-character for the case of $\ell = 3$ in \ref{admissible_char_l=3}, we found that it does not fall into either one of the types in the classification since it possesses alternating positive and negative signs (we checked this to order $\mathcal{O}(q^{2001})$). The higher-order coefficients, however, still remain as integers.

\subsection*{\texorpdfstring{$\ell = 6$}{Lg}}
The next instance when $j_{7}^{\mu_{1}}$ contains non-negative integers is with $\mu_{1} = 8$ which corresponds to the case $\ell = 6$ with central charge $c = 192$. The ansatz for the admissible solution, in this case, takes the following form
\begin{align}
    \begin{split}
        \chi^{(\ell = 6)}(\tau) =& j_{7}^{8}\left(j_{7} - 4\right)^{-7}(\tau)\\
        =& q^{-1}\left(1 + 24 q + 450 q^2 + 7176 q^3 + 103035 q^4 + \ldots\right).
    \end{split}
\end{align}
\subsection{\texorpdfstring{$\Gamma_{0}^{+}(7)$}{Γ0(7)+}}
\begin{figure}[htb!]
    \centering
    \includegraphics[width = 14cm]{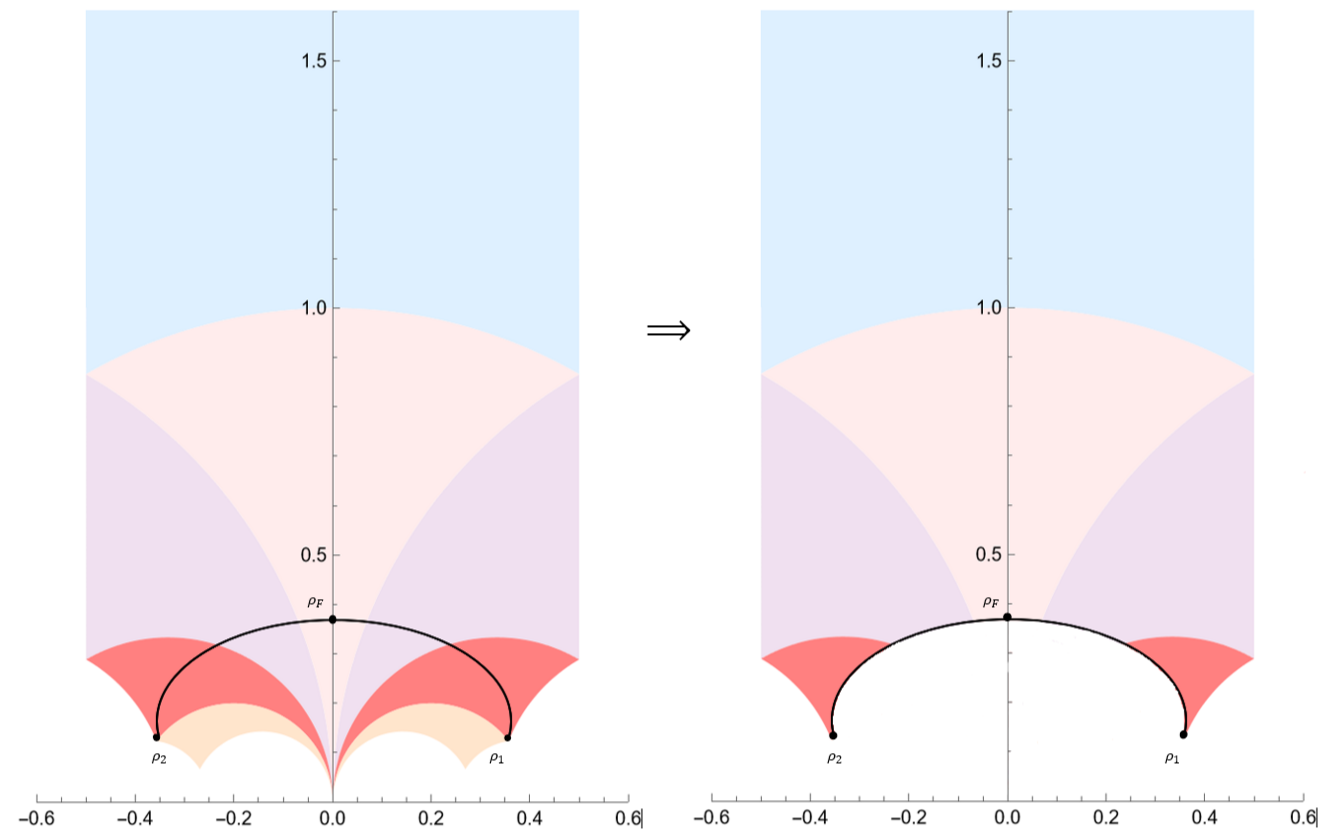}
    \caption{Left: The fundamental domain $X_{0}(7)$ for $\Gamma_{0}(7)$ that is symmetric under the Fricke involution $W_{7}$. Right: A fundamental domain for $\Gamma_{0}^{+}(7)$ is obtained by quotienting $W_{7}$ as $X_{0}(7)/W_{7}$ which corresponds to only considering the part above the arc. The marked points are the following: $\rho_{1} = \tfrac{5 + i\sqrt{3}}{14}$ and $\rho_{2} = \tfrac{-5 + i\sqrt{3}}{14}$ are elliptic points and $\rho_{F} = \tfrac{i}{\sqrt{7}}$ is the fixed point of $W_{7}$. Notice that $\Gamma_{0}^{+}(7)$ only has one cusp at $i\infty$.}
    \label{fig:Fricke_fundamental_domain}
\end{figure}

We now consider the curve $X_{0}^{+}(7) = \Gamma_{0}^{+}(7)\backslash\mathbb{H}^{2*}$. The fundamental domain of the Fricke group $\Gamma_{0}^{+}(7)$ is easy to obtain and is done by moding out the Fricke involution from the rational curve $X_{0}(N)$, i.e. $X_{0}(N)/W_{N}$ for a positive integer $N$. For the cases of $N \in\{5,7,11\}$ in $\Gamma^{+}_{0}(N)$, one has to apply cuts along hyperbolic lines emanating from the points $k = e^{\frac{i\pi}{3}}$ and $-\tfrac{1}{k}$ \cite{Fricke_folding}. This procedure is shown in figure \ref{fig:Fricke_fundamental_domain}. The curve $X^{+}_{0}(7)$ possesses the following Hauptmodul.
\begin{align}
        j_{7^{+}}(\tau) =& \left(\frac{E_{2}^{(7)}(\tau)}{\eta(\tau) \eta(7\tau)}\right)^{3}.
\end{align}
\noindent Using the relation established in \ref{level_7_eisenstein_2} and rearranging the eta products, we can rewrite this in terms $\theta_{7}(\tau)$, $j_{7}(\tau)$, and $\mathbf{k}(\tau)$ as follows
\begin{align}\label{Hauptmodul_Fricke_7}
    \begin{split}
        j_{7^{+}}(\tau) =& \theta_{7}^{3}(\tau) j_{7}(\tau)\mathbf{k}^{-1}(\tau) = \left[\left(\frac{\eta(\tau)}{\eta(7\tau)}\right)^{2} + 7\left(\frac{\eta(7\tau)}{\eta(\tau)}\right)^{2}\right]^{2} - 1.\\
        =& q^{-1} + 9 + 51 q + 204 q^2 + 681 q^3 + 1956 q^4 + \ldots 
    \end{split}
\end{align}
From this, we see that $j_{7^{+}}$ is just a function of $j_{7}$ and hence, we get the relation
\begin{align}\label{j_7+_and_j_7}
    j_{7^{+}}(\tau) = j_{7}(\tau) + 49j_{7}^{-1}(\tau) + 13.    
\end{align}
The valence formula for $\Gamma_{0}^{+}(7)$ is given by
\begin{equation}\label{valence_formula_Fricke}
    \nu_{\infty}(f) + \frac{1}{2}\nu_{\rho_{F}}(f) + \frac{1}{2}\nu_{k_{2}}(f) + \frac{1}{3}\nu_{\rho_{2}}(f) + \sum\limits_{\substack{p\in\ X_{0}^{+}(7)\\ p\neq \rho_{F},k_{2},\rho_{2}}}\nu_{p}(f) = \frac{k}{3},
\end{equation}
where $\rho_{F} = \tfrac{i}{\sqrt{7}}$, $k_{2} = -\tfrac{1}{2} + \tfrac{i}{2\sqrt{7}}$, and $\rho_{2} = \tfrac{-5 + i\sqrt{3}}{14}$ are identified in figure \ref{fig:Fricke_fundamental_domain}.
\noindent
From \ref{W_n_asymptotic}, we have $\nu_{i\infty}(W_{n}) = -\tfrac{nc}{24} + \sum\limits_{i}\Delta_{i}$. To find the number of zeros inside the fundamental domain $\mathcal{F}_{7^{+}}$, we first fix the index $\left[\text{SL}(2,\mathbb{Z}):[\Gamma_{0}^{+}(7)\right]$. The group $\Gamma_{0}(7)$ has a discriminant $D = 7$ and hence, only has elliptic fixed points of orders $2$ and $3$ thus making the signature $(g,\mu,\nu_{2},\nu_{3},\nu_{4},\nu_{6}, \nu_{\infty}) = (0,8,0,2,0,0,2)$. Using theorem \ref{thm:genus_Fricke_index} with $(N,h,m)=(7,1,7)$, we have
\begin{align}
    \chi(\Gamma_{0}^{+}(7)) = \frac{7}{6}\prod\limits_{p\vert 7}\frac{p+1}{2p} = \frac{2}{3},\ \ 
    \left[\Gamma_{0}^{+}(7):\Gamma_{0}(7)\right] = 2,
\end{align}
i.e. $\Gamma_{0}(7)$ is an index-$2$ subgroup of $\Gamma_{0}^{+}(7)$. With this, we find the index of $\Gamma_{0}^{+}(7)$ in $\text{SL}(2,\mathbb{Z})$ to be
\begin{align}
    \left[\text{SL}(2,\mathbb{Z}):\Gamma_{0}^{+}(7)\right] = \frac{\left[\text{SL}(2,\mathbb{Z}):\Gamma_{0}(7)\right]}{\left[\Gamma^{+}_{0}(7):\Gamma_{0}(7)\right]} = \frac{8}{2} = 4.
\end{align}
The number of zeros in $\mathcal{F}_{7^{+}}$ is computed to be $\# =  2\ell\cdot\tfrac{4}{12} = \tfrac{2}{3}\ell.$
With this result the valence formula \ref{valence_formula_Fricke} gives us
\begin{align}\label{central_charge_valence_Fricke}
    -\frac{nc}{24} + \sum\limits_{i}\Delta_{i} + \frac{2}{3}\ell = \frac{n(n-1)}{3},
\end{align}
where the contribution of the zeros reads
\begin{align}
    \frac{1}{3}\ell = \frac{1}{2}\nu_{\rho_{F}}(W_{n}) + \frac{1}{2}\nu_{k_{2}}(W_{n}) + \frac{1}{3}\nu_{\rho_{2}}(W_{n}) + \sum\limits_{\substack{p\in\ X_{0}^{+}(7)\\ p\neq \rho_{F},k_{2},\rho_{2}}}\nu_{p}(W_{n}),
\end{align}
Since there is no modular form in $\Gamma_{0}^{+}(7)$ of weight $2$, we rule out $\ell  = 1$ thus making $\ell$ a non-negative integer that is either zero or greater or equal to two, $\ell = 0,2,3\ldots$ The dimension of the space of modular forms of $\Gamma_{0}^{+}(7)$ is
\begin{equation}
    \begin{split}
        \text{dim}\ \mathcal{S}_{k}(\Gamma_{0}^{+}(7)) =&  2\left\lfloor\left.\frac{k}{4}\right\rfloor\right. + \left\lfloor\left.\frac{k}{3}\right\rfloor\right. - \frac{k}{2},\\
        \mathcal{M}_{k}(\Gamma_{0}^{+}(7)) =& 1 + \text{dim}\ \mathcal{S}_{k}(\Gamma_{0}^{+}(7)),
    \end{split}
\end{equation}
for $k\in2\mathbb{Z}$ and $k>2$. The space of modular forms of $\Gamma_{0}^{+}(7)$ has the following basis decomposition
\begin{align}\label{Fricke_basis}
    \begin{split}
     \mathcal{M}_{k}(\Gamma^{+}_{0}(7)) =& E^{(7^{+})}_{\overline{k}}(\tau)\left(\mathbb{C}\left(\theta_{7}^{3}\right)^{n}\oplus\mathbb{C}\left(\theta_{7}^{3}\right)^{n-1}\Delta_{7}\oplus\mathbb{C}\left(\theta_{7}^{3}\right)^{n-2}\left(\Delta_{7}\right)^{2}\oplus\ldots\oplus\mathbb{C}\left(\Delta_{7}\right)^{n}\right),\\
    \Delta_{7}(\tau) \equiv& \Delta_{7}^{\infty}(\tau)\Delta_{7}^{0}(\tau) = \eta^{3}(\tau)\eta^{3}(7\tau) =  \mathbf{k}(\tau)\mathbf{t}(\tau),
    \end{split}
\end{align}
where $n = \text{dim}\mathcal{M}_{k}(\Gamma^{+}_{0}(7)) - 1$, and $E^{(7^{+})}_{\overline{k}}(\tau)$ is defined as follows
\begin{align}
    \begin{split}
     E^{(7^{+})}_{\overline{k}} = \begin{cases}
     1,\ & k \equiv 0\ (\text{mod}\ 12),\\
     \theta_{7}^{2}\tfrac{\Delta_{7,10}}{\Delta_{7^{+},4}\Delta_{7}},\ & k\equiv 2\ (\text{mod}\ 12),\\
     \theta_{7},\ & k\equiv 4\ (\text{mod}\ 12),\\
     \tfrac{\Delta_{7,10}}{\Delta_{7^{+},4}\Delta_{7}},\ & k\equiv 6\ (\text{mod}\ 12),\\
     \theta_{7}^{2},\ &k\equiv 8\ (\text{mod}\ 12),\\
     \theta_{7}\tfrac{\Delta_{7,10}}{\Delta_{7^{+},4}\Delta_{7}},\ & k\equiv 10\ (\text{mod}\ 12).
    \end{cases}
    \end{split}
\end{align}
where the $\Delta_{7^{+},4}$ is a cusp form for $\Gamma^{+}_{0}(7)$ of weight $4$ defined in table \ref{tab:table_Delta} and $\Delta_{7,10}$ is a cusp form for $\Gamma_{0}^{(7)}$ of weight $10$.
For $2k \equiv 0\ (\text{mod}\ 4)$ and an arbitrary prime $p$, the theta series $\Theta_{k,p}\in\Gamma_{0}^{+}(p)$, is defined as follows \cite{Krieg}
\begin{align}
        \Theta_{k,p}(\tau)\equiv \frac{1}{1+(-p)^{\frac{k}{2}}}\left(E_{k}(\tau) + \left(-p\right)^{\frac{k}{2}} E_{k}(p\tau)\right),\ \tau\in\mathbb{H}^{2}.
\end{align}
For the case $k=2$ and $p=7$ we have
\begin{align}\label{Theta_2,7-theta_7}
    \Theta_{2,7}(\tau) = \frac{1}{6}\left(7E_{2}(7\tau) - E_{2}(\tau)\right) = \theta^{2}_{7}(\tau).
\end{align}
These forms are defined as follows
\begin{align}
    \begin{split}
        \Delta_{7^{+},4}(\tau) \equiv& \frac{5}{16}\left(\Theta_{2,7}^{2}(\tau) - \Theta_{4,7}(\tau)\right) = \theta_{7}(\tau)j_{7}^{-1}(\tau)\mathbf{k}(\tau)\\
        =& q - q^2 - 2 q^3 - 7 q^4 +\ldots\\
        \Delta_{7,10}(\tau) \equiv& \frac{559}{690}\left(\frac{41065}{137592}\left(E^{(7^{+})}_{4}(\tau)E^{(7^{+})}_{6}(\tau) - E^{(7^{+})}_{10}(\tau)\right) - E^{(7^{+})}_{6}(\tau)\Delta_{7,4}(\tau)\right)\\
        =& q^2 - 11 q^3 - 6 q^4 +\ldots
    \end{split}
\end{align}
\begin{table}[htb!]
    \centering
    \begin{tabular}{||c|c|c|c|c|c||}
        \hline
        $k$ &  $r_{k}$ & $\Delta_{7^{+},r_{k}}(\tau)$ & $\text{dim}\ \mathcal{S}_{r_{k}}(\Gamma_{0}^{+}(7)$ & Behaviour of cusp form & Label\\ [0.5ex]
        \hline\hline
        $12\mathbb{Z}$ & 0 & 1 &  0 &$\mathcal{O}(1)$ & -\\
        $4$ & $4$ & $\theta_{7}(\tau)j_{7}^{-1}(\tau)\mathbf{k}(\tau)$ &  1 & $q  - q^{2} - 2q^{3} - 7q^{4} + \ldots$ & $7.4.a.a$\\
        $6$ & $6$ & $\sum\limits_{n=1}^{\infty}a_{n}\left(j_{7}^{-1}(\tau)\mathbf{k}(\tau)\right)^{n}$ & 1 & $q - 10q^{2} - 14q^{3} + 68q^{4} +\ldots$ & $7.6.a.a$\\ [0.5ex]
        \hline
    \end{tabular}
    \caption{The table shows the expression of the cusp form $\Delta^{2}_{7^{+},r_{k}}(\tau)$ for a specific weight $r_{k}$, the dimension of the space of cusp forms of the Fricke group of level $7$ of weight $r_{k}$ and the corresponding label found in Stein's tables \cite{Stein}.}
    \label{tab:table_Delta}
\end{table}

\subsubsection{Single character solutions}
Since $n=1$ and $\Delta_{i} = 0$ for the vacuum state, using \ref{central_charge_valence_Fricke} we find
\begin{align}\label{c=16l}
    c = 16\ell.
\end{align}

\subsection*{\texorpdfstring{$\ell = 0$}{Lg}}
For the simple case when there are no poles, $\ell = 0$, we have to make choices for the functions $\omega_{0}(\tau)$ and $\omega_{2}(\tau)$. Since the spaces $\mathcal{M}_{0}(\Gamma_{0}^{+}(7))$ and $\mathcal{M}_{2}(\Gamma_{0}(7))$ are one-dimensional, the only solution is the trivial one, $f(\tau) = \chi^{(\ell = 1)}(\tau) = 1$.

\subsection*{\texorpdfstring{$\ell = 1$}{Lg}}
If an admissible theory exists for $\ell = 1$, then it would have a central charge $c = 16$. The MLDE in this case takes the following form
\begin{align}
    \left[\omega_{2}(\tau)\mathcal{D} + \omega_{4}(\tau)\right]f(\tau) = 0.
\end{align}
The space $\mathcal{M}_{4}\left(\Gamma_{0}^{+}(7)\right)$ is two-dimensional and possesses the following basis decomposition
\begin{align}
    \mathcal{M}_{4}\left(\Gamma_{0}^{+}(7)\right) = \theta_{7}(\tau)\left(\mathbb{C}\theta_{7}^{3}\oplus\mathbb{C}\Delta_{7}\right).
\end{align}
We have $\omega_{2}(\tau) = 1$ but require the ratio $\tfrac{\omega_{4}(\tau)}{\omega_{2}(\tau)}$ to be a weight two modular form. Hence, we consider the ratio of the modular forms from the space of weight $6$ and $4$ to obtain the unique choice
\begin{align}
    \frac{\omega_{4}(\tau)}{\omega_{2}(\tau)} = \left(\mu_{1}\frac{\Delta_{7,10}\theta_{7}^{3}}{\Delta_{7^{+},4}\Delta_{7}\left(\mu_{3}\theta_{7}^{4} + \mu_{4}\theta_{7}\Delta_{7}\right)} + \mu_{2}\frac{\Delta_{7,10}\theta_{7}^{3}}{\Delta_{7^{+},4}\left(\mu_{3}\theta_{7}^{4} + \mu_{4}\theta_{7}\Delta_{7}\right)}\right)(\tau),
\end{align}
where $\mu_{1} = \tfrac{2}{3}$. Now, matching $q$-series with the ansatz $\chi(\tau) = j_{7^{+}}^{\tfrac{2}{3}}(\tau)$, we find $\mu_{2} = 0$, $\mu_{3} = 1$, and $\mu_{4} = 0$. The final form of the MLDE reads
\begin{align}
    \left[\mathcal{D} + \mu_{1}\left(\frac{\Delta_{7,10}}{\Delta_{7^{+},4}\Delta_{7}\theta_{7}}\right)(\tau)\right]f(\tau) = 0,
\end{align}
with the admissible solution
\begin{align}
    \chi^{(\ell = 1)}(\tau) =& j_{7^{+}}^{\tfrac{2}{3}}(\tau)\\
    =& q^{-\tfrac{2}{3}}\left(1 + 6 q + 25 q^2 + 70 q^3 + 180 q^4 + \ldots\right).
\end{align}

\subsection*{\texorpdfstring{$\ell = 2,4,5$}{Lg}}
Redoing the calculations at $\ell = 2,4,5$, we find that the following admissible solutions corresponding to central charges $c = 32, 64, 80$ respectively,
\begin{align}
    \begin{split}
        \chi^{(\ell = 2)}(\tau) =& j_{7^{+}}^{\tfrac{4}{3}}(\tau)\\
        =& q^{-\tfrac{2}{3}}\left(1 + 12 q + 86 q^2 + 440 q^3 + 1825 q^4 + \ldots\right),\\
        \chi^{(\ell = 4)}(\tau) =& j_{7^{+}}^{\tfrac{8}{3}}(\tau)\\
        =& q^{-\tfrac{8}{3}}\left(1 + 24 q + 316 q^2 + 2944 q^3 + 21606 q^4 + \ldots\right),\\
        \chi^{(\ell = 5)}(\tau) =& j_{7^{+}}^{\tfrac{10}{3}}(\tau)\\
        =& q^{-\tfrac{10}{3}}\left(1 + 30 q + 485 q^2 + 5510 q^3 + 49030 q^4 + \ldots\right).
   \end{split}
\end{align}

\subsection*{\texorpdfstring{$\ell = 3$}{Lg}}
If an admissible theory exists for $\ell = 3$, then it would have a central charge $c = 48$. The MLDE in this case takes the following form
\begin{align}
    \left[\omega_{6}(\tau)\mathcal{D} + \omega_{8}(\tau)\right]f(\tau) = 0.
\end{align}
By matching, we find $\mu = \tfrac{c}{24} = 2$. From \ref{Fricke_basis}, we obtain the following basis decompositions for $k=6$ and $k = 8$,
\begin{align}
    \begin{split}
        \mathcal{M}_{6}(\Gamma_{0}^{+}(7)) =& \frac{\Delta_{7,10}(\tau)}{\Delta_{7^{+},4}(\tau)\Delta_{7}(\tau)}\left(\mathbb{C}\theta_{7}^{3}\oplus \mathbb{C}\Delta_{7}\right),\\
        \mathcal{M}_{8}(\Gamma_{0}^{+}(7)) =& \theta_{7}^{2}(\tau)\left(\mathbb{C}\theta_{7}^{6}\oplus \mathbb{C}\theta_{7}^{3}\Delta_{7}\oplus \mathbb{C}\Delta_{7}^{2}\right).
    \end{split}
\end{align}
We make the following choices for the ratio $\tfrac{\omega_{8}(\tau)}{\omega_{6}(\tau)}$
\begin{align}
    \begin{split}
    \frac{\omega_{8}(\tau)}{\omega_{6}(\tau)} =& \left(\frac{\mu_{1}\theta_{7}^{8} + \mu_{2}\left(\theta_{7}^{5}\Delta_{7}\right) +\mu_{3}\left(\theta_{7}^{2}\Delta_{7}^{2}\right)}{\mu_{4}\frac{\Delta_{7,10}\theta_{7}^{3}}{\Delta_{7^{+},4}\Delta_{7}} + \mu_{5}\frac{\Delta_{7,10}}{\Delta_{7^{+},4}}}\right)(\tau),
    \end{split}
\end{align}
where $\mu_{1} = 2$. Matching series with the ansatz $\chi(\tau) = j_{7^{+}}^{2}(\tau)$, we find $\mu_{2} = -52$,  $\mu_{3} = -54$, $\mu_{4} = 1$, and $\mu_{5} = 0$. Thus, the admissible solution reads
\begin{align}
    \begin{split}
        \chi^{(\ell = 3)}(\tau) =& j_{7^{+}}^{2}(\tau)\\
        =& q^{-2}\left(1 + 18 q + 183 q^2 + 1326 q^3 + 7635 q^4 + \ldots\right).
    \end{split}
\end{align}

\subsection*{\texorpdfstring{$\ell = 6$}{Lg}}
If an admissible theory exists for $\ell = 6$, then it would have a central charge $c = 96$. The MLDE in this case takes the following form
\begin{align}
    \left[\omega_{12}(\tau)\mathcal{D} + \omega_{14}(\tau)\right]f(\tau) = 0.
\end{align}
By matching, we find $\mu = \tfrac{c}{24} = 4$. From \ref{Fricke_basis}, we obtain the following basis
\begin{align}
    \begin{split}
        \mathcal{M}_{12}(\Gamma_{0}^{+}(7)) =& \mathbb{C}\theta_{7}^{12}\oplus\mathbb{C}\theta_{7}^{9}\Delta_{7}\oplus \mathbb{C}\theta_{7}^{6}\Delta_{7}^{2}\oplus\mathbb{C}\theta_{7}^{3}\Delta_{7}^{3}\oplus\mathbb{C}\Delta_{7}^{4},\\
        \mathcal{M}_{14}(\Gamma_{0}^{+}(7)) =& \frac{\theta_{7}^{2}\Delta_{7,10}}{\Delta_{7^{+},4}\Delta_{7}}\left(\mathbb{C}\theta_{7}^{9}\oplus\mathbb{C}\theta_{7}^{6}\Delta_{7}\oplus\mathbb{C}\theta_{7}^{3}\Delta_{7}^{2}\oplus\mathbb{C}\Delta_{7}^{3}\right).
    \end{split}
\end{align}
We make the following choices for $\omega_{12}(\tau)$ and $\omega_{14}(\tau)$
\begin{align}
    \begin{split}
        \omega_{12} =& \mu_{1}\theta_{7}^{12} + \mu_{2}\theta_{7}^{9}\Delta_{7} + \mu_{3}\theta_{7}^{6}\Delta_{7}^{2} + \mu_{4}\theta_{7}^{3}\Delta_{7}^{3} + \mu_{5}\Delta_{7}^{4},\\
        \omega_{14} =& \frac{\theta_{7}^{2}\Delta_{7,10}}{\Delta_{7^{+},4}\Delta_{7}}\left(\mu_{6}\theta_{7}^{9} + \mu_{7}\theta_{7}^{6}\Delta_{7} + \mu_{8}\theta_{7}^{3}\Delta_{7}^{2} + \mu_{9}\Delta_{7}^{3}\right).
    \end{split}
\end{align}
Repeating the method we used at $\ell = 6$ in the previous sections to fix the free parameters, we find the following admissible solution
\begin{align}\label{single_char_Fricke}
    \begin{split}
        \begin{split}
        \chi^{(\ell = 6)}(\tau) =& j_{7^{+}}^{4}(\tau)\\
        =& q^{-4}\left(1 + 36q + 690q^{2} + 9240q^{3} + 96495q^{4} + \ldots\right).
        \end{split}
    \end{split}
\end{align}
We can now express the $\Gamma_{0}^{+}(7)$ characters for $\ell\leq 6$ as follows
\begin{align}
    \begin{split}
        \chi(\tau) =& j_{7^{+}}^{w_{\rho}}(\tau),\\
        c =& 24w_{\rho} = 16\ell,
    \end{split}
\end{align}
where $w_{\rho}\in\left\{0,\tfrac{2}{3},\tfrac{4}{3},2,\tfrac{8}{3},\tfrac{10}{3},4\right\}$ corresponding to $\ell\in\{0,1,2,3,4,5,6\}$ respectively.
\subsection{\texorpdfstring{$\Gamma_{0}(49)$}{Γ0(49)}}
\begin{figure}[htb!]
    \centering
    \begin{tikzcd}
	&& {\mathfrak{X}=X(7)} \\
	&& {} \\
	{X_{1}(7)} & {} && {\mathfrak{E}\cong X_{0}(49)\subset X_{0}^{+}(49)} \\
	\\
	&& {X_{0}(7)\subset X_{0}^{+}(7)} \\
	\\
	&& {X(1)\cong\mathbb{P}^{1}(\mathbb{C})}
	\arrow["7", from=3-4, to=5-3]
	\arrow["3", from=1-3, to=3-4]
	\arrow["7"', from=1-3, to=3-1]
	\arrow["3"', from=3-1, to=5-3]
	\arrow["8", from=5-3, to=7-3]
\end{tikzcd}
\caption{The numbers accompanying the maps indicate the degree of covering which is equal to the index as shown in \ref{degree_index}. The index of any other covering can be computed by taking the product of the individual index of each map. Recall that $\mathbb{P}^{1}(\mathbb{C}) = \mathbb{C}\cup \{i\infty\}$ denotes the Riemann sphere.}
\label{fig:commutative_diagram}
\end{figure}
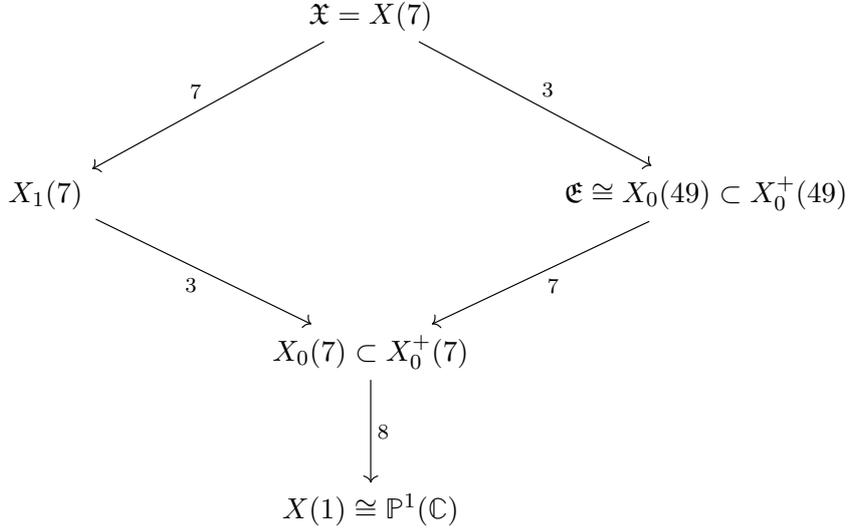

We present a preliminary introduction to the curve $X_{0}(49)$ and the Hecke group $\Gamma_{0}(49)$ following \cite{Ramanujan,Modular_tower}. Up to an isomorphism, there exists only one elliptic curve $\mathfrak{E}$ defined over $\mathbb{C}$ with complex multiplication by $\mathbb{Z}[\alpha]$. A morphism of the Klein quartic $\mathfrak{X}$ into $\mathfrak{E}$. The modular curve $X_{0}(49) = \Gamma_{0}(49)\backslash\mathbb{H}^{2*}$ is isomorphic to $\mathfrak{E}$ with the isomorphism defined as follows
\begin{align}
    G:\ X_{0}(49)\to \mathfrak{E},
\end{align}
where $G(\tau) = (\mathbf{u}(\tau),\mathbf{v}(\tau))$, and 
\begin{align}
    \begin{split}
        \mathbf{u}(\tau) =& \frac{\eta(49\tau)}{\eta(\tau)},\\
        =& 0 + 0q + q^{2} + q^3 + 2 q^4 + \ldots,\\
        \mathbf{v}(\tau) =& \left(\frac{2\mathbf{t} - 7\textbf{u}\left(1 + 5\textbf{u} + 7\textbf{u}^{2}\right)}{1 + 7\textbf{u} + 7\textbf{u}^{2}}\right)(\tau)\\
        =&2 - 13 q + 98 q^2 - 606 q^3 + 3878 q^4 + \ldots
    \end{split}
\end{align}
$X_{0}(49)$ is one of the curves in the recursive tower corresponding to the elliptic modular curves $X_{0}(7^{n+2})$. The curve $X_{0}(49)$ is elliptic with the following affine equation 
\begin{align}\label{parameterizations}
    \begin{split}
    y^{2} =& 4x^{3} + 21x^{2} + 28x,\\
    x(\tau) =& \frac{\eta(\tau)}{\eta(49\tau)},\ \ y(\tau) = 2\left(\frac{\mathcal{E}_{21}(\tau)}{\mathcal{E}_{7}(\tau)} + \frac{\mathcal{E}_{7}(\tau)}{\mathcal{E}_{14}(\tau)} - \frac{\mathcal{E}_{14}(\tau)}{\mathcal{E}_{21}(\tau)}\right) - 3x(\tau) - 4,
    \end{split}
\end{align}
where $\mathcal{E}_{g}(\tau)$ is called the generalized Dedekind $\eta$-function defined as follows
\begin{align}
    \mathcal{E}_{g}(\tau) \equiv q^{\frac{N}{2}B\left(\frac{g}{N}\right)}\prod\limits_{m=1}^{\infty}\left(1 - q^{(m-1)N+g}\right)\left(1 - q^{mN-g}\right),\ \ \tau\in\mathbb{C},\ \text{Im}\ \tau>0,
\end{align}
where $N\in\mathbb{Z}_{+}$ indicates the level, $g$ is an arbitrary real number that is not congruent to $0\ (\text{mod}\ N)$, and $B(x) = x^{2} - x + \tfrac{1}{6}$. This function satisfies
\begin{align}
    \mathcal{E}_{g+N} = \mathcal{E}_{-g} = -\mathcal{E}_{g}.
\end{align}
With $\gamma = \left(\begin{smallmatrix} a & b\\ cN& d\end{smallmatrix}\right)\in\Gamma_{0}(N)$, we have
\begin{align}
    \begin{split}
        \mathcal{E}_{g}(\tau + b) =& e^{\pi ibNB\left(\frac{g}{N}\right)}\mathcal{E}_{g}(\tau),\ \text{for}\ c=0,\\
        \mathcal{E}_{g}(\gamma(\tau)) =& \epsilon(a,bN,c,d)e^{\pi i\frac{g^{a}ab}{N - gb}}\mathcal{E}_{g}(\tau),
    \end{split}
\end{align}
where $\epsilon(a,b,c,d)$ is defined as in theorem \ref{thm:Apostol_eta}. The commutative diagram  that describes the relations among the various coverings of the Klein quartic is shown in \ref{fig:commutative_diagram}.
\\
\\
The degree of coverings is nothing but the index which is found to be the following
\begin{align}\label{degree_index}
    \begin{split}
        &\text{deg}\left(X(7)\to X_{1}(7)\right) = \left[\overline{\Gamma}_{1}(7):\overline{\Gamma}(7)\right] = 7,\\
        &\text{deg}\left(X_{1}(7)\to X_{0}(7)\right) = \left[\overline{\Gamma}_{0}(7):\overline{\Gamma}_{1}(7)\right] =3,\\
        &\text{deg}\left(X_{0}(7)\to X(1)\right) = \left[\text{PSL}(2,\mathbb{Z}):\overline{\Gamma}_{0}(7)\right] = 8,\\
        &\text{deg}\left(X_{0}(49)\to X_{0}(7)\right) = \left[\overline{\Gamma}_{0}(7):\overline{\Gamma}_{0}(49)\right] = 7,\\
        &\text{deg}\left(X(7)\to X_{0}(49)\right) = \left[\overline{\Gamma}_{0}(49):\overline{\Gamma}(7)\right] = \frac{\left[\text{PSL}(2,\mathbb{Z}):\overline{\Gamma}(7)\right]}{\left[\text{PSL}(2,\mathbb{Z}):\overline{\Gamma}_{0}(7)\right]\left[\overline{\Gamma}_{0}(7):\overline{\Gamma}_{0}(49)\right]} = 3,
    \end{split}
\end{align}
\noindent
where we used \ref{index}, \ref{index_subgroups}, and \ref{index_relations}. Coverings $X_{1}(7)\overset{24}{\to} X(1)$ and $X_{0}(7)\overset{8}{\to} X(1)$ are subcoverings of $\mathfrak{X}$ whose expressions can be found in \cite{Kenji}. The covering map $X_{0}(49)\to X_{0}(7)$ is given by \cite{Modular_tower}
\begin{align}
    \begin{split}
        j_{7}(7\tau) =& \frac{1}{2}\left[A(x(\tau))y(\tau) + B(x(\tau)\right],\\
        A(x) =& x^{2} + 7x + 7,\ B(x) = 7x^{3} + 35x^{2} + 29x + 16.
    \end{split}
\end{align}
Since $X_{0}^{+}(7)\subset X_{0}(7)$, and the Hauptmodules $j_{7^{+}}$ and $j_{7}$ are related by \ref{j_7+_and_j_7}, we find the covering $X_{0}(49)\to X_{0}^{+}(7)$ to be
\begin{align}\label{X49_X_7+_covering}
    j_{7^{+}}(7\tau) = 13 + \frac{98}{A(x(\tau))y(\tau) + B(x(\tau)} + \frac{1}{2}\left[A(x(\tau))y(\tau) + B(x(\tau)\right].
\end{align}
Using \ref{index_subgroups}, we find the index of $\Gamma_{0}(49)$ in $\text{SL}(2\mathbb{Z})$ to be $\left[\text{SL}(2,\mathbb{Z}),\Gamma_{0}(49)\right] = 56$. Using this we find that the number of zeros inside the fundamental domain $\mathcal{F}_{49}$ is given by 
\begin{align}
    \# =  2\ell\cdot\frac{56}{12} = \frac{28}{3}\ell.
\end{align}
This leads to the Riemann-Roch theorem for the group $\Gamma_{0}(49)$,
\begin{align}\label{Riemann_Roch_49}
    \sum\limits_{i=0}^{\infty}\alpha_{i} = \frac{14}{3}n(n-1) - \frac{28}{3}\ell.
\end{align}
\\
\begin{figure}[htb!]
    \centering
    \includegraphics[width = 15.5 cm]{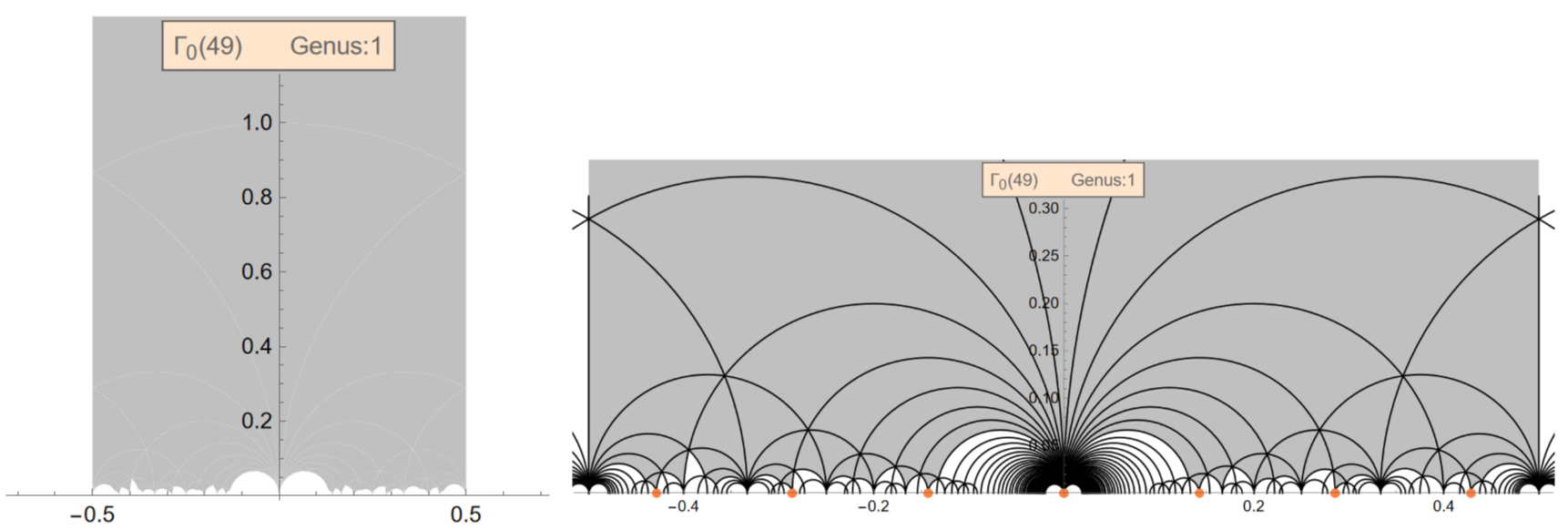}
    \caption{The fundamental domain $\mathcal{F}_{49}\cup \{i\infty\}$ without tesselation (left) and zoomed in with cusps indicated (right). There are $8$ cusps in total located at points $i\infty,0,\tfrac{1}{7}, \tfrac{2}{7},\tfrac{3}{7},\tfrac{4}{7},\tfrac{5}{7},\tfrac{6}{7}$.}
    \label{fig:fundamental_domain_49}
\end{figure}
\noindent
Using the expressions in \ref{thm:eliptic_points}, we fix
\begin{align}\label{49_data}
    \nu_{2}(\Gamma_{0}(49)) = 0,\  \nu_{3}(\Gamma_{0}(49)) = 2,\ \nu_{\infty}(\Gamma_{0}(49)) = 8,
\end{align}
where $\nu_{\infty}(\Gamma_{0}(49))$ was found using the genus formula given in theorem  \ref{thm:genus} and the fact that $X_{0}(49)$ is a genus $1$ modular curve. We find the dimension of the space of modular forms $\mathcal{M}_{k}(\Gamma_{0}(49))$ to be
\begin{align}\label{dim_49}
    \text{dim}\ \mathcal{M}_{k}(\Gamma_{0}(49)) = 2\left(\left\lfloor\left.\frac{k}{3}\right\rfloor\right. + 2k\right).
\end{align}
We shall only focus on the Eisenstein series part in the decomposition $\mathcal{M}_{k}(\Gamma_{0}(49)) = E_{k}(\Gamma_{0}(49)) \oplus \mathcal{S}_{k}(\Gamma_{0}(49))$. A standard modular form of weight $2$ for $\Gamma_{0}(49))$ is the Eisenstein series \cite{Ramanujan}
\begin{align}
    \begin{split}
        E^{(49)}_{2}(\tau) =& \frac{1}{4}\left(E_{2}^{(7)} + 2E_{2}^{(7)}(7\tau)\right) = 2 + \sum\limits_{n=1}^{\infty}\frac{n}{1-q^{n}}(q^{n} - 49q^{49n})\\
        =& 2 + q + 3 q^2 + 4 q^3 + 7 q^4 + \ldots
    \end{split}
\end{align}

\subsubsection{Single character solutions}
From the Riemann-Roch formula \ref{Riemann_Roch_49}, for $n = 1$ and $\Delta = 0$, we find
\begin{align}\label{c=224l}
    c = 224\ell.
\end{align}

\subsection*{\texorpdfstring{$\ell = 0$}{Lg}}
For the simple case when there are no poles, $\ell = 0$, we have to make choices for $\omega_{0}(\tau)$ and $\omega_{2}(\tau)$. We have $\omega_{0}(\tau) = 1$ and although $\mathcal{M}_{2}(\Gamma_{0}(49))$ is eight-dimensional, we shall restrict our focus to only the Eisenstein series part of the basis decomposition. Thus, we set $\omega_{2}(\tau) = \mu E_{2}^{(49)}(\tau)-1$. The MLDE reads
\begin{align}
    \left[\mathcal{D} + \mu\left( E_{2}^{(49)}(\tau)-1\right)\right]f(\tau) = 0.
\end{align}
Matching gives us $\mu = \tfrac{c}{24} = 0$ and hence, the only solution is the trivial one, i.e. $f(\tau) = \chi^{(\ell = 0)}(\tau) = 1$.

\subsection*{\texorpdfstring{$\ell = 3$}{Lg}}
Matching fixes $\mu = 28$ and with $\omega_{6}(\tau) = \mu_{1}\left(E_{2}^{(49)}(\tau) - 1\right)^{3}$ and $\omega_{8}(\tau) = \mu_{2}\left(E_{2}^{(49)}(\tau) - 1\right)^{4}$, we find the MLDE to be
\begin{align}
    \left[\mathcal{D} + \mu \left(E_{2}^{(49)}(\tau) - 1\right)\right]f(\tau) = 0,
\end{align}
where we have defined $\mu \equiv \tfrac{\mu_{2}}{\mu_{1}}$. The solution, $f(\tau) = \chi^{(\ell = 3)}(\tau)$, has the following $q$-series expansion
\begin{align}
    \chi^{(\ell = 3)}(\tau) = 1 - 28 q + 350 q^2 - 2520 q^3 + 11025 q^4 - 26180 q^5 + 4158 q^6 + \ldots
\end{align}
Surprisingly this can be obtained by taking the twenty-eighth power of the function $q^{\tfrac{49}{24}}x(\tau) = q^{\tfrac{49}{24}}j_{7}(\tau)j_{7}(7\tau)$ defined in \ref{parameterizations}, i.e.
\begin{align}\label{quasi_Gamma_0_49}
    \begin{split}
        \chi^{(\ell = 3)}(\tau) =& \left(q^{\tfrac{49}{24}}x(\tau)\right)^{28} = \left(\prod\limits_{n=1}^{\infty}\frac{1-q^{n}}{1-q^{49n}}\right)^{28}\\
        =& 1 - 28 q + 350 q^2 - 2520 q^3 + 11025 q^4 - 26180 q^5 + 4158 q^6+ \ldots
    \end{split}
\end{align}
After performing a check to the order $\mathcal{O}(q^{2001})$, we conclude that $\chi^{(\ell = 3)}(\tau)$ is a quasi-character that is not of type I or II since the coefficients possess alternating positive and negative signs.

\subsection{Ghostbusting at level 49}
\begin{figure}
    \centering
   \begin{tikzcd}
	&&& {\mathfrak{X}=X(7)} \\
	\\
	&&&&&& {\mathfrak{E}\cong X_{0}(49)} \\
	{X_{1}(7)} \\
	\\
	\\
	&&&&&& {X_{0}^{+}(49)} \\
	{X_{0}(7)} \\
	&&&&&&&&& {X(1)\cong\mathbb{P}^{1}(\mathbb{C})} \\
	&&& {X_{0}^{+}(7)}
	\arrow["7"'{pos=0.4}, from=1-4, to=4-1]
	\arrow["3"{pos=0.4}, from=1-4, to=3-7]
	\arrow["3"'{pos=0.4}, from=4-1, to=8-1]
	\arrow["7"'{pos=0.4}, color={rgb,255:red,92;green,92;blue,214}, from=3-7, to=8-1]
	\arrow["8"{pos=0.4}, color={rgb,255:red,214;green,92;blue,92}, from=8-1, to=9-10]
	\arrow["2"{pos=0.4}, from=10-4, to=8-1]
	\arrow["2"'{pos=0.4}, from=7-7, to=3-7]
	\arrow["4"{pos=0.4}, color={rgb,255:red,214;green,92;blue,92}, from=10-4, to=9-10]
	\arrow["28"'{pos=0.4}, color={rgb,255:red,214;green,92;blue,92}, from=7-7, to=9-10]
	\arrow["56"'{pos=0.4}, color={rgb,255:red,214;green,92;blue,92}, from=3-7, to=9-10]
	\arrow["7"{pos=0.4}, from=10-4, to=7-7]
	\arrow["24"{pos=0.4}, color={rgb,255:red,214;green,92;blue,92}, from=4-1, to=9-10]
	\arrow["168"{pos=0.4}, color={rgb,255:red,214;green,92;blue,92}, from=1-4, to=9-10]
	\arrow["14"'{pos=0.4}, color={rgb,255:red,92;green,92;blue,214}, from=7-7, to=8-1]
	\arrow["21"'{pos=0.4}, color={rgb,255:red,92;green,92;blue,214}, from=1-4, to=8-1]
\end{tikzcd}
    \label{fig:commutative_diagram_expanded}
    \caption{The extended commutative diagram with the degrees of each sub covering.}
\end{figure}
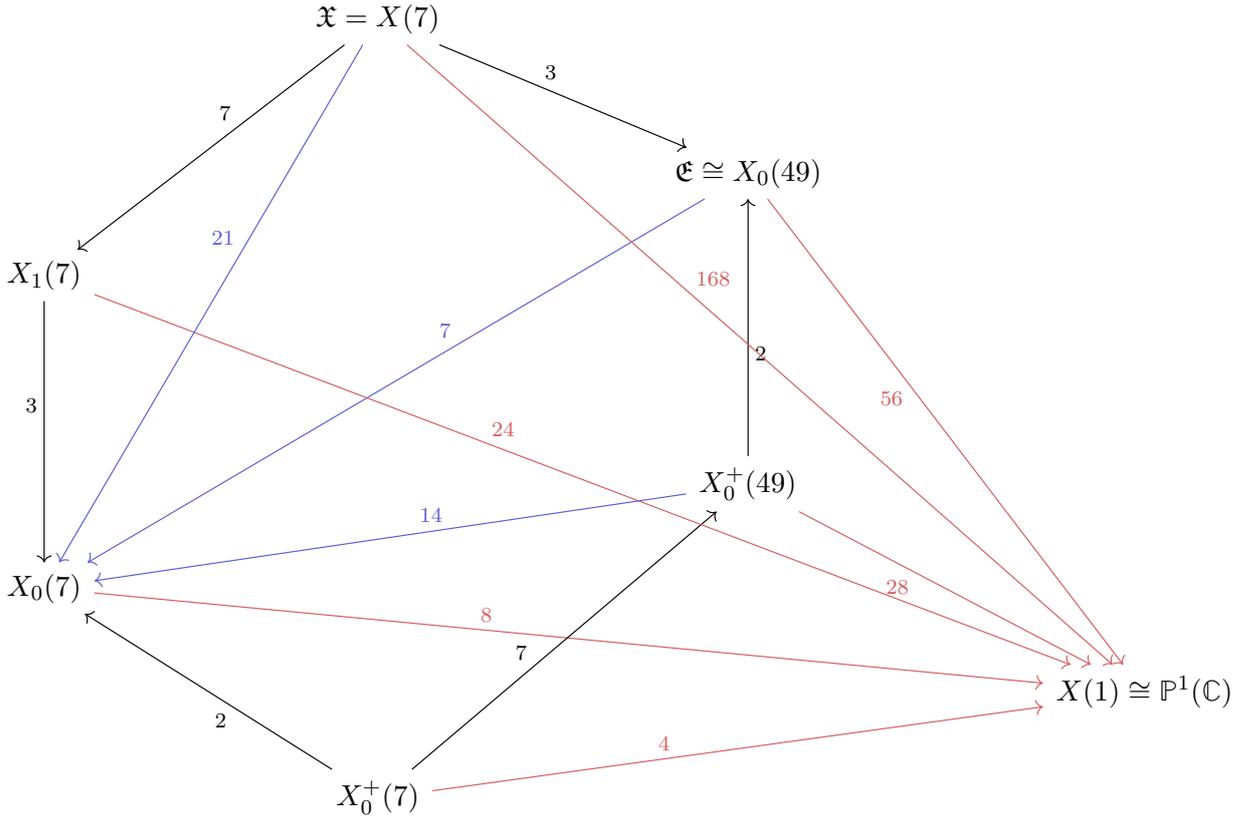

The groups $\Gamma^{+}_{0}(25)$, $\Gamma_{0}^{+}(49)$, and $\Gamma_{0}^{+}(50)$ are called the Conway-Norton ghosts \cite{Kok,Conway}. These three are genus zero groups that do not possess a product formula, i.e. there exist no eta-products corresponding to these subgroups. focusing on the Fricke group of level $49$ from theorem \ref{thm:genus_Fricke_index}, we find
\begin{align}
    \chi(\Gamma_{0}^{+}(49)) = \frac{49}{6}\prod\limits_{p\vert 49}\frac{p+1}{2p} = \frac{14}{3}.
\end{align}
The data for groups $\Gamma_{0}^{+}(49)$ and $\Gamma_{0}(49)$ is shown in table \ref{tab:data_level_49,49+}. 
\begin{table}[htb!]
    \centering
    \begin{tabular}{||c|c|c|c|c|c|c|c||}
        \hline
        $\Gamma'$ & $\nu_{2}(\Gamma')$ & $\nu_{3}(\Gamma')$ &
        $\nu_{4}(\Gamma')$ & $\nu_{6}(\Gamma')$ &
        $\nu_{\infty}(\Gamma')$ & $\chi(\Gamma')$ & $g(\Gamma')$\\[0.5ex]
        \hline\hline
        $\Gamma_{0}^{+}(49)$ & 4 & 1 & 0 & 0 & 4 & $\frac{14}{3}$ & 0\\
        $\Gamma_{0}(49)$ & 0 & 2 & 0 & 0 & 8 & $\frac{28}{3}$ & 1\\ [1ex]
        \hline
    \end{tabular}
    \caption{The data of the Hecke group $\Gamma_{0}(49))$ was computed in \ref{49_data} and that of the Fricke group $\Gamma_{0}^{+}(49)$ obtained from \cite{Kok}. We used \ref{index_euler_char} to compute $\chi(\Gamma_{0}(49))$.}
    \label{tab:data_level_49,49+}
\end{table}
Using this data in \ref{genus_Fricke_eqn}, we find the index of $\Gamma_{0}(49)$ in $\Gamma_{0}^{+}(49)$ to be
\begin{align}
    \left[\Gamma_{0}^{+}(49):\Gamma_{0}(49)\right] =  2,
\end{align}
Using \ref{index_subgroups}, we find the index of $\Gamma_{0}^{+}(49)$ in $\text{SL}(2,\mathbb{Z})$ to be
\begin{align}
    \mu = \left[\text{SL}(2,\mathbb{Z}):\Gamma_{0}^{+}(49)\right] = \frac{\left[\text{SL}(2,\mathbb{Z}):\Gamma_{0}(49)\right]}{\left[\Gamma_{0}^{+}(49):\Gamma_{0}(49)\right]} = \frac{56}{2} = 28. 
\end{align}
The expanded commutative diagram that describes the relations among the various coverings of the Klein quartic is shown in \ref{fig:commutative_diagram_expanded}.
\\
\noindent
The number of zeros inside the fundamental domain $\mathcal{F}_{49^{+}}$ is given by
\begin{align}
    \# = 2\ell\cdot \frac{28}{12} = \frac{14}{3}\ell.
\end{align}
This leads to the following Riemann-Roch theorem for the group $\Gamma_{0}^{+}(49)$,
\begin{align}\label{Riemann_Roch_49+}
    \sum\limits_{i=0}^{\infty}\alpha_{i} = \frac{7}{3}n(n-1) - \frac{14}{3}\ell.
\end{align}
Akin to what we did previously, we shall only consider $E_{k}$ in $\mathcal{M}_{k}(\Gamma_{0}(49)) = E_{k}(\Gamma_{0}(49)) \oplus\mathcal{S}_{k}(\Gamma_{0}(49))$. We couldn't find any literature discussing the theory of modular forms for $\Gamma_{0}^{+}(49)$ which leaves us to guess suitable candidates. A first guess for the standard modular form of weight $2$ for $\Gamma_{0}^{+}(49))$ would be the level $49$ Eisenstein series defined as follows
\begin{align}
    \begin{split}
        E^{(49^{+})}_{2}(\tau) =& \frac{1}{50}\left(E_{2}(\tau) + 49E_{2}(49\tau)\right)\\
        =& 2 + q + 3 q^2 + 4 q^3 + 7 q^4 + \ldots
    \end{split}
\end{align}
This, however, turns out to be a quasi-modular form. There exists a morphism between the curve $X_{0}(N^{2})$ and the curve $X(N)$ (see proposition $8.1$ in \cite{Ramanujan} and references therein for the case when $N = 7$). This can also be seen from the coinciding index formulae in \ref{index_relations}. Since $X_{0}(49)\subset X_{0}^{+}(49)$, we use this as a hint and loosely choose instead the modular forms of the group $\Gamma(7)$ as a first approximation. The dimension of the space of modular forms of weight $k$ for any congruence subgroup of level $N$, $\Gamma(N)$, is given by
\begin{align}
    \text{dim}\ \mathcal{M}_{k}(\Gamma(N)) = \frac{N(k-1)+6}{24}N^{2}\prod\limits_{p\vert N}\left(1 - p^{-2}\right),\ N>2,\ k\geq 1.
\end{align}
In the case of $N = 7$, we find
\begin{align}
    \text{dim}\ \mathcal{M}_{k}(\Gamma(7)) = 14k-2.
\end{align}

\subsubsection{Single character solutions}
From the Riemann-Roch formula \ref{Riemann_Roch_49+}, for $n = 1$ and $\Delta = 0$, we find
\begin{align}\label{c=112l}
    c = 112\ell.
\end{align}

\subsection*{\texorpdfstring{$\ell = 0$}{Lg}}
For the simple case when there are no poles, $\ell = 0$, we have to make choices for $\omega_{0}(\tau)$ and $\omega_{2}(\tau)$. Since $\mathcal{M}_{0}(\Gamma(7))$ is one-dimensional, $\omega_{0}(\tau) = 1$ and $\mathcal{M}_{2}(\Gamma(7))$ is twenty-six-dimensional, we shall restrict our focus to only a simple weight $2$ modular form and hence, we set $\omega_{2}(\tau) = \mu \mathfrak{a}(\tau)$. The MLDE reads
\begin{align}
    \left[\mathcal{D} + \mu \mathfrak{a}(\tau)\right]f(\tau) = 0.
\end{align}
Matching gives us $\mu = \tfrac{c}{24} = 0$ and hence, the only solution is the trivial one, i.e. $f(\tau) = \chi^{(\ell = 0)}(\tau) = 1$.

\subsection*{\texorpdfstring{$\ell = 3$}{Lg}}
Matching fixes $\mu = 14$. From \cite{mod_forms_principal_level_7}, we have
\begin{align}
    \begin{split}
        \mathfrak{a}(\tau) = \frac{\alpha}{168}\left(343E_{2}(7\tau) - \sum\limits_{n=0}^{6}E_{2}\left(\frac{\tau + n}{7}\right)\right)\in\mathcal{M}_{2}(\Gamma(7)),
    \end{split}
\end{align}
where $\alpha$ is a constant that we set to $\alpha = \tfrac{1}{2}$. The $q$-series expansion of this weight $2$ modular form reads
\begin{align}
    \mathfrak{a}(\tau) = 1 + 4 q + 12 q^2 + 16 q^3 + 28 q^4 + 24 q^5 + 48 q^6 + 4 q^7 + \ldots
\end{align}
With choices $\omega_{6}(\tau) = \mu_{1}\mathfrak{a}^{3}(\tau)$ and $\omega_{8}(\tau) = \mathfrak{a}^{4}(\tau)$, we find the MLDE to be
\begin{align}
    \left[\mathcal{D} + \mu \mathfrak{a}(\tau)\right]f(\tau) = 0,
\end{align}
where we have defined $\mu \equiv \tfrac{\mu_{2}}{\mu_{1}}$. The solution, $f(\tau) = \chi^{(\ell = 3)}(\tau)$, has the following $q$-series expansion
\begin{align}
    \chi^{(\ell = 3)}(\tau) =1 - 56 q + 1484 q^2 - 24640 q^3 + 285670 q^4 - 2433760 q^5 +  15542296 q^6 + \ldots
\end{align}
Surprisingly this can be obtained by taking the fifty-sixth power of the function $q^{\tfrac{49}{24}}x(\tau)$ defined in \ref{parameterizations}, i.e.
\begin{align}\label{quasi_Gamma_0_49+}
    \begin{split}
        \chi^{(\ell = 3)}(\tau) =& \left(q^{\tfrac{49}{24}}x(\tau)\right)^{56} = \left(\prod\limits_{n=1}^{\infty}\frac{1-q^{n}}{1-q^{49n}}\right)^{56}\\
        =& 1 - 56 q + 1484 q^2 - 24640 q^3 + 285670 q^4 - 2433760 q^5 +  15542296 q^6 + \ldots
    \end{split}
\end{align}
\\
After a check to order $\mathcal{O}(q^{2001})$, we conclude that $\chi^{(\ell = 3)}(\tau)$ is a quasi-character that is not of type I or II since the coefficients possess alternating positive and negative signs. 

\section{\texorpdfstring{$\mathbf{\Gamma_{0}^{+}(11)}$}{Γ0(11)+}}\label{sec:Gamma_0_11+}
The Fricke group of level $11$ is generated by the $T$-matrix, the Atkin-Lehner involution of level $11$, $W_{11} = \left(\begin{smallmatrix}0 & -\tfrac{1}{\sqrt{11}}\\ \sqrt{11} & 0\end{smallmatrix}\right)$, $S^{-1}T^{-3}S^{-1}T^{-4}S$, and $S^{-1}T^{-4}S^{-1}T^{-3}S$, i.e. $\Gamma_{0}^{+}(11) = \langle\left(\begin{smallmatrix}1 & 1\\ 0 & 1\end{smallmatrix}\right), W_{3}, \left(\begin{smallmatrix}4 & 1\\ 11 & 3\end{smallmatrix}\right), \left(\begin{smallmatrix}3 & 1\\ 11 & 4\end{smallmatrix}\right)\rangle$. A non-zero $f\in\mathcal{M}_{k}^{!}(\Gamma_{0}^{+}(11))$ satisfies the following valence formula
\begin{align}
    \nu_{\infty}(f) + \frac{1}{2}\nu_{\tfrac{i}{\sqrt{11}}}(f) + \frac{1}{2}\nu_{\rho_{11,1}}(f) + \frac{1}{2}\nu_{\rho_{11,2}}(f) + \frac{1}{2}\nu_{\rho_{11,3}}(f) +  \sum\limits_{\substack{p\in\Gamma_{0}^{+}(11)\backslash\mathbb{H}^{2}\\ p\neq \tfrac{i}{\sqrt{11}},\rho_{11,1}, \rho_{11,2}, \rho_{11,3}}}\nu_{p}(f) =   \frac{k}{2},
\end{align}
where $\tfrac{i}{\sqrt{11}}$, $\rho_{11,1} = -\tfrac{1}{2}+\tfrac{i}{2\sqrt{11}}$, $\rho_{11,2} = -\tfrac{1}{3}+\tfrac{i}{3\sqrt{11}}$, and $\rho_{11,3} = \tfrac{1}{3}+\tfrac{i}{3\sqrt{11}}$ are elliptic points and this group has a cusp at $\tau = \infty$ (see \cite{Junichi} for more details). Using theorem \ref{thm:genus_Fricke_index}, with $(N,h,m)=(11,1,11)$, we have
\begin{align}
    \chi(\Gamma_{0}^{+}(11)) = \frac{11}{6}\prod\limits_{p\vert 11}\frac{p+1}{2p} = 1,\ \ 
    \left[\Gamma_{0}^{+}(11):\Gamma_{0}(11)\right] = 2,
\end{align}
i.e. $\Gamma_{0}(11)$ is an index-$2$ subgroup of $\Gamma_{0}^{+}(11)$. With this, we find the index of $\Gamma_{0}^{+}(11)$ in $\text{SL}(2,\mathbb{Z})$ to be
\begin{align}
    \left[\text{SL}(2,\mathbb{Z}):\Gamma_{0}^{+}(11)\right] = \frac{\left[\text{SL}(2,\mathbb{Z}):\Gamma_{0}(11)\right]}{\left[\Gamma^{+}_{0}(11):\Gamma_{0}(11)\right]} = \frac{12}{2} = 6.
\end{align}
Hence, we find the number of zeros in the fundamental domain $\mathcal{F}_{11^{+}}$ to be $\# = 2\ell\cdot \tfrac{\mu}{12} = \ell$. The Riemann-Roch theorem takes the following form
\begin{align}
    \begin{split}
        \sum\limits_{i=0}^{n-1}\alpha_{i} =& -\frac{nc}{24} + \sum\limits_{i}\Delta_{i}\\
        =& \frac{1}{6}n(n-1) - \ell.
    \end{split}
\end{align}
From this for $n = 1$, we find the following relation between the number of zeros in the fundamental domain and the central charge
\begin{align}
    c = 24\ell.
\end{align}
The space of modular forms of $\Gamma_{0}^{+}(11)$ has the following basis decomposition
\begin{align}\label{basis_decomposition_Gamma_0_11+}
    \begin{split}
    \mathcal{M}_{4n}(\Gamma_{0}^{+}(11)) =& \mathbb{C}\left(E_{k,11^{'}}\right)^{2n}\oplus\left(E_{k,11^{'}}\right)^{2n-1}\Delta_{11}\oplus\ldots\oplus\mathbb{C}\Delta_{11}^{2n}\\
    \mathcal{M}_{4n+6}(\Gamma_{0}^{+}(11)) =& E_{4,11^{'}}\left[\mathbb{C}\left(E_{k,11^{'}}\right)^{2n+1}\oplus\left(E_{k,11^{'}}\right)^{2n}\Delta_{11}\oplus\ldots\oplus\mathbb{C}\Delta_{11}^{2n+1}\right]
    \end{split}
\end{align}
where 
\begin{align}
    \begin{split}
        E_{2,11^{'}}(\tau) \equiv& \frac{11E_{2}(11\tau) - E_{2}(\tau)}{10}\in\mathcal{M}_{2}(\Gamma_{0}(11))\\
        =& \frac{1}{5}\left(5 + 12 q + 36 q^2 + 48 q^3 + 84 q^4 + \ldots\right),\\
        \Delta_{11}(\tau) \equiv& \left(\eta(\tau)\eta(11\tau)\right)^{2}\in\mathcal{S}_{2}(\Gamma_{0}(11))\\
        =& q - 2 q^2 - q^3 + 2 q^4 + \ldots,\\
        E_{k,11}^{\infty}(\tau) \equiv& \frac{11^{k}E_{k}(11\tau) - E_{k}(\tau)}{11^{k} - 1},\ k\geq 4,\\
        E_{4,11^{'}}(\tau) \equiv& \left(\frac{-1525\left(E_{2,11^{'}}\right)^{2} + 4320E_{2,11^{'}}\Delta_{11} + 2016\left(\Delta_{11}\right)^{2} + 3050E_{4,11}^{\infty}}{1525}\right)(\tau)\\
        =& \frac{1}{305}\left(305 - 599 q - 5397 q^2 - 16796 q^3 - 43793 q^4 + \ldots\right).
    \end{split}
\end{align}
$E_{k,11}^{\infty}$ is the Eisenstein series of weight $4$ corresponding to the cusp at $\infty$ for $\Gamma_{0}(11)$ and $E_{4,11^{'}}$ is a semimodular form of weight $4$ for $\Gamma_{0}^{+}(11)$. We find that there exist no admissible solutions at the single character level since all the coefficients turn out to be fractional. This is due to the type of coefficients in the $q$-series expansion of $E_{2,11^{'}}$ which forms the ring of modular forms at this level as shown in the basis decomposition \ref{basis_decomposition_Gamma_0_11+}. Based on previous calculations, we can expect solutions of the form $\chi(\tau) = j_{11^{+}}^{w_{\rho}}(\tau)$, but this ansatz would not hold due to the nature of the Hauptmodul which not only contains a fractional coefficient but also, unlike those at levels $N = 2,3,5,7$, does not map the fundamental domain to the real line. The Hauptmodul in $\Gamma_{0}^{+}(11)$ is defined as follows
\begin{align}
    j_{11^{+}}(\tau) \equiv& \left(\frac{E_{4,11^{'}}}{\Delta_{11}}\right)(\tau)\\
    =& q^{-1} + \frac{22}{5} + 17q + 47q^{2} + 116q^{3} + +252 q^4 + \ldots
\end{align}
which does not take real values on some arcs of the fundamental domain (see \cite{Junichi}). One might be tempted to conclude that the admissibility of solutions might be solely based on the fact of the incapability of the Hauptmodul to map $\mathcal{F}_{p^{+}}$ to purely the real line. We will consider the next prime divisor level, where this mapping isn't an issue, to show that there can be other issues that might plague the solutions. 

\section{\texorpdfstring{$\mathbf{\Gamma_{0}^{+}(13)}$}{Γ0(13)+}}\label{sec:Gamma_0_13+}
\begin{figure}[htb!]
    \centering
    \includegraphics[width = 16cm]{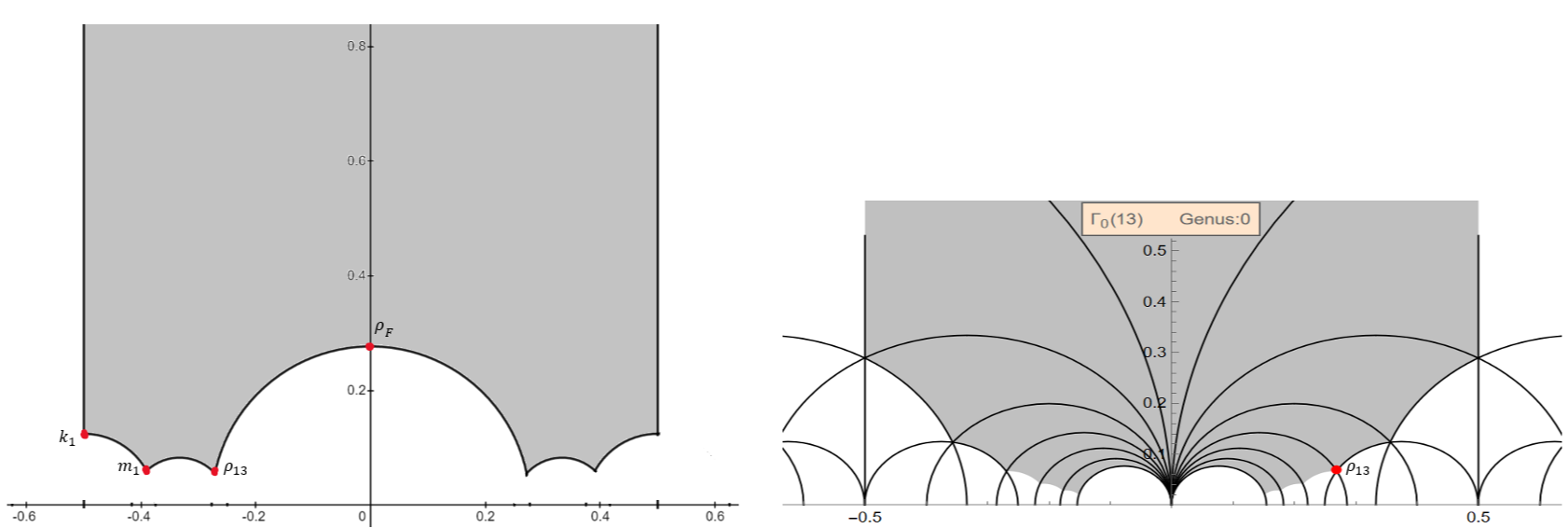}
    \caption{Fundamental domains of level $13$ groups. (Left) $\mathcal{F}_{13^{+}}$ obtained from Fricke folding of $\mathcal{F}_{13}$ (right). The marked (elliptic) points are $\rho_{F} = \tfrac{i}{\sqrt{13}}$ (also called the Fricke involution point),$\rho_{13} = \tfrac{\pm7 + i\sqrt{3}}{26}$, $k_{1} = -\tfrac{1}{2} + \tfrac{i}{2\sqrt{13}}$, and \& $m_{1} = \tfrac{-5 + i}{13}$.}
    \label{fig:fundamental_domain_13+}
\end{figure}

We have attributed the failure of obtaining admissible solutions in $\Gamma_{0}^{+}(11)$ to the non-integral coefficient associated with order $\mathcal{O}(q^{0})$ in its Hauptmodul and the fact that the Fundamental domain $\mathcal{F}_{11^{+}}$ is not mapped to the real line. Let us move ahead to the next prime divisor level, $p = 13$, and check if this group passes the criteria we have assumed to find admissible solutions. For the group $\Gamma_{0}^{+}(13)$, from \ref{J_function_McKay_Fricke}, we find
\begin{align}
    j_{13^{+}}(\tau) \equiv& \left(\frac{\eta(\tau)}{\eta(13\tau)}\right)^{2} + 13\left(\frac{\eta(13\tau)}{\eta(\tau)}\right)^{2}\\
    =& q^{-1} - 2 + 12 q + 28 q^2 + 66 q^3 + 132 q^4 + \ldots
\end{align}
This clearly possesses integral coefficients. From \ref{number_of_elliptic_points}, we find that the Hecke group $\Gamma_{0}(13)$ has $\epsilon_{\rho} = 2$ and $\epsilon_{i} = 2$ number of elliptic points with the $\rho$-elliptic points located at $\rho_{13} = \tfrac{\pm 7 + i\sqrt{3}}{26}$. Let us examine the behaviour of this at the cusp $i\infty$, at the elliptic points, $\rho_{13}$ (which is the point at which we start Fricke folding from $\mathcal{F}_{13}$) and other elliptic points (see figure \ref{fig:fundamental_domain_13+}). We have
\begin{align}
    j_{13^{+}}(i\infty) = \infty,\ j_{13^{+}}(\rho_{13}) = 7,\
    j_{13^{+}}(m_{1,2}) \approx 7.19615,\
    j_{13^{+}}(k_{1,2}) \approx 7.211,\ j_{13^{+}}(\rho_{F})\approx 7.211.
\end{align}
\noindent
Hence, we have the following map between the fundamental domain $\mathcal{F}_{13^{+}}$ and the complex plane,
\begin{align}
    j_{13^{+}}:\ \partial\mathcal{F}_{13^{+}}\backslash\left\{z\in\mathbb{H}^{2}; \ \text{Re}(z) = \pm\tfrac{1}{2}\right\}\to \left[7,7.211\right]\subset \mathbb{R}.
\end{align}
Although $\Gamma_{0}^{+}(13)$ seems to be an ideal candidate to probe for single character solutions, there turns out to be other issues that prevent the solutions from being admissible. Here the issues arise in the form of the basis decomposition. We will expand on this now. A non-zero $f\in\mathcal{M}_{k}^{!}(\Gamma_{0}^{+}(13))$ satisfies the following valence formula \cite{Junichi_extended}
\begin{align}
    \nu_{\infty}(f) + \frac{1}{2}\nu_{\tfrac{i}{\sqrt{13}}}(f) + \frac{1}{2}\nu_{\rho_{13,1}}(f) + \frac{1}{2}\nu_{\rho_{13,2}}(f) + \frac{1}{3}\nu_{\rho_{13,3}}(f) +  \sum\limits_{\substack{p\in\Gamma_{0}^{+}(13)\backslash\mathbb{H}^{2}\\ p\neq \tfrac{i}{\sqrt{13}},\rho_{13,1}, \rho_{13,2}, \rho_{13,3}}}\nu_{p}(f) =   \frac{7k}{12},
\end{align}
where $\tfrac{i}{\sqrt{13}}$, $\rho_{13,1} = -\tfrac{1}{2}+\tfrac{i}{2\sqrt{13}}$, $\rho_{13,2} = -\tfrac{5}{13}+\tfrac{i}{\sqrt{13}}$, and $\rho_{13,3} = -\tfrac{7}{26}+\tfrac{i\sqrt{3}}{26}$ are elliptic points (which is exactly equal to our construction above) and this group has a cusp at $\tau = \infty$. The dimension of the space of modular forms reads
\begin{align}
    \begin{split}
        \text{dim}\ \mathcal{S}_{k}(\Gamma_{0}^{+}(13)) = \begin{cases}
        \left\lfloor\left.\frac{k}{3}\right\rfloor\right. + 3\left\lfloor\left.\frac{k}{4}\right\rfloor\right. - \frac{k}{2},\ &k\geq 4,\\
        0,\ & k\leq 2,
        \end{cases}\\
        \text{dim}\ \mathcal{M}_{k}(\Gamma_{0}^{+}(13)) = 1 + \text{dim}\ \mathcal{S}_{k}(\Gamma_{0}^{+}(13)).
    \end{split}
\end{align}
The space of modular forms of $\Gamma_{0}^{+}(13)$ has the following basis decomposition
\begin{align}\label{modular_form_space_Gamma_0_13+}
    \mathcal{M}_{k}(\Gamma_{0}^{+}(13)) =\mathcal{S}_{k}(\Gamma_{0}^{+}(13))\oplus \mathbb{C}E_{k}^{(13^{+})}(\tau),
\end{align}
where
\begin{align}
    E^{(13^{+})}_{k}(\tau) = \frac{13^{\tfrac{k}{2}}E_{k}(13\tau) + E_{k}(\tau)}{13^{\tfrac{k}{2}} + 1},\ \text{for}\ k\geq 4.
\end{align}
Furthermore, we also see that we can cook up a modular form of weight $2$ as follows
\begin{align}
    \begin{split}
        E_{2,13^{'}}(\tau) =& \frac{13E_{2}(13\tau) - E_{2}(\tau)}{12}\in\mathcal{M}_{2}(\Gamma_{0}(13))\\
        =& 1 + 2 q + 6 q^2 + 8 q^3 + 14 q^4 + \ldots,\\
        E^{2}_{2,13^{'}}(\tau) =& \frac{13E_{2}(13\tau) - E_{2}(\tau)}{12}\in\mathcal{M}_{4}(\Gamma_{0}^{+}(13))\\
        =& 1 + 4 q + 16 q^2 + 40 q^3 + 96 q^4 \ldots,\\
    \end{split}
\end{align}
From all the MLDE calculations thus far, we have noticed that including the cusp form in our choice of $\omega_{2\ell}(\tau)$ yields inadmissible solutions. Hence, we shall only focus on the space of the Eisenstein series in \ref{modular_form_space_Gamma_0_13+}. Similar to the construction in \ref{Fricke_basis}, we can construct a following basis decomposition for the space of modular forms in $\Gamma_{0}^{+}(13)$ for every weight $k\in2\mathbb{Z}$. For example, when $k = 2$, from $\text{dim}\ \mathcal{M}_{2}(\Gamma_{0}^{+}(13)) = 1$ and we have $\mathcal{M}_{2}(\Gamma_{0}^{+}(13)) =1$. Next, for $k = 4$, $\text{dim}\ \mathcal{M}_{4}(\Gamma_{0}^{+}(13)) = 2$ and we have
\begin{align}
    \mathcal{M}_{4}(\Gamma_{0}^{+}(13)) = E_{4}^{(13^{+})}\oplus\mathbb{C}\left(E_{2,13^{'}}^{2} - E_{4}^{(13^{+})}\right).
\end{align}
For the case of $k = 6$, we find that we cannot construct a cusp form polynomially and hence, we have $\mathcal{M}_{6}(\Gamma_{0}^{+}(13)) = E_{6}^{(13^{+})}$. Next, when $k = 8$, $\text{dim}\ \mathcal{M}_{8}(\Gamma_{0}^{+}(13)) = 4$, and we have 
\begin{align}
    \begin{split}
        \mathcal{M}_{8}(\Gamma_{0}^{+}(13)) = E_{8}^{(13^{+})}&\oplus\left[\mathbb{C}E_{2,13^{'}}^{2}\left(E_{2,13^{'}}^{2} - E_{4}^{(13^{+})}\right)\oplus \mathbb{C}E_{4}^{(13^{+})}\left(E_{2,13^{'}}^{2} - E_{4}^{(13^{+})}\right)\right.\\
        &\oplus\left. \mathbb{C}\left(E_{8}^{(13^{+})} - \left(E_{2}^{(13^{+})}\right)^{2}\right)\oplus \mathbb{C}\left(E_{8}^{(13^{+})} - E_{4}^{(13^{+})}\right)\right].
    \end{split}
\end{align}
Similar constructions can be made for $k = 10,12$, and $14$. Using theorem \ref{thm:genus_Fricke_index}, with $(N,h,m)=(13,1,13)$, we have
\begin{align}
    \chi(\Gamma_{0}^{+}(13)) = \frac{13}{6}\prod\limits_{p\vert 13}\frac{p+1}{2p} = \frac{7}{6},\ \ 
    \left[\Gamma_{0}^{+}(13):\Gamma_{0}(13)\right] = 2,
\end{align}
i.e. $\Gamma_{0}(13)$ is an index-$2$ subgroup of $\Gamma_{0}^{+}(13)$. With this, we find the index of $\Gamma_{0}^{+}(13)$ in $\text{SL}(2,\mathbb{Z})$ to be
\begin{align}
    \left[\text{SL}(2,\mathbb{Z}):\Gamma_{0}^{+}(13)\right] = \frac{\left[\text{SL}(2,\mathbb{Z}):\Gamma_{0}(13)\right]}{\left[\Gamma^{+}_{0}(13):\Gamma_{0}(13)\right]} = \frac{14}{2} = 7.
\end{align}
Hence, we find the number of zeros in the fundamental domain $\mathcal{F}_{13^{+}}$ to be $\# = 2\ell\cdot \tfrac{\mu}{12} = \tfrac{7}{6}\ell$. The Riemann-Roch theorem takes the following form
\begin{align}
    \begin{split}
        \sum\limits_{i=0}^{n-1}\alpha_{i} =& -\frac{nc}{24} + \sum\limits_{i}\Delta_{i}\\
        =& \frac{7}{12}n(n-1) - \frac{7}{6}\ell.
    \end{split}
\end{align}
From this for $n = 1$, we find the following relation between the number of zeros in the fundamental domain and the central charge
\begin{align}
    c = 28\ell.
\end{align}
It turns out that the all of the choices $\tfrac{\omega_{2\ell + 2}(\tau)}{\omega_{2\ell}(\tau)}$ for different $\ell\leq 6$ lead to solutions of MLDE that contain fractional coefficients. This is due to the fact that the $q$-series expansions of the Eisenstein series $E_{k}^{(13^{+})}(\tau)$ and the cusp form expansion for each $k$ contain fractional coefficients that don't turn out to conspire to yield integral coefficients in the final solution as we previously saw for the case of $\Gamma_{0}^{+}(7)$. It could be that there are admissible solutions at $\ell\geq 6$ or that a more careful analysis of the space of modular forms of $\Gamma_{0}^{+}(13)$ reveals solutions, but we don't pursue the latter due to lack of sufficient mathematical literature.

\section{Other groups}\label{sec:other_groups}
Investigation for single character solutions in the Hecke groups of levels $N\leq 12$ did not reveal any admissible solutions except at level $N = 7$. In the previous section, we also found quasi-characters in $\Gamma_{0}(7)$. We see this repeat in the solutions of the $\Gamma_{0}(2)$ which we then use to construct the admissible solutions of $\Gamma_{0}^{+}(2)$. After this, we consider $\Gamma_{0}(3)$ as an example and present the single character analysis to elucidate some of the issues that make the solution inadmissible in the other Hecke group levels. What about Fricke groups of other levels? It turns out that there are no admissible single character solutions for $\Gamma_{0}^{+}(N)$ with $N\leq 12$ except at $N = 2,3,5,7$ which are all genus zero Moonshine groups and are the first four prime divisors of $\mathbb{M}$. 

\subsection{\texorpdfstring{$\Gamma_{0}(2)$}{Γ0(2)}}\label{sec:Gamma_0_2}
The Hecke group of level $2$ is generated by the $T$-matrix and $S^{-1}T^{-2}S$, i.e. $\Gamma_{0}(2) = \langle\left(\begin{smallmatrix}1 & 1\\ 0 & 1\end{smallmatrix}\right), \left(\begin{smallmatrix}1 & 0\\ 2 & 1\end{smallmatrix}\right)\rangle$. A non-zero $f\in\mathcal{M}_{k}^{!}(\Gamma_{0}(2))$ satisfies the following valence formula
\begin{align}
    \nu_{\infty}(f) + \nu_{0}(f) + \frac{1}{2}\nu_{\rho_{2}}(f) + \sum\limits_{\substack{p\in\Gamma_{0}(2)\backslash\mathbb{H}^{2}\\p\neq\rho_{2}}}\nu_{p}(f) = \frac{k}{4},
\end{align}
where $\rho_{2}$ is an elliptic point and this group has two cusps at $\tau = 0$ and $\tau = \infty$ (see \cite{Junichi} for more details). The dimension of the space of modular forms for $\Gamma_{0}(2)$ is given by (see pg. 90 of \cite{dimension_formulae})
\begin{align}
    \text{dim}\ \mathcal{M}_{k}(\Gamma_{0}(2)) \leq 1 + \frac{k}{4}.
\end{align}
Using \ref{index_subgroups}, we find the index of $\Gamma_{0}(2)$ in $\text{SL}(2,\mathbb{Z})$ to be $\mu_{0} = 3$ and hence, we find the number of zeros in the fundamental domain $\mathcal{F}_{2}$ to be $\# = 2\ell\cdot \tfrac{\mu}{12} = \tfrac{1}{2}\ell$. The Riemann-Roch theorem takes the following form
\begin{align}
    \begin{split}
        \sum\limits_{i=0}^{n-1}\alpha_{i} =& -\frac{nc}{24} + \sum\limits_{i}\Delta_{i}\\
        =&\frac{1}{4}n(n-1) - \frac{1}{2}\ell,
    \end{split}
\end{align}
From this for $n = 1$, we find the following relation between the number of zeros in the fundamental domain and the central charge
\begin{align}
    c = 12\ell.
\end{align}
We pause here to comment on the shared Riemann-Roch relations obtained here and that in the case of the theta group, $\Gamma_{\theta}$ (see \cite{Bae_Hecke_2}). There exists three groups between $\Gamma(1) = \text{SL}(2,\mathbb{Z})$ and $\Gamma(2)$- the unique index-$2$ normal subgroup $\Gamma^{2}$, the Hecke congruence group $\Gamma_{0}(2)$ and its conjugate $\Gamma^{0}(2)$, and $\Gamma_{\theta}$. All the groups $\Gamma_{0}(2)$, $\Gamma^{0}(2)$, and $\Gamma_{\theta}$ are index-$3$ congruence subgroups of $\text{SL}(2,\mathbb{Z})$, have unique definitions and different elliptic points. The valence formula for a weight $k$ modular function $f$ of arbitrary congruence subgroup $\Gamma$ of level $N$ is given by \cite{Diamond}
\begin{align}\label{general_valence_formula}
    \sum\limits_{\tau\in\Gamma\backslash\mathbb{H}}\frac{\nu_{\tau}(f)}{n_{\Gamma}(\tau)} + \sum\limits_{\rho\in\text{Cusps}(\Gamma)}v_{\rho,\Gamma}(f) = \frac{k\overline{\mu}}{12},
\end{align}
where $n_{\Gamma}(\tau) = \vert\text{stab}_{\overline{\Gamma}}(\tau)\vert$ and $\overline{\mu}$ is the projective index defined in \ref{index_subgroups}. Since the right-hand-side of the formula, $\tfrac{k\overline{\mu}}{12}$ is what is used to obtain the Riemann-Roch-like relation, we observe the number of zeros in the fundamental domain in the case of $\Gamma_{0}(2)$ and $\Gamma_{\theta}$ to be $\# = \tfrac{1}{2}\ell$, and hence, we find common Riemann-Roch relations for the two groups. The space of modular forms of $\Gamma_{0}(2)$ has the following basis decomposition
\begin{align}\label{basis_Gamma_0_2}
    \begin{split}
        \mathcal{M}_{4n}(\Gamma_{0}(2)) =& \mathbb{C}\left(\Delta_{2}^{\infty}\right)^{n}\oplus \mathbb{C}\left(\Delta_{2}^{\infty}\right)^{n-1}\Delta_{2}^{0}\oplus\ldots\oplus\mathbb{C}\left(\Delta_{2}^{0}\right)^{n},\\
        \mathcal{M}_{4n+2}(\Gamma_{0}(2)) =& E_{2,2^{'}}\mathcal{M}_{4n}(\Gamma_{0}(2)),
    \end{split}
\end{align}
where $E_{2,2^{'}}(\tau)$ is a modular forms of weight $2$, $\Delta_{2}^{\infty}(\tau)$ and $\Delta_{2}^{0}(\tau)$ are weight $4$ modular forms of $\Gamma_{0}(2)$ defined as follows
\begin{align}
    \begin{split}
        E_{2,2^{'}}(\tau) \equiv& 2E_{2}(2\tau) - E_{2}(\tau)\\
        =& 1 +24 q + 24 q^2 + 96 q^3 + 24 q^4 + \ldots,\\
        \Delta_{2}^{\infty}(\tau) \equiv& \left(\frac{\eta^{2}(2\tau)}{\eta(\tau)}\right)^{8}\\
        =& q + 8 q^2 + 28 q^3 + 64 q^4 + \ldots,\\
        \Delta_{2}^{0}(\tau) \equiv& \left(\frac{\eta^{2}(\tau)}{\eta(2\tau)}\right)^{8}\\
        =& 1 - 16 q + 112 q^2 - 448 q^3 + 1136 q^4 + \ldots
    \end{split}
\end{align}
The Hauptmodul of $\Gamma_{0}(2)$ is defined as follows
\begin{align}\label{j_2}
    \begin{split}
        j_{2}(\tau) \equiv& \frac{\Delta_{2}^{0}}{\Delta_{2}^{\infty}} = \left(\frac{\eta(\tau)}{\eta(2\tau)}\right)^{24}\\
        =& q^{-1} - 24 +276 q - 2048 q^2 + 11202 q^3 - 49152 q^4  +\ldots
    \end{split}
\end{align}

\subsection*{\texorpdfstring{$\ell = 2$}{Lg}}
Following the approach we took in the case of $\Gamma_{0}(7)$, we see that $\tfrac{c}{24}\ell$, which counts the number of zeros in the fundamental domain is an integer in the case of $\Gamma_{0}(2)$ only when $\ell\in\{0, 2,4,6\}$. Thus, we shall analyze only these cases. When $\ell = 2$, an admissible theory would have a central charge $c = 24$. Spaces $\mathcal{M}_{6}(\Gamma_{0}(2))$ and $\mathcal{M}_{4}(\Gamma_{0}(2))$ are both two-dimensional and we make the following choice for the ratio $\tfrac{\omega_{6}(\tau)}{\omega_{4}(\tau)}$,
\begin{align}
        \frac{\omega_{6}(\tau)}{\omega_{4}(\tau)} =& \mu E_{2,2^{'}}(\tau).
\end{align}
Choosing the second ratio, the MLDE reads
\begin{align}\label{lab_rat_l=2}
    \left[\Tilde{\partial} + \mu E_{2,2^{'}}(\tau)\right]f(\tau) = 0,
\end{align}
where $\mu = 1$. The solution to this MLDE equals the Hauptmodul $j_{2}(\tau)$,
\begin{align}
    \begin{split}
    \chi^{(\ell = 2)}(\tau) =& j_{2}(\tau)\\
    =& q^{-1}\left(1 - 24q + 276q^{2} - 2048q^{3} + 11202q^{4} + \ldots\right).
    \end{split}
\end{align}
Although there are negative coefficients, all remain to be integral. Hence, this is a quasi-character that is of neither type I nor type II. 

\subsection*{\texorpdfstring{$\ell = 4$}{Lg}}
An admissible theory would have a central charge $c = 48$. The spaces $\mathcal{M}_{8}(\Gamma_{0}(2))$ and $\mathcal{M}_{10}(\Gamma_{0}(2))$ are both three-dimensional respectively and we make the following choice for the ratio $\tfrac{\omega_{10}(\tau)}{\omega_{8}(\tau)}$,
\begin{align}\label{omega_10_omega_8}
        \frac{\omega_{10}(\tau)}{\omega_{8}(\tau)} =& \mu E_{2,2^{'}}(\tau) = \frac{\omega_{6}(\tau)}{\omega_{4}(\tau)},
\end{align}
where $\mu = 2$. The solution to the MLDE turns out to be equal to the square of the Hauptmodul, i.e.
\begin{align}
    \begin{split}
    \chi^{\ell = 4}(\tau) =& j_{2}^{2}(\tau)\\
    =& q^{-2}\left(1 - 48 q + 1128 q^2 - 17344 q^3 + 196884 q^4 + \ldots\right).
    \end{split}
\end{align}

\subsection*{\texorpdfstring{$\ell = 6$}{Lg}}
An admissible theory would have a central charge $c = 48$. The spaces $\mathcal{M}_{12}(\Gamma_{0}(2))$ and $\mathcal{M}_{14}(\Gamma_{0}(2))$ are both four-dimensional and the choice we make for the ratio $\tfrac{\omega_{14}(\tau)}{\omega_{12}(\tau)}$ is the same as that in \ref{omega_10_omega_8} with $\mu = 3$. The solution to the MLDE turns out to be equal to the cube of the Hauptmodul, i.e.
\begin{align}
    \begin{split}
    \chi^{\ell = 6}(\tau) =& j_{2}^{3}(\tau)\\
    =&q^{-3}\left(1 - 72 q + 2556 q^2 - 59712 q^3 + 1033974 q^4 + \ldots\right).
    \end{split}
\end{align}
This too is a quasi-character. We can bunch all the solutions in $\Gamma_{0}(2)$ and write
\begin{align}
    \begin{split}
    &\chi(\tau) = j_{2}^{w_{\rho}}(\tau),\\
    &c = 24w_{\rho} = 12\ell,
    \end{split}
\end{align}
where $w_{\rho}\in\left\{0,1,2,3\right\}$ corresponding to the solutions at $\ell\in\{0, 2,4,6\}$ respectively. Comparing \ref{Hauptmodul_Gamma_0_2+} and \ref{j_2}, we find the relation
\begin{align}
    j_{2^{+}}(\tau) = \left(j_{2} +4096j_{2}^{-1} + 128\right)(\tau).
\end{align}
Using this relation, we see that we can build admissible characters out of the $\Gamma_{0}(2)$ characters. Denoting $\chi_{2}$ and $\chi_{2^{+}}$ to be level-$2$ Hecke and Fricke characters respectively, we have
\begin{align}
    \begin{split}
        \chi^{(\ell = 4)}_{2^{+}} =& \chi^{(\ell = 2)}_{2} + 4096\left(\chi^{(\ell = 2)}_{2}\right)^{-1} + 128,\\
        \chi^{(\ell = 6)}_{2^{+}} =& \left(\chi^{(\ell = 2)}_{2} + 4096\left(\chi^{(\ell = 2)}_{2}\right)^{-1} + 128\right)\cdot\left(\chi^{(\ell = 4)}_{2^{+}}\right)^{\tfrac{1}{2}}.
    \end{split}
\end{align}

\subsection*{Rademacher series expansions}
The $S$-transform of a character $\chi_{i}(\tau)$ can be found as follows
\begin{align}
    \chi_{i}(S(\tau)) = \sum\limits_{j=0}^{1}S_{ij}\chi_{j}(\tau).
\end{align}
In the simple case when $w_{\rho} = 1$, we find
\begin{align}\label{chi_S_matrix}
    \chi_{0}(S(\tau)) = 2^{12}\left(\frac{\eta(2\tau)}{\eta\left(\frac{\tau}{2}\right)}\right)^{24}\chi_{0}(\tau),
\end{align}
where we used \ref{Dedekind_S_transform} and where the subscript $0$ denotes the vacuum solution. Following \cite{Ramesh_Mukhi}, it is possible to determine the type of quasi-character by finding the sign of $S_{00}^{-1}$ that shows up in the asymptotic expansion of the Rademacher expansion of the character. For a modular form $\chi_{i}(\tau) = q^{\alpha_{i}}\sum_{n=0}^{\infty}a_{i}(n)q^{i}$, the leading behaviour of the coefficient $a_{j}(n)$ of a quasi-character is found to be (see \cite{Farey_tail} and appendix A of \cite{Ramesh_Mukhi} for details)
\begin{align}
    a_{j}(n) = S_{j0}^{-1}a_{0}(0)e^{4\pi\sqrt{\tfrac{c}{24}(n + \alpha_{j})}} + \ldots,
\end{align}
and whose sign in the asymptotic limit entirely depends on the sign of $S_{j0}^{-1}a_{0}(0)$. For the vacuum character, $a_{0}(0) = 1$ and thus, it is the sign of $S_{00}^{-1}$ that determines type I or type II quasi-character behaviours. From \ref{chi_S_matrix}, we read off the S-matrix element $S_{00}$ and find its inverse to be (up to a positive multiplicative constant)
\begin{align}
    \begin{split}
    S^{-1}_{00} =& \left(\frac{\eta\left(\frac{\tau}{2}\right)}{\eta(2\tau)}\right)^{24}\\
    =& q^{-\tfrac{3}{2}}\left(1 - 24q^{\tfrac{1}{2}} + 252q - 1472q^{\tfrac{3}{2}} + 4854 q^{2} - 6624 q^{\tfrac{5}{2}} -10696 q^{3} + 49152 q^{\tfrac{7}{2}} +2601q^{4} + \ldots\right).
    \end{split}
\end{align}
The sign keeps fluctuating, not always alternatively (as can be seen in the repeated negative signs for the coefficients associated with $q^{\tfrac{5}{2}}$ and $q^{3}$ for example), or in other words, it possesses no definite sign which would indicate the type of quasi-character. Thus, we see that the character of the level $2$ Hecke group is a quasi-character of neither type I nor type II. This also holds for the characters corresponding to $w_{\rho} = 2,3$, and for the quasi-characters we found for the groups $\Gamma_{0}(49)$ and $\Gamma_{0}^{+}(49)$ in \ref{quasi_Gamma_0_49} and \ref{quasi_Gamma_0_49+} respectively.

\subsection{\texorpdfstring{$\Gamma_{0}(3)$}{Γ0(3)}}
The Hecke group of level $3$ is generated by the $T$-matrix and $S^{-1}T^{-3}S^{-1}$, i.e. $\Gamma_{0}(3) = \langle\left(\begin{smallmatrix}1 & 1\\ 0 & 1\end{smallmatrix}\right), -\left(\begin{smallmatrix}1 & 0\\ 3 & 1\end{smallmatrix}\right)\rangle$. A non-zero $f\in\mathcal{M}_{k}^{!}(\Gamma_{0}(3))$ satisfies the following valence formula
\begin{align}
    \nu_{\infty}(f) + \nu_{0}(f) + \frac{1}{3}\nu_{\rho_{3}}(f) + \sum\limits_{\substack{p\in\Gamma_{0}(3)\backslash\mathbb{H}^{2}\\p\neq\rho_{3}}}\nu_{p}(f) = \frac{k}{3},
\end{align}
where $\rho_{3}$ is an elliptic point and this group has two cusps at $\tau = 0$ and $\tau = \infty$ (see \cite{Junichi} for more details).
Using \ref{index_subgroups}, we find the index of $\Gamma_{0}(3)$ in $\text{SL}(2,\mathbb{Z})$ to be $\mu_{0} = 4$ and hence, we find the number of zeros in the fundamental domain $\mathcal{F}_{3}$ to be $\# = 2\ell\cdot \tfrac{\mu}{12} = \tfrac{2}{3}\ell$. The Riemann-Roch theorem takes the following form
\begin{align}
    \begin{split}
        \sum\limits_{i=0}^{n-1}\alpha_{i} =& -\frac{nc}{24} + \sum\limits_{i}\Delta_{i}\\
        =&\frac{1}{3}n(n-1) - \frac{2}{3}\ell,
    \end{split}
\end{align}
where we used $\mu_{0} = 4$. For $n = 1$, we find the following relation between the number of zeros in the fundamental domain and the central charge
\begin{align}
    c = 16\ell.
\end{align}
The dimension of the space of modular forms reads
\begin{align}
    \text{dim}\ \mathcal{M}_{k}(\Gamma_{0}(3)) = \left\lfloor\left.\frac{k}{3}\right\rfloor\right. + 1,
\end{align}
where $k\geq 2$ for $k$ even and $k\geq 3$ for $k$ odd. The space of modular forms of $\Gamma_{0}(3)$ has the following basis decomposition
\begin{align}\label{basis_decomposition_Gamma_0_3}
    \begin{split}
        \mathcal{M}_{k}(\Gamma_{0}(3)) =& E_{k-3n,3^{'}}\left[\mathbb{C}\left(\Delta_{3}^{\infty}\right)^{n}\oplus \mathbb{C}\left(\Delta_{3}^{\infty}\right)^{n-1}\Delta_{3}^{0}\oplus \ldots\oplus \mathbb{C}\left(\Delta_{3}^{0}\right)^{n}\right],\\
        n=& \text{dim}\ \mathcal{M}_{k}(\Gamma_{0}(3)) - 1,
    \end{split}
\end{align}
where $E_{0,3^{'}}\equiv 1$, $E_{1,3^{'}}(\tau)\equiv E_{2,3^{'}}^{\tfrac{1}{2}}(\tau)$, and the forms $\Delta_{3}^{0}(\tau)$ \& $\Delta_{3}^{\infty}(\tau)$ are weight $3$ semi-modular forms of $\Gamma_{0}(3)$ with the following $q$-series expansions
\begin{align}
    \begin{split}
        \Delta_{3}^{0}(\tau) =& \frac{\eta^{9}(\tau)}{\eta^{3}(3\tau)}\\
        =& 1 - 9 q + 27 q^2 - 9 q^3 - 117 q^4,\\
        \Delta_{3}^{\infty}(\tau) =& \frac{\eta^{9}(3\tau)}{\eta^{3}(\tau)}\\
        =& q +3 q^2 + 9 q^3 + 13 q^4 + \ldots
    \end{split}
\end{align}
The Hauptmodul of $\Gamma_{0}(3)$ is defined as follows
\begin{align}
    \begin{split}
        j_{3}(\tau) =& \left(\frac{\Delta_{3}^{0}}{\Delta_{3}^{\infty}}\right)(\tau) = \left(\frac{\eta(\tau)}{\eta(3\tau)}\right)^{12}\\
        =& q^{-1} - 12 + 54q -76q^{2} - 243q^{3} + \ldots
    \end{split}
\end{align}

\subsection*{\texorpdfstring{$\ell = 0$}{Lg}}
Let us consider the simple case with no poles $\ell = 0$ and a single character. We consider
\begin{align}\label{E_1,3'_choices}
    \begin{split}
        \omega_{0}(\tau) =& 1,\\
        \omega_{2}(\tau) =& \mu E_{1,3^{'}}(\tau)\\
        =& \mu\left(1 + 6 q + 6 q^3 + 6 q^4 + \ldots\right).
    \end{split}
\end{align}
The MLDE with these choices reads
\begin{align}
    \left[\Tilde{\partial} + \mu E_{1,3^{'}}(\tau)\right]f(\tau) =  0.
\end{align}
Plugging in the behaviour of $f_{0}(\tau)$, we find $\mu = \tfrac{c}{24} = 0$. Hence, the only trivial solution is $f_{0}(\tau) = 1$. 

\subsection*{\texorpdfstring{$\ell = 3$}{Lg}}
Now, $\tfrac{c}{24}\ell$ is an integer in the case of $\Gamma_{0}(3)$ only when $\ell\in\{0, 3,6\}$. Since the Hauptmodul $j_{2}(\tau)$, like $j_{2}(\tau)$, contains negative coefficients, the solutions we obtain at $\ell = 3,6$ will also turn out to be quasi-characters. We postpone the detailed study of these characters and their relation to the Fricke characters for future work.

\subsection{\texorpdfstring{$\Gamma_{1}(N)$}{Lg}}
We have not addressed the groups $\Gamma_{1}(N)$. Although we do not perform explicit calculations here, we provide some comments regarding these groups. When $N\leq 2$, we have the isomorphism $\Gamma_{0}(N)\cong\Gamma_{1}(N)$, and both the groups share the same space of modular forms.  When $N\leq 4$, the fundamental domains and groups data match for groups $\Gamma_{0}(N)$ and $\Gamma_{1}(N)$. Hence, the level $N = 5$ proves to be an ideal starting point. The genus $0$ group $\Gamma_{1}(5)$ is an index-$12$ group in $\text{SL}(2,\mathbb{Z})$ and possesses no elliptic points unlike $\Gamma_{0}(5)$ which possesses two elliptic points of order $2$. Another difference is that $\Gamma_{1}(5)$ has four cusps while $\Gamma_{0}(5)$ has only two. This difference in group data certainly affects the valence formula and the central charge relation obtained via the Riemann-Roch theorem. The group $\Gamma_{1}(5)$ has four cusps at points $0$, $\tfrac{1}{2}$, $\tfrac{2}{5}$, and $\infty$. The valence formula for a non-zero $f\in\mathcal{M}_{k}^{!}(\Gamma_{1}(5))$ reads
\begin{align}
     \sum\limits_{\tau\in\Gamma_{1}(5)\backslash\mathbb{H}}\frac{\nu_{\tau}(f)}{n_{\Gamma_{1}(3)}(\tau)} + \sum\limits_{\rho\in\text{Cusps}(\Gamma_{1}(3))}v_{\rho,\Gamma_{1}(5)}(f) = k,
\end{align}
where we used $\overline{\mu}_{1} = 12$. The dimension of the space of modular forms of $\Gamma_{1}(3)$ reads \cite{Schimannek}
\begin{align}
    \begin{split}
        \text{dim}\ \mathcal{M}_{k}(\Gamma_{1}(3)) = \begin{cases}
        k + 1,\ & k\geq 0,\\
        0,\ &\text{otherwise}. 
        \end{cases}
    \end{split}
\end{align}
The Riemann-Roch theorem is found to be the following
\begin{align}
     \begin{split}
        \sum\limits_{i=0}^{n-1}\alpha_{i} =& -\frac{nc}{24} + \sum\limits_{i}\Delta_{i}\\
        =& n(n-1) - 2\ell.
    \end{split}
\end{align}
where we used $\mu = 12$ to obtain the number of zeros in the fundamental domain $\mathcal{F}_{\Gamma_{1}(3)}$. For $n = 1$, we find the following relation between the number of zeros of the fundamental domain and the central charge
\begin{align}
    c = 48\ell.
\end{align}
Similar relation can be derived for other $\Gamma_{1}(N)$ with $N\geq 5$ and studying the theory of modular forms in these groups would help us set up MLDEs and probe for admissible solutions. 
\section{Kissing numbers}\label{sec:Kissing}
The kissing number of packing of spheres in any dimension refers to the number of non-overlapping spheres that can touch one sphere. It is usually denoted by $\tau$ but we shall denote it by $\mathscr{K}$ to avoid confusion with the torus modulus. For example, in the one-dimensional case where the densest packing is the trivial one (i.e. all spheres arranged on the line) with $\mathbb{Z}$ being the lattice and the kissing number, in this case, is $\mathscr{K} = 2$ since each sphere (dot) is touched by the sphere in front of it and the sphere behind it. In two dimensions, the densest packing is given by the $A_{2}$ lattice (the hexagonal lattice) where each sphere (circle) is surrounded by six other spheres thus making the kissing number $\mathscr{K} = 6$. In three dimensions, the densest packing is given by the $A_{3}$ lattice with kissing number $\mathscr{K} = 12$. Establishing $\mathscr{K}$ in dimensions $d>3$ is highly non-trivial; in fact, it has been exactly determined only in $d = 1, 2, 3, 4, 8$, and $24$. Surprisingly, the large dimensions $8$ and $24$ happen to be special since the densest packing is exactly known thanks to the recent works of Viazovska \cite{Viazovska_d=8} and Cohn et al. \cite{Cohn_d=24}. Viazovska, in her Pi day paper (Mar 14, 2016), provided construction of so-called magic functions in $d = 8$ that helped prove that the $E_{8}$ lattice with a kissing number $\mathscr{K} = 240$ gives the densest packing in eight dimensions. The paper by Cohn et al. swiftly followed this discovery with the construction of magic functions in $d = 24$ that helped prove that the Leech lattice $\Lambda_{24}$ with a kissing number $\mathscr{K} = 196560$ gives the densest packing in twenty-four dimensions. The range of kissing numbers for packing at various dimensions is shown in table \ref{tab:kissing}.
\\
\begin{table}[htb!]
    \centering
    \begin{tabular}{||c|c|c|c||}
    \hline
    Dimension $d$ & Kissing number $\mathscr{K}$ & Densest packing & Densest packing kissing number $\mathscr{K}_{d}$\\ [0.5ex]
    \hline\hline
    1 & 2 & $\mathbb{Z}$ & 2\\
    2 & 6 & $A_{2}$ & 6\\
    3 & 12 & $A_{3}$ & 12\\
    4 & 24-25 & $D_{4}$ & 24\\
    5 & 40-46 & $D_{5}$ & 40\\
    6 & 72-82 & $E_{6}$ & 72\\
    7 & 126-140 & $E_{7}$ & 126\\
    8 & 240 & $E_{8}$ & 240\\
    12 & 840-1416 & $K_{12}$ & 840\\
    16 & 4320-8313 & $\Lambda_{16}$ & 4320\\
    24 & 196560 & $\Lambda_{24}$ & 196560\\
    48 & 52416000-? & $P_{48p}, P_{48q}$ & 52416000\\
    72 & ?-(6218175600)-? & $\Gamma_{72}$ \cite{Nebe} & 6218175600 \\
    96 & ?-(565866362880)-? & $\eta\left(P_{48q}\right)$ & ?\\
    128 & 1260230400-? & $BW_{128}$ & 1260230400\\
    128 & 1260230400-? & $MW_{128}$ \cite{MW} & 218044170240\\ [1ex]
    \hline
    \end{tabular}
    \caption{The table shows the kissing numbers that correspond to the densest packing at various dimensions \cite{sphere_packing_conway}. The dimensions where $\mathscr{K}$ is a range are the best-known bounds in literature with “?" indicating an unknown bound. $\Gamma_{72}$ denotes the Nebe lattice, $BW_{128}$ denotes the Barnes-Wall lattice, and $MW_{128}$ denotes the Mordell-Weil lattice. Kissing numbers for $d = 72, 96$ were obtained from \cite{72_96}.}
    \label{tab:kissing}
\end{table}
\\
Kabatiansky and Levenshtein found that the $d$-dimensional kissing number is bounded above by \cite{Kissing_bound}
\begin{align}
    \mathscr{K}\leq 2^{0.4041d\left(1 + \mathcal{O}(1)\right)}.
\end{align}
Using this we determine the upper bounds in dimensions $d = 48, 72, 96, 128$ to be 
\begin{align}
    \begin{split}
        d = 48:&\ \mathscr{K} = 690269,\\
        d = 72:&\ \mathscr{K} = 573492747,\\
        d = 96:&\ \mathscr{K} = 476471841459,\\
        d = 128:&\ \mathscr{K} =372148540169517.
    \end{split}
\end{align}
Clearly, this is not true for $d = 48, 72, 96$ which makes sense since these are valid for $d>115$. Information about the packing radius $\rho$, the kissing number $\mathscr{K}$ and the density $\Delta$ can be obtained from the $q$-series expansion of the $\Theta$-functions of lattices as follows
\begin{align}
   \begin{split}
    &\Theta_{\Lambda}(\tau) = 1 + \mathscr{K}q^{4\rho^{2}} + \ldots,\\
    &\Delta = \lim\limits_{r\to \infty}\left(\frac{\rho}{r}\right)^{n}\sum\limits_{m\leq r^{2}}N(m).
   \end{split}
\end{align}
From \ref{partition_theta_lattice}, we immediately conclude that series expanding the product of the partition function and $\eta^{d}(\tau)$, the first non-zero coefficient in the expansion would be the kissing number of a specific lattice with the highest packing.

\subsection{Lattice constructions}
As an example, consider the $c = 24$ Monster CFT. Following \cite{Jankiewicz}, to obtain the $q$-series expansion of the $\Theta$-function, if use the partition function $\mathcal{Z}(\tau) = J(\tau) + a$ with $a\in\mathbb{Z}_{+}$ we obtain 
\begin{align}
    \eta^{24}(\tau)(J(\tau) + a) = 1 + (a-24)q + (197136 - 24 a)q^{2} + \ldots
\end{align}
To rid ourselves of the vestigial constant at order $\mathcal{O}(q)$, we instead consider the partition function $\mathcal{Z}(\tau) = J(\tau) + 24+ a$ which yields
\begin{align}
        \eta^{24}(\tau)\left(J(\tau) + 24 + a\right) =& 1 + aq + \left(196560 - 24 a\right)q^{2} + \left(16773120 + 252 a\right)q^{3} + \ldots
\end{align}
The lattice corresponding to the extremal $\Theta$-function is obtained by setting $a = 0$ and this reveals the kissing number and lattice radius to be $\mathscr{K} = 196560$ and $\rho = \tfrac{1}{\sqrt{2}}$ respectively. This data corresponds to the Leech lattice $\Lambda_{24}$. In $d = 24$, the value of $a_{1}$ was constrained while in the other cases of $k>1$, we would need to constrain $k$ independent parameters. Repeating this procedure for a $k = 2$, $c = 48$ CFT and a $k = 4$, $c = 96$ CFT, we find lattice data $(\mathscr{K},\rho)$ to be $\left(52416000,\tfrac{\sqrt{3}}{2}\right)$ and $\left(19028983034880,\tfrac{\sqrt{5}}{2}\right)$ respectively (see appendix \ref{appendix:C}). From the data in table \ref{tab:kissing}, we see that the $c = 48$ case is the $\Theta$-series of the $P_{48}$ lattice while the $c = 96$ case could correspond to the lattice $\eta\left(P_{48q}\right)$ but this cannot be confirmed since the data corresponding to this lattice is not known. Let us redo this calculation for other Fricke groups. We found earlier that the character of an admissible single character theory in $\Gamma_{0}^{+}(2)$ with central charge $c = 24$ is expressed as $\chi(\tau) = j_{2^{+}}(\tau)$. With $\mathcal{Z}(\tau) =j_{2^{+}}(\tau)$ as the partition function, we obtain the following $q$-series expansion of the $\Theta$-function
\begin{align}\label{theta_Gamma_0_2+}
    \eta^{24}(\tau)\left(j_{2^{+}}(\tau) - 80 + a\right) = 1 + aq + (4048 - 24a)q^{2} + (-4096 + 252 a)q^{3} + \ldots
\end{align}
Setting $a = 0$, we find the kissing number and lattice radius to be $\mathscr{K} = 4048$ and $\rho = \tfrac{1}{\sqrt{2}}$ respectively. Redoing the same for the admissible single character theory in $\Gamma_{0}^{+}(3)$ with central charge $c = 24$ and character $\chi(\tau) = j_{3^{+}}(\tau)$, we find lattice data $\left(459, \tfrac{1}{\sqrt{2}}\right)$. For the $c = 48$ and character $\chi(\tau) = j_{3^{+}}^{2}(\tau)$, we obtain the lattice data $\left(918,\tfrac{1}{\sqrt{2}}\right)$. In the case of solutions with central charge $c = 48$ in the group $\Gamma_{0}^{+}(5)$ and central charge $c = 96$ in the group $\Gamma_{0}^{+}(7)$, we find lattice data $\left(14000,\tfrac{\sqrt{3}}{2}\right)$ and $\left(80426640,\tfrac{\sqrt{5}}{2}\right)$ respectively (see appendix \ref{appendix:C}). Although we were not able to find any lattice with these specifications in the literature (see \cite{Lattice_database}), we did find that the $\Theta$-series corresponding to the $\Gamma_{0}^{+}(2)$ is related to the odd Leech lattice. To see this let us first write the $q$-series expansion \ref{theta_Gamma_0_2+} to higher order
\begin{align}\label{2A_lattice}
    \begin{split}
    \Theta_{2^{+}}(\tau) =& 1 + \sum\limits_{m=1}^{\infty}N_{2^{+}}(m)q^{m}\\
    =& 1 + 4048q^{2} - 4096 q^3 + 1104 q^4 - 1130496 q^5 + 9404608 q^6 - 45785088 q^7\\
    +& 193108048 q^8 - 726630400 q^9 + 2333677152 q^{10} + \ldots
    \end{split}
\end{align}
The $\Theta$-series of the odd Leech lattice, denoted by $O_{24}$, reads \cite{odd_Leech}
\begin{align}
    \begin{split}
        \Theta_{O_{24}}(\tau) = 1 +& 4096q^{\tfrac{3}{2}} + 98256q^{2} + 1130496q^{\tfrac{5}{2}} + 8384512q^{3} + 45785088q^{\tfrac{7}{2}} + 199066704q^{4}\\
        +& 726630400q^{\tfrac{9}{2}} + 2314125312q^{5} + \ldots
    \end{split} 
\end{align}
Comparing the above expansions, we observe that the coefficients of the odd Leech lattice of half-integral order sit in \ref{2A_lattice} as coefficients of integral order. We find the following relation 
\begin{align}
    \Theta_{O_{24}}(\tau)= \left(1 + 4048q + 99360 q^{2} + 17789120q^{3} + \ldots\right) - \sum\limits_{m=1}^{\infty}N_{2^{+}}(\tau)q^{\tfrac{m}{2}}
\end{align}
The coefficients in the parenthesis can be built out of the character degrees of $T_{23A}(\tau)$, the McKay-Thompson series of class $23A$ for $\mathbb{M}$, as follows (see \cite{MT_23A})
\begin{align}
    \begin{split}
        1 =& \mathbf{1} = \chi_{1},\\
        4048 =&\mathbf{4048}= \chi_{22},\\
        99360 =& \mathbf{1}\oplus \mathbf{10}\oplus \textbf{772} \oplus \textbf{1837} \oplus \textbf{96536}\\
        =& \chi_{1} + \chi_{10} + \chi_{14} + \chi_{17} + \chi_{34},\\
        17789120 =& \textbf{13}\oplus \textbf{47}\oplus \textbf{1379}\oplus\textbf{5232}\oplus\textbf{33783}\oplus\textbf{258327} \oplus\textbf{2219086}\oplus\textbf{15271253}\\
        =&\chi_{3} + \chi_{6} + \chi_{16} + \chi_{21} + \chi_{29} + \chi_{39} + \chi_{51} + \chi_{63},\\
        &\vdots
    \end{split}
\end{align}
The explicit expression for the McKay-Thompson series of class $23A$ can be found in \cite{MT_23A_explicit}. Although it would be interesting to investigate if the $\Theta$-series of the theories in $\Gamma_{0}^{+}(3)$ and $\Gamma_{0}^{+}(5)$ can be related to a $q$-series expansions of lattices in $48$-dimensions, and if that in $\Gamma_{0}^{+}(7)$ can be related to a $q$-series expansion of a lattice in $96$-dimensions, the expressions for the $\Theta$-series of all the lattices in these dimensions is not known. To get an idea of the type of $\Theta$-series in $48$-dimensions, we can follow the construction by Elkies \cite{Elkies_theta} and write the $\Theta$-series associated to an even self-dual lattice $\Lambda$ as follows\footnote{we note that there is a typo in \cite{Elkies_theta} where the summand has $E^{n-24m}_{4}(\tau)$ instead of $E^{\tfrac{n-24m}{8}}_{4}(\tau)$.}
\begin{align}
    \Theta_{\Lambda}(\tau) = E^{\tfrac{n}{8}}_{4}(\tau) + \sum\limits_{m=1}^{\left\lfloor\left.\tfrac{n}{24}\right\rfloor\right.}c_{m}E^{\tfrac{n-24m}{8}}_{4}(\tau)\Delta^{m}(\tau),
\end{align}
where $E_{k}(\tau)$ is the $\text{SL}(2,\mathbb{Z})$ Eisenstein series of weight $k$ and $\Delta(\tau)$ is the modular discriminant defined as $\Delta(\tau)\equiv \tfrac{E_{4}^{3}(\tau) - E_{6}^{2}(\tau)}{1728}$. For $n=48$, we obtain the following series
\begin{align}\label{Theta_48,96}
    \begin{split}
        \Theta_{\Lambda_{48}}(\tau) =& E_{4}^{3}(\tau) - 1440\left(E_{4}^{3}\Delta\right)(\tau) + 125280\Delta^{2}(\tau)\\
        =& 1 + 52416000 q^3 + 39007332000 q^4 + 6609020221440 q^5 + \ldots.
    \end{split}
\end{align}
This is just \ref{Theta_series_d=48_J} which can be realized by writing out $J(\tau)$ in terms of the Eisenstein series and the discriminant. The odd lattice expansion is found by \cite{Elkies_theta}
\begin{align}
    \Theta_{O_{n}}(\tau) = \Theta_{\mathbb{Z}}^{n}(\tau) - \sum\limits_{m=1}^{\left\lfloor\left.\tfrac{n}{8}\right\rfloor\right.}c_{m}\Theta_{\mathbb{Z}}^{n-8m}(\tau)\Delta_{+}^{m}(\tau),
\end{align}
where
\begin{align}
    \begin{split}
        \Theta_{\mathbb{Z}}(\tau) \equiv& \sum\limits_{m=-\infty}^{\infty}q^{-\tfrac{m^{2}}{2}} = 1 + 2q^{\tfrac{1}{2}} + 2q^{2} + 2q^{\tfrac{9}{2}} + \ldots,\\
        \Delta_{+}(\tau) \equiv& \frac{1}{16}\left(\Theta_{\mathbb{Z}}^{8} - E_{4}\right)(\tau) = q^{\tfrac{1}{2}} - 8q + 28q^{\tfrac{3}{2}} - 64q^{2} + \ldots
    \end{split}
\end{align}
\begin{align}\label{Theta_48_odd}
   \begin{split}
    \Theta_{O_{48}}(\tau) =& \Theta_{\mathbb{Z}}^{48}(\tau) - 96\left(\Theta_{\mathbb{Z}}^{40}\Delta_{+}\right)(\tau) + 2400\left(\Theta_{\mathbb{Z}}^{32}\Delta_{+}^{2}\right)(\tau) - 12800\left(\Theta_{\mathbb{Z}}^{24}\Delta_{+}^{3}\right)(\tau) + 2400\left(\Theta_{\mathbb{Z}}^{16}\Delta_{+}^{4}\right)(\tau)\\
    -& 393216\left(\Theta_{\mathbb{Z}}^{8}\Delta_{+}^{45}\right)(\tau)\\
    =& 1 + 35638784 q^3 + 805306368 q^{\tfrac{7}{2}} + 20082632352 q^4 +  290984034304 q^{\tfrac{9}{2}}\\
    +& 3306660132864 q^5 + 29598230249472 q^{\tfrac{11}{2}} + \ldots
   \end{split}
\end{align}
Now, consider the $\Theta$-series of $\Gamma_{0}^{+}(3)$ character theory,
\begin{align}
   \begin{split}
    \Theta_{3^{+}}(\tau) =& 1 + \sum\limits_{m=1}^{\infty}N_{3^{+}}(m)q^{m}\\
    =& 1 + 918 q^2 - 11088 q^3 + 258795 q^4 - 5198688 q^5 + 53125200 q^6 - 317715696 q^7\\
    &+ 1322211897 q^8 - 4667071680 q^9 + 13629975948 q^{10} + \ldots
   \end{split}
\end{align}
Comparing this to \ref{Theta_48_odd}, we can write
\begin{align}
    \begin{split}
    \Theta_{O_{48}}(\tau) = 1 +& \left(N_{3^{+}}(2) - 918\right)q^{2} + \left(N_{3^{+}}(3) + 35649872\right)q^{3} +\left(N_{3^{+}}(4) + 20082373557\right)q^{4} + \ldots\\
    +&\left(805306368q^{\tfrac{7}{2}} + 290984034304q^{\tfrac{9}{2}} + 29598230249472q^{\tfrac{11}{2}} +  \ldots\right)
    \end{split}
\end{align}
The coefficients in the parenthesis that add to the coefficients $N_{3^{+}}(m)$ of each integral order can be built out of the character degrees of the McKay-Thompson series of class $32A$ for the Monster group (see \cite{MT_32A}) as follows
\begin{align}
    \begin{split}
        1 =& \textbf{1} = \chi_{0},\\
        918 =& \textbf{918} = \chi_{22},\\
        35649872 =& 2\cdot\textbf{1}\oplus\textbf{12}\oplus\textbf{318}\oplus\textbf{201849}\oplus\textbf{2551542}\oplus\textbf{32896149}\\
        =& 2\chi_{0} + \chi_{5} + \chi_{17} + \chi_{55} + \chi_{75} + \chi_{98},\\
        20082373557 =& \textbf{3}\oplus\textbf{4}\oplus\textbf{30}\oplus\textbf{153}\oplus\textbf{1119}\oplus\textbf{35829}\oplus\textbf{394228}\oplus\textbf{3629136}\oplus\textbf{182518500}\\
        \oplus&\textbf{19895794555}\\
        =& \chi_{1} + \chi_{3} + \chi_{8} + \chi_{14} + \chi_{23} + \chi_{43} + \chi_{60} + \chi_{78} + \chi_{115} + \chi_{168},\\
        &\vdots
    \end{split}
\end{align}
The $\Theta$-series corresponding to the two single character solutions of $\Gamma_{0}^{+}(5)$ at dimensions $n = 24$ and $n = 48$ read
\begin{align}
    \begin{split}
        \Theta_{5^{+}, 24}(\tau) =& 1 - 190 q^2 + 2120 q^3 - 11625 q^4 + 36120 q^5 - 45250 q^6 -  122640 q^7\\
        +& 674175 q^8 - 1047000 q^9 - 670320 q^{10},\\
        \Theta_{5^{+}, 48}(\tau) =& 1 + 14000q^{3} - 143250q^{4} + \ldots
    \end{split}
\end{align}
Upon expressing the coefficients of the odd Leech lattice and $O_{48}$ in terms of the coefficients of the above $\Theta$-series, we could not find any relations akin to the one found in the case of $\Gamma_{0}^{+}(2)$.
\section{Conclusions and outlook}\label{sec:conclusion}
In this paper, we have studied the theory of modular forms for the Hecke and Fricke groups of levels $N\leq 12$ using which we set up MLDEs and explicitly solved them to find admissible character-like solutions to an MLDE of order one. We have found that only the Fricke groups of levels $p = 2,3,5,7$, which happen to be the first four prime divisors of $\mathbb{M}$, possess these admissible solutions while the Hecke groups, except at levels $N = 2,7$, do not possess any admissible solutions. All the character solutions turn out to be expressible in terms of the powers of the Hauptmodules of the respective groups. These results are summarized in the table below.
\begin{table}[htb!]
\centering
\begin{tabular}{||c|c|c|c|c|c||}
\hline
$\Gamma_{0}^{+}(p)$ & $\chi(\tau)$ & $w_{\rho}$ & $x$ & $\ell$ & $c$\\ [0.5ex]
\hline\hline
$\Gamma_{0}^{+}(2)$ & $j_{2^{+}}^{w_{\rho}} + x\delta_{\ell,4}$ & $\left\{0,\tfrac{1}{4}, \tfrac{1}{2}, \tfrac{3}{4}, 1, \tfrac{5}{4}, \tfrac{3}{2}\right\}$ & $x\geq-104$ & $0\leq \ell\leq 6$ & $0\leq c\leq 36$\\[0.5ex] \hline
$\Gamma_{0}^{+}(3)$ & $j_{3^{+}}^{w_{\rho}} + x\delta_{\ell,3}$ & $\left\{0,\tfrac{1}{3}, \tfrac{2}{3}, 1, \tfrac{4}{3}, \tfrac{5}{3}, 2\right\}$ & $x\geq-108$ & $0\leq \ell\leq 6$ & $0\leq c\leq 48$\\[0.5ex] \hline
$\Gamma_{0}^{+}(5)$ & $j_{5^{+}}^{w_{\rho}}$ & $\left\{0,1, 2\right\}$ & $-$ & $0,2,4$ & $0\leq c\leq 48$\\[0.5ex] \hline
$\Gamma_{0}^{+}(7)$ & $j_{7^{+}}^{w_{\rho}}$ & $\left\{0,\tfrac{2}{3}, \tfrac{4}{3}, 2, \tfrac{8}{3}, \tfrac{10}{3}, 4\right\}$ & $-$ & $0\leq \ell\leq 6$ & $0\leq c\leq 96$\\[1ex] \hline
\hline
\end{tabular}
\end{table}
\noindent
We found that upon twinning the baby Monster characters with the characters of conjugacy classes $2C$ and $5A$ of $\mathbb{M}$, the character solutions corresponding to $\ell = 4$ in $\Gamma_{0}^{+}(2)$ and $\ell = 2$ in $\Gamma_{0}^{+}(5)$ respectively, can be expressed as a bilinear combination of Ising characters and the twinned characters. For the Hecke groups of levels $N = 2,7,49$ and for the Fricke group of level $49$, we found quasi-characters that are neither of type I nor type II according to the classification discussed in section \ref{sec:MLDE}. After a brief discussion on the sphere packing problem, we explicitly constructed $\Theta$-series for these character solutions and found a relation between $\Theta_{2^{+}}(\tau)$ and $\Theta_{O_{24}}(\tau)$. It would be interesting to investigate if the single character solutions or bilinear combinations of them actually correspond to well-known CFTs. Another point of interest would be to further investigate the Fricke groups of higher levels and those of levels equal to the prime divisors of $\mathbb{M}$. However, there are some subtleties here that we will discuss now.  

\subsection{Classification of admissible single character solutions in \texorpdfstring{$\Gamma_{0}^{+}(p)$}{Γ0(p)+}}
It would be interesting to check whether only Fricke groups of levels equal to the prime divisors of the Monster groups yield admissible single character solutions. The remaining cases to investigate are $N \in\{13,17,19,23,29,31,41,$ $59,71\}$. If the first four prime divisor levels ($N = 2,3,5,7$) indicate anything, it is the following trend for the character in $(n,\ell) = (1,\ell)$ theories
\begin{align}
    \chi^{(\ell)}(\tau) = j_{p^{+}}^{w_{\rho}}(\tau),\\
    c = 24w_{\rho} = 4\ell\mu.
\end{align}
From this we get $w_{\rho} = \tfrac{\mu}{6}\ell$. Since the index $\mu$ can be computed for each of the Fricke groups under consideration (see table \ref{tab:Fricke_indices} for $p<23$ and \ref{tab:correct_data} for $p\geq23$), we obtain the following set of values that $w_{\rho}$ can take
\begin{align}
    w_{\rho}\in\ell\cdot\left\{\frac{1}{4},\frac{1}{3},\frac{1}{2},\frac{1}{3},1,\frac{7}{6},\frac{3}{2},\frac{5}{3},2,\frac{5}{2},\frac{8}{3},\frac{7}{2},5,6\right\}.
\end{align}
Consider now the case of $\Gamma_{0}^{+}(2)$, the expression $\omega_{\rho} = \tfrac{\ell}{4}$ for $\ell\leq 6$ yields the correct set $\omega_{\rho}\in\left\{0,\tfrac{1}{4},\tfrac{1}{2}, \tfrac{3}{4}, 1, \tfrac{5}{4}, \tfrac{3}{2}\right\}$. This however does not work for $N = 11$ and $N = 13$ since we saw  the Hauptmodul $j_{11^{+}}(\tau)$ possess a fractional coefficient and the basis decomposition of the space of modular forms of $\Gamma_{0}^{+}(13)$ yielded non-integrable solutions when used in the MLDE for $\ell\leq 6$ respectively. Hence, we rule out the cases $w_{\rho} = \ell$ and $w_{\rho} = \tfrac{7}{6}\ell$. To rule out other $w_{\rho}$, we should not only check each group's Hauptmodul but also the type of coefficients the modular forms each group possesses.
\\
\begin{table}[htb!]
    \centering
    \resizebox{\textwidth}{!}{\begin{tabular}{||c|c|c|c|c|c|c|c|c|>{\columncolor[RGB]{230, 242, 255}}c|>{\columncolor[RGB]{230, 242, 255}}c|>{\columncolor[RGB]{230, 242, 255}}c|>{\columncolor[RGB]{230, 242, 255}}c|>{\columncolor[RGB]{230, 242, 255}}c|>{\columncolor[RGB]{230, 242, 255}}c||}
    \hline
    $p$ & 2 & 3 & 5 & 7 & 11 & 13 & 17 & 19 & 23 & 29 & 31 & 41 & 59 & 71\\ [0.5ex]
    \hline\hline
    $\nu_{2}(\Gamma_{0}(p))$  & 1 & 0 & 2 & 0 & 0 & 2 & 2 & 0 & 0 & 2 & 0 & 2 & 0 & 0\\
    $\nu_{3}(\Gamma_{0}(p))$ & 0 & 1 & 0 & 2 & 0 & 2 & 0 & 2 & 0 & 0 & 2 & 0 & 0 & 0\\
    $\nu_{1}(\Gamma_{0}(p)), \mu$ & 3 & 4 & 6 & 8 & 12 & 14 & 18 & 20 & 24 & 30 & 32 & 42 & 60 & 72\\
    $g$ & 0 & 0 & 0 & 0 & 1 & 0 & 1 & 1 & 2 & 2 & 2 & 3 & 5 & 6\\
    $\chi_{0}(\Gamma_{0}^{+}(p))$ & $\tfrac{1}{4}$ & $\tfrac{1}{3}$ & $\tfrac{1}{2}$ & $\tfrac{2}{3}$ & $1$ & $\tfrac{7}{6}$ & $\tfrac{3}{2}$ & $\tfrac{5}{3}$ & $2$ & $\tfrac{5}{2}$ & $\tfrac{8}{3}$ & $\tfrac{7}{2}$ & $5$ & $6$\\
    $\mu_{0f}$ & 2 & 2 & 2 & 2 & 2 & $2$ & $2$ & $2$ & $1$ & $1$ & $1$ & $1$ & $1$ & $1$\\
    $\mu_{0}^{+}$ & $\tfrac{3}{2}$ & $2$ & $3$ & $4$ & $6$ & $7$ & $9$ & $10$ & $24$ & $30$ & $32$ & $42$ & $60$ & $72$\\
    c & $6\ell$ & $8\ell$ & $12\ell$ & $16\ell$ & $24\ell$ & $28\ell$ & $36\ell$ & $40\ell$ & $96\ell$ & $120\ell$ & $128\ell$ & $168\ell$ & $240\ell$ & $288\ell$\\ [1ex]
    \hline
    \end{tabular}}
    \caption{Elements required for calculation of the index $\mu_{0}^{+} = \left[\text{SL}(2,\mathbb{Z}):\Gamma_{0}^{+}(p)\right]$ in theorem \ref{thm:genus_Fricke_index}. Here, $\mu = \left[\text{SL}(2,\mathbb{Z}:\Gamma_{0}(p)\right]$, $g$ is the genus of group, $\mu_{0f} = \left[\Gamma_{0}^{+}(p):\Gamma_{0}(p)\right]$,
    $\nu_{\infty} = 2$ for all $\Gamma_{0}(p)$, and $\nu_{4}(\Gamma_{0}(p)) = \nu_{6}(\Gamma_{0}(p)) = 0$ for all $p<13$, $\left(\nu_{4},\nu_{6}\right) = (2,1)$ for $p = 13$, $\left(\nu_{4},\nu_{6}\right) = (0,6)$ for $p = 17$, and $\left(\nu_{4},\nu_{6}\right) = (0,4)$ for $p = 19$ (these were found by comparing with the valence formulae for levels $p = 13,17,19$ in \cite{Junichi_extended}). The last row shows the central charges obtained via the Riemann-Roch theorem. The shaded part of the table contains index data, computed with the assumption that elliptic points of orders $4$ and $6$ are null-valued, yielding incorrect valence formulae. The reason for this discrepancy is explained below.}
    \label{tab:Fricke_indices}
\end{table}
\\
For the cases $p = 13,17,19$, the right-hand side of the valence formula is $\tfrac{7k}{12}, \tfrac{3k}{4}, \tfrac{5k}{6}$ respectively \cite{Junichi_extended}. Assuming $\nu_{4} =\nu_{6} = 0$ for these levels yields an index $\left[\Gamma_{0}^{+}(p): \Gamma_{0}(p)\right] = 1$, from which we obtain indices $\overline{\mu}_{0}^{+}(\Gamma_{0}^{+}(13)) = 14$, $\overline{\mu}_{0}^{+}(\Gamma_{0}^{+}(17)) = 18$, and $\overline{\mu}_{0}^{+}(\Gamma_{0}^{+}(19)) = 20$. Using this data in the general valence formula \ref{general_valence_formula} gives us $\tfrac{7k}{6}, \tfrac{3k}{2}, \tfrac{5k}{3}$ which differ from the original values by a factor of $\tfrac{1}{2}$. This difference stems from the fact that all Hecke groups of levels $p\geq 13$ in \ref{tab:Fricke_indices} are found to be index-$1$ subgroups of their Fricke counterparts while they should really be index-$2$ subgroups to obtain desired results. To account for this change, we find the number of elliptic points of orders $4$ and $6$ for all levels $p\geq 13$ assuming $\mu_{0f} = 2$. The correct values of central charges for Hecke groups of prime divisor levels $23\leq p\leq 71$ are shown in \ref{tab:correct_data}.
\\
\begin{table}[htb!]
    \centering
    \begin{tabular}{||c|c|c|c|c|c|c||}
    \hline
    $p$ & 23 & 29 & 31 & 41 & 59 & 71\\ [0.5ex]
    \hline\hline
    $\mu_{0}^{+}$ & 12 & 15 & 16 & 21 & 30 & 36\\
    $c$ & $48\ell$ & $60\ell$ & $64\ell$ & $84\ell$ & $120\ell$ & $144\ell$\\ [1ex]
    \hline
    \end{tabular}
    \caption{The correct index and central charge data for Hecke groups of levels $23\leq p\leq 71$.}
    \label{tab:correct_data}
\end{table}

\subsection{Relation to Klein's j-function}
It is easy to see that the relation between the Hauptmodul $j_{2}(\tau)$ and the $j$-function is as follows
\begin{align}\label{j-Hecke_2}
    j(\tau) = \left(\frac{256 + j_{2}}{j_{2}^{2}}\right)(\tau).
\end{align}
Now, from \ref{Hauptmodul_Gamma_0_2+} and \ref{j_2} we can write
\begin{align}
    j_{2^{+}}(\tau) = \left(j_{2} + 4096j_{2}^{-1} + 128\right)(\tau).
\end{align}
Inverting this, we find an expression for $j_{2}(\tau)$ in terms of $j_{2^{+}}(\tau)$ and plugging this into \ref{j-Hecke_2}, we obtain the following relation between the Fricke Hauptmodul of level $2$ and the $j$-function
\begin{align}\label{j_to_Fricke_2}
    j(\tau) = \frac{1}{2}\left[\left(\chi_{2^{+}} - 81\right)\sqrt{\chi_{2^{+}}(256 - \chi_{2^{+}})} + \left(\chi^{2}_{2^{+}} - 207\chi_{2^{+}} + 3456\right)\right](\tau),
\end{align}
where $\chi_{2^{+}}(\tau) = j_{2^{+}}(\tau)$ is the vacuum character at $\ell = 4$. We can draw similar relations between the Hauptmodules $j_{3^{+}}(\tau)$ and $j_{5^{+}}(\tau)$ and the $j$-function, although they would be a bit more complicated. By the map \ref{j_to_Fricke_2}, we see that we can use the Fricke characters to build the $c = 24$ CFTs classified by Schellekens. Thus, single character solutions in Fricke groups of prime divisor levels of $\mathbb{M}$ with $c = 24$ can be used to build the CFTs classified by Schellekens in \cite{Schellekens}.

\subsection{Other candidate congruence groups}
\begin{figure}[htb!]
    \centering
    \begin{tikzcd}
	&& {X_{1}(2,14)} \\
	\\
	{X_{1}(14)} & {X_{1}(2,14)/A_{3}} & {X_{1}(2,14)/H} & {X_{1}(2,14)/H'} & {X_{1}(2,14)/C_{3}} \\
	\\
	{X_{1}(7)} && {X_{*}(7)} && {X_{1}(2,14)/(A_{3}\times C_{3})}
	\arrow["2"', from=1-3, to=3-1]
	\arrow["3"', from=3-1, to=5-1]
	\arrow["3"', from=5-1, to=5-3]
	\arrow["3", from=3-2, to=5-5]
	\arrow["2", from=5-5, to=5-3]
	\arrow["3", from=1-3, to=3-4]
	\arrow["3", from=3-4, to=5-5]
	\arrow["3"', from=1-3, to=3-2]
	\arrow["2"', from=3-2, to=5-1]
	\arrow["3"', from=1-3, to=3-3]
	\arrow["3", from=1-3, to=3-5]
	\arrow["3", from=3-5, to=5-5]
	\arrow["3", from=3-3, to=5-5]
\end{tikzcd}
    \caption{A new sequence of related modular curves.}\label{fig:sequence}
\end{figure}
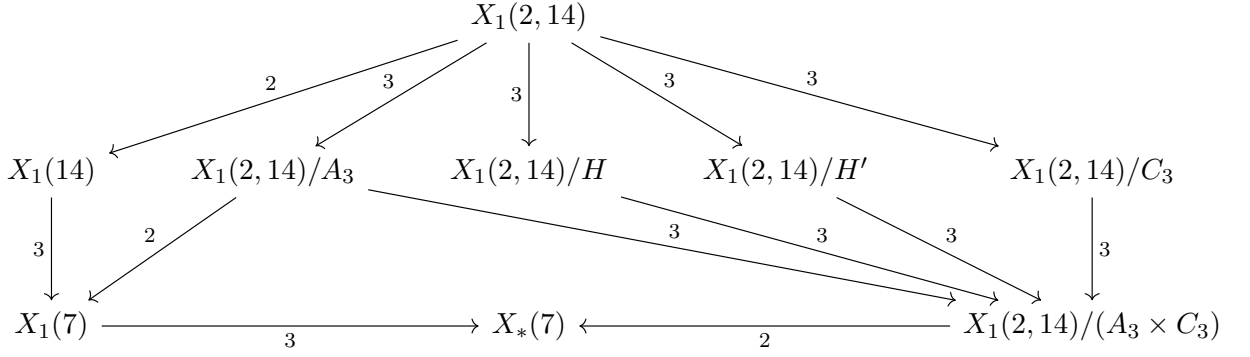

Another possible way to investigate other admissible solutions and CFT candidate theories would be to build them from congruence groups concocted out of the prime divisor levels. Let us consider the cases of the level $7$ and level $49$ groups. For example, from \cite{concocted_group}, consider the intersection of $\Gamma_{0}(49)$ and $\Gamma_{1}(7)$ defined as follows
\begin{align}
    \Gamma\equiv \Gamma_{0}(49)\cap\Gamma_{1}(7) = \left\{\left.\begin{pmatrix}
    a & b\\ c & d
    \end{pmatrix}\right\vert c\equiv 0\ (\text{mod}\ 49)\ \&\ a,d\equiv 1\ (\text{mod}\ 7)\right\}.
\end{align}
The index of this group in $\text{SL}(2,\mathbb{Z})$ is
\begin{align}
    \mu_{\Gamma} = \left[\text{SL}(2,\mathbb{Z}):\Gamma\right] = 168.
\end{align}
We can use this to find the number of zeros in the fundamental domain $\mathcal{F}_{\Gamma}$ to be
\begin{align}
    \# = 2\ell \cdot \frac{168}{12} = 28.
\end{align}
This gives us sufficient knowledge to write the Riemann-Roch theorem for $\Gamma$,
\begin{align}
    \sum\limits_{i=1}^{\infty}\alpha_{i} = 14n(n-1) - 28\ell.
\end{align}
Single character theories in this group would possess the following central charge
\begin{align}
    c = 14\ell.
\end{align}
This tells us that we could search for a $k = 7$, $c = 168$ theory when $\ell = 7$. Computations with the modular form of weight $1$ given in \cite{concocted_group} yielded fractional coefficients in the character expansion which forces us to rule out the existence of $\ell = 7$ single character theories in $\Gamma$. However, this doesn't rule out the existence of two-character and three-character theories. As another example, we refer to the work done in \cite{fixing_group}, where the genus $0$ fixing groups generated from $\Gamma_{0}(49)$ are considered. These fixing groups, for $g = 7+$ and $h\in\text{He}$ (the Held sporadic group), are defined as follows
\begin{align}
    \begin{split}
        \Tilde{\Gamma}^{(1)}_{h,g} =& \left\langle\left.\Gamma_{0}(49), \begin{pmatrix}
        3 & \tfrac{2}{7}\\
        -35 & -3
        \end{pmatrix}\right\rangle\right.,\\
        \Tilde{\Gamma}^{(2)}_{h,g} =& \left\langle\left.\Gamma_{0}(49), \begin{pmatrix}
        3 & -\tfrac{2}{7}\\
        35 & -3
        \end{pmatrix}\right\rangle\right.
    \end{split}
\end{align}
The Hauptmoduls of these fixing groups which possess irrational $q$-series expansions read
\begin{align}
    \begin{split}
        j_{\Tilde{\Gamma}^{(1)}}(\tau) =& \frac{\eta(\tau) \eta\left(\tau + \frac{3}{7}\right) \eta\left(\tau + \frac{5}{7}\right) \eta\left(\tau + \frac{6}{7}\right)}{\eta^{4}(7\tau)} + \frac{1}{2} + i\frac{\sqrt{7}}{2},\\
        j_{\Tilde{\Gamma}^{(2)}}(\tau) =& \frac{\eta(\tau) \eta\left(\tau + \frac{1}{7}\right) \eta\left(\tau + \frac{2}{7}\right) \eta\left(\tau + \frac{4}{7}\right)}{\eta^{4}(7\tau)} + \frac{1}{2} + i\frac{\sqrt{7}}{2},
    \end{split}
\end{align}
which correspond to the $7E$ and $7D$ conjugacy classes of the Held sporadic group respectively. Although we don't do MLDE computations with these fixing groups here, it would be interesting to investigate candidate theories in these classes of groups. Other interesting groups to investigate would be those that show up in the tower $X_{0}(7^{2+n})$, i.e. $\Gamma_{0}(49)\to \Gamma_{0}(343)\to \ldots$ and the corresponding Fricke counterparts. From \ref{c=224l} and \ref{c=112l}, we see that the central charge is already rather large for the groups at level $49$ and it would only be larger at level $343$. Nevertheless, following this tower could yield candidate $c = 24k$ single-character theories with large $k$ values. Also, higher-character theories with smaller central charge values might exist. Following the construction given by Elkies \cite{shifting_base_X0(Npn)}, we can change the base of the tower from $X_{0}(p^{n}) = \Gamma_{0}(p^{n})\backslash\mathbb{H}^{2*}$ where $p\in\mathbb{P}$ to $X_{0}(Np^{n}) = \left(\Delta\cap\Gamma_{0}(p^{n})\right)\backslash\mathbb{H}^{2*}$, where $\Delta$ is a congruence subgroup of $\text{PGL}(2,\mathbb{Q})$ with a prime modulus. For example, with $(N,p,n) = (1,7,1)$, we get $X_{0}(14)$ which is a genus $1$ modular curve. Following the construction given by Bruin et al. \cite{X0(14)}, we can search for other genera zero modular curves. First, we look to modular curves with group $\Gamma = \Gamma_{1}(2,14) = \Gamma(2)\cap\Gamma_{1}(7)$. We define $\Gamma_{*}(2)$ to be a subgroup of $\text{SL}(2,\mathbb{Z})$, with its index with respect to $\Gamma(2)$ being $\left[\Gamma_{*}(2):\Gamma(2)\right] = 3$, as follows
\begin{align}
    \Gamma_{*}(2) = \left\{\Gamma\in\text{SL}(2,\mathbb{Z})\vert \gamma \equiv\mathbb{1}_{2},\left(\begin{smallmatrix}
    1 & 1\\ 1 & 0
    \end{smallmatrix}\right),\left(\begin{smallmatrix}
    0 & 1\\ 1 & 1
    \end{smallmatrix}\right)(\text{mod}\ 2)\right\}.
\end{align}
Similarly, we also define the group $\Gamma_{*}(7)$ as follows
\begin{align}
    \Gamma_{*}(7) = \left\{\left.\begin{pmatrix}
    a & b\\ c & d
    \end{pmatrix}\right\vert a,d\equiv 1,2,4\ (\text{mod}\ 7),\  c\equiv 0\ (\text{mod}\ 7)\right\}.
\end{align}
We now have the sequence shown in figure \ref{fig:sequence}.
\noindent
Here $X_{1}(2,14)$ is a genus $4$ modular curve, $X_{1}(14)$ is a genus $1$ curve that is isomorphic to the following elliptic curve
\begin{align}
    y^{2} + xy + y = x^{3} - 11x + 12,
\end{align}
and has a weight $2$ modular form given by the following $\eta$-product
\begin{align}
   \begin{split}
        f(\tau) =& \eta(\tau)\eta(2\tau)\eta(7\tau)\eta(14\tau)\\
        =& q - q^2 - 2 q^3 + q^4 + 2 q^6 + q^7 - q^8 + q^9 + \ldots
   \end{split}
\end{align}
$X_{1}(2,14)/A_{3}$ and $X_{1}(2,14)/C_{3}$ are genus $2$ curves.  $X_{1}(2,14)/\left(A_{3}\times C_{3}\right)$, $X_{1}(2,14)/H\cong\mathbb{P}^{1}(\mathbb{Q})$ and $X_{1}(2,14)/H'\cong\mathbb{P}^{1}(\mathbb{Q})$ are a genus zero curves where $\left(A_{3}\times C_{3}\right) = \tfrac{\Gamma_{*}(2)\cap\Gamma_{*}(7)}{\Gamma_{1}(2,14)}$. It would also be interesting to further climb up the modular ladders corresponding to the Fricke groups of prime levels $2,3,5$ and investigate for admissible solutions although we suspect that there would be single character solutions since none of the higher groups are of prime level. Lastly, we mention that the groups of level $7$ possess a connection to maximally symmetric Riemann surfaces. These are also Hurwitz surfaces to the maximal nature of the automorphisms for a given genus. The Hurwitz bound for a surface of genus $g$ is $84(g-1)$ and the allowable values the genera can take follow the sequence $3,7,14,17,\ldots$ (see \cite{Hurwitz_sequence}). The $g = 3$ Riemann surface, called the Klein quartic, has the automorphism group $\text{PSL}(2, \mathbb{F}_{7})$ of order $168$ which is equal to the index of $\overline{\Gamma}(7)$ (see appendix \ref{appendix:A}). The isomorphic cousins of this group are $\Gamma_{0}(7)$ and $\Gamma_{0}^{+}(7)$. In fact, for $p>3$, $\text{PSL}(2,\mathbb{F}_{p})\cong\Gamma(1)\backslash\Gamma(p)$. If we considered the maximally symmetric surface of genus $g = 14$, the automorphism groups of the corresponding Hurwitz surface is
$\text{PSL}(2,\mathbb{F}_{13})$ whose order is $1092$ which happens to be the index of $\overline{\Gamma}(13)$. Thus, the isomorphic cousins of $\text{PSL}(2,\mathbb{F}_{13})$ are $\Gamma_{0}(13)$ and $\Gamma_{0}^{+}(13)$. It would be interesting to explore spectral theoretic connections to character theories in Fricke groups.

\subsection{Higher character solutions}
A more systematic analysis  of two- and three-character MLDE for the Fricke groups for levels $p = 2,3,5,7$ and associated modular towers, i.e. $\Gamma_{0}^{+}(p^{2+n})$, would be pertinent to understand better the classification of CFTs. We reserve this for future work\footnote{This is now complete, see \cite{Naveen-Arpit}}.

\subsection{Lattice relations}
We found that the coefficients of $\Theta_{O_{24}}(\tau)$ can be expressed in terms of those of $\Theta_{2^{+}}(\tau)$ and $T_{23A}(\tau)$. We also found $\Theta_{3^{+}}(\tau)$, $\Theta_{5^{+}}(\tau)$, and $\Theta_{7^{+}}(\tau)$ but the relating these to other (odd or even unimodular) lattices turns out to be highly non-trivial. However, it is interesting to point out that the Schwarz derivative of the Hauptmodul of the group $\Gamma_{1}(5)$, whose normalizer is $\Gamma_{0}^{+}(5)$, yields the $q$-series expansion of the $\Theta$-series of $\mathbf{Q}_{8}(1)$ which is an $8$-dimensional $5$-modular lattice \cite{Schwarz_lattice}. Taking the Schwarz derivative of $j_{5^{+}}(\tau)$ yields the following $q$-series expansion
\begin{align}
    S(j_{5^{+}}(\tau)) = 1 + 1608 q^2 + 36480 q^3 + 832344 q^4 + \ldots
\end{align}
It would be interesting to explore if this matches the $\Theta$-series of a well-known lattice. It would also be interesting to investigate if the $\Theta$-series corresponding to $c = 24$ solution $\Gamma_{0}^{+}(5)$ is related to a $\Theta$-series of a particular odd unimodular lattice in $24$-dimensions. A list of such lattices was found by Borcherds \cite{sphere_packing_conway} (see chap. $17$).

\acknowledgments
I'd like to thank Mikhail Isachenkov for multiple discussions, for encouragement to pursue my own interests, and for the constant support from the inception of this work until its completion. Thanks to Sunil Mukhi for productive discussions and suggestions on the work. I'd also like to thank Chethan Krishnan, Arpit Das, Prashanth Raman, and Mikhail Isachenkov for their insightful comments on the draft as well as Erik Verlinde and Miranda Cheng for enlightening discussions. Finally, thanks to the theory group at UvA for creating a stimulating environment where I got to discuss this work in depth with peers.

\appendix
\section{Mathematical preliminaries}\label{appendix:A}
\subsection*{Definitions}
The special linear group over the integers $\text{SL}(2,\mathbb{Z})$ is defined as the set of all $2\times 2$ matrices with real entries and unit determinant, i.e.
\begin{equation}
    \text{SL}(2,\mathbb{Z}) = \left\{\left.\begin{pmatrix}
    a & b\\ c& d
    \end{pmatrix}\right\vert a,b,c,d\in\mathbb{R},ad-bc=1\right\}.
\end{equation}
The projective special linear group  over the integers or the modular group $\text{PSL}(2,\mathbb{Z})$ is nothing but the special linear group quotiented by $\mathbb{Z}_{2}$, i.e.
\begin{equation}
    \text{PSL}(2,\mathbb{Z}) = \text{SL}(2,\mathbb{Z})/\mathbb{Z}_{2} = \left\{\left.\begin{pmatrix}
    a & b\\ c& d
    \end{pmatrix}\right\vert a,b,c,d\in\mathbb{R},ad-bc=1\right\}/\{\pm 1\}.
\end{equation}
Since $\text{PSL}(2,\mathbb{Z})$ and $\text{SL}(2,\mathbb{Z})$ are equal up to a quotient $\mathbb{Z}_{2}$, both can be loosely referred to as the modular group. $\text{GL}(2,\mathbb{R})$ is the general linear group over the reals and is the group of all $2\times 2$ invertible matrices with coefficients in the ring $\mathbb{R}$ with $\text{SL}(2,\mathbb{R})$ being its subgroup with unit determinant. Let $\reallywidehat{\mathbb{C}} \equiv \mathbb{C}\cup \{\infty\}$ denote the Riemann sphere, which is also sometimes denoted by $\mathbb{P}^{1}(\mathbb{C})$. Let us consider an element $A\in\text{SL}(2,\mathbb{Z})$, where our ring now is the integers and the element $A$ is a $2\times 2$ matrix of the form
\begin{equation}
    A = \begin{pmatrix}
    a & b\\ c & d
    \end{pmatrix}.
\end{equation}
The action of $A$ on $\reallywidehat{\mathbb{C}}$ is defined by M\"obius or fractional linear transformations in the usual way
\begin{equation}
  A(z)\equiv \frac{az + b}{cz + d},
\end{equation}
and
\begin{align}\label{action_SL2Z}
\begin{split}
    A(\infty) = \lim\limits_{y\to \infty}\frac{a(x+iy) + b}{c(x+iy) + d} = \begin{cases}
    \frac{a}{c},\ &c\neq 0\\
    0,\ &c=0
    \end{cases}
\end{split}
\end{align}
The modular group contains the following three matrices
\begin{equation}
    I = \mathbb{1}_2,\ S = \begin{pmatrix}
    0 & -1\\ 1 & 0
    \end{pmatrix},\ T = \begin{pmatrix}
    1 & 1\\ 0 & 1
    \end{pmatrix},
\end{equation}
where $\mathbb{1}_{2}$ is the $2\times 2$ identity matrix. It is to be noted that different definitions are followed for the $S$ matrix with the convention of placing the $-1$ in different off-diagonal positions. We either denote $\left(\begin{smallmatrix}0 &-1\\ 1&\ 0\end{smallmatrix}\right)$ or $\left(\begin{smallmatrix}\ 0 &1\\ -1&0\end{smallmatrix}\right)$ to be the $S$ matrix and we shall stick to the former in this paper. The $S$ and the $T$ matrices generate the modular group. The discrete subgroups of $\text{PSL}(2,\mathbb{R})$ are denoted by $\overline{\Gamma}$ with the convention that we denote $\overline{\Gamma}(1) = \text{PSL}(2,\mathbb{Z})$ and the discrete subgroups of $\text{SL}(2,\mathbb{Z})$ are denoted by $\Gamma$ with the convention that $\Gamma(1) = \text{SL}(2,\mathbb{Z})$. These subgroups are related as $\overline{\Gamma} = \Gamma/\pm I$. The discrete subgroup of level $2$, for example, is defined as follows 
\begin{align}
    \begin{split}
    \Gamma(2) =& \left\{A\in\text{SL}(2,\mathbb{Z})\vert A\equiv \mathbb{1}_{2}\ (\text{mod}\ 2)\right\}\\
    \Gamma_{\theta} =& \left\{A\in\text{SL}(2,\mathbb{Z})\vert A\equiv \mathbb{1}_{2}\ \text{or}\  \mathbb{1}_{2}\ (\text{mod}\ 2)\right\},
    \end{split}
\end{align}
where $\text{SL}(2,\mathbb{Z})\supset \Gamma_{\theta}\supset \Gamma(2)$, and $\Gamma_{\theta}$ is called the theta group. The discrete subgroup $\Gamma(2)$ for example, is generated by $T^{2} = \left(\begin{smallmatrix}1 & 2\\ 0 & 1\end{smallmatrix}\right)$ and $ST^{2}S = \left(\begin{smallmatrix}-1 & \ 0\\ \ 2 & -1\end{smallmatrix}\right)$ which is denoted as $\Gamma(2) = \langle T^{2},ST^{2}S\rangle.$
There also exist special subgroups of the discrete group $\Gamma$ that we are interested in working with. These are called the congruence subgroups of level $N$ or the level $N$ principal congruence subgroups and are denoted by $\Gamma(N)$ with $N\in\mathbb{N}$. These are defined to be the identity mod $N$, i.e.
\begin{align}
    \begin{split} 
    \Gamma(N) =& \{\gamma\in\Gamma(1)\vert\ \gamma = \mathbb{1}_{2}\ (\text{mod}\ N)\}\\
    =& \left.\left\{\begin{pmatrix}
    a & b\\ c & d
    \end{pmatrix}\in\text{SL}(2,\mathbb{Z})\right\vert a\equiv d\equiv 1\  (\text{mod}\ N),\ b\equiv c\equiv 0\  (\text{mod}\ N)\right\},
   \end{split}
\end{align}
where $\Gamma(1) = \text{PSL}(2,\mathbb{Z})$ as noted earlier. A congruent subgroup of level $N$ is also a congruent subgroup of any multiple of $N$, i.e. $N' = \ell N$ for all $\ell\in\mathbb{N}$. The index of principal congruence subgroup $\Gamma(N)$ in $\text{SL}(2,\mathbb{Z})$ and in $\text{PSL}(2,\mathbb{Z})$ is given by
\begin{equation}\label{index}
      \begin{split}
          \mu =& \left[\text{SL}(2,\mathbb{Z}):\Gamma(N)\right]= N^{3}\prod\limits_{p\vert N}(1-p^{-2}),\\
          \overline{\mu} =& \left[\text{PSL}(2,\mathbb{Z}):\overline{\Gamma}(N)\right] = \begin{cases}
          \frac{1}{2}N^{3}\prod\limits_{p\vert N}(1-p^{-2}),\ N\geq 3,\\
          6,\ \ \ \ \ \ \ \ \ \ \ \ \ \ \ \ \ \ \ \ \ \ N=2.
          \end{cases}
      \end{split}
\end{equation}
where $p\vert N$ denotes the prime divisors of $N$. Every congruence subgroup $\Gamma$ has a finite index in $\text{SL}(2,\mathbb{Z})$. There are several other special subgroups of $\Gamma$ of which the two most relevant for our discussion are $\Gamma_{0}(N)$ and $\Gamma_{1}(N)$, where the former is defined to be the subgroup of $\Gamma$ with matrices which have lower left entry $0$ mod $N$ and the latter is defined to the subgroup of $\Gamma$ with matrices that are identity except possibly in the upper right corner mod $N$. Explicitly, we have
\begin{align}\label{Hecke_subgroup}
    \begin{split}
    \Gamma_{0}(N) =& \left\{\left.\begin{pmatrix}
    a & b\\ c & d
    \end{pmatrix}\in\text{SL}(2,\mathbb{Z})\right\vert c\equiv 0\ (\text{mod}\ N)\right\} = \left\{\begin{pmatrix}
    * & *\\ 0 & *
    \end{pmatrix}\ \text{mod}\ N\right\},\\
    \Gamma_{1}(N) =& \left\{\begin{pmatrix}
    1 & *\\ 0 & 1
    \end{pmatrix}\text{mod}\ N\right\}.
    \end{split}
\end{align}
Here “$*$” stands for “unspecified”. These definitions satisfy
\begin{align}
    \Gamma(N)\subset\Gamma_{1}(N)\subset\Gamma_{0}(N)\subset\text{SL}(2,\mathbb{Z}).
\end{align}
We shall refer to $\Gamma_{0}(N)$ as Hecke subgroups from here on. When $N=1$, we have
\begin{equation}
    \text{SL}(2,\mathbb{Z}) = \Gamma_{0}(1) = \Gamma_{1}(1) = \Gamma(1).
\end{equation}
The index of the Hecke subgroup $\Gamma_{0}(N)$ in $\text{SL}(2,\mathbb{Z})$ and in $\text{PSL}(2,\mathbb{Z})$, and the index of the congruence subgroup $\Gamma_{1}(N)$ in $\text{SL}(2,\mathbb{Z})$ and in $\text{PSL}(2,\mathbb{Z})$ read \cite{Cohen_modular_forms}
\begin{align}\label{index_subgroups}
      \begin{split}
          \mu_{0} =& \left[\text{SL}(2,\mathbb{Z}):\Gamma_{0}(N)\right]= N\prod\limits_{p\vert N}(1+p^{-1}),\\
          \overline{\mu}_{0} =& \left[\text{PSL}(2,\mathbb{Z}):\overline{\Gamma}_{0}(N)\right] =  \begin{cases}
          N\prod\limits_{p\vert N}(1+p^{-1}),\ N\geq 3,\\
          3,\ \ \ \ \ \ \ \ \ \ \ \ \ \ \ \ \ \ \ N=2,
          \end{cases}\\
          \mu_{1} =& \left[\text{SL}(2,\mathbb{Z}):\Gamma_{1}(N)\right]= \frac{N^{2}}{2}\prod\limits_{p\vert N}(1-p^{-2}),\\
          \overline{\mu}_{1} =& \left[\text{PSL}(2,\mathbb{Z}):\overline{\Gamma}_{1}(N)\right]= \begin{cases}
          \frac{N^{2}}{2}\prod\limits_{p\vert N}(1-p^{-2}),\ N\geq 3,\\
          3,\ \ \ \ \ \ \ \ \ \ \ \ \ \ \ \ \ \ \ \ N=2.
          \end{cases}
      \end{split}
\end{align}
Additionally, we also have the following relations
\begin{align}\label{index_relations}
    \begin{split}
    &\left[\Gamma_{1}(N):\Gamma(N)\right] = N,\\
    &\left[\overline{\Gamma}_{1}(N):\overline{\Gamma}(N)\right] = N,\ N>2\\
    &\left[\Gamma_{1}(N):\Gamma(N^{2})\right] = N,\\
    &\left[\overline{\Gamma}_{1}(N):\overline{\Gamma}(N^{2})\right] = N,\ N>2\\
    &\left[\Gamma_{0}(N):\Gamma_{1}(N)\right] = \phi(N),\\
    &\left[\overline{\Gamma}_{0}(N):\overline{\Gamma}_{1}(N)\right] = \frac{1}{2}\phi(N),\ N>2\\
    &\left[\text{PSL}(2,\mathbb{Z}):\overline{\Gamma}_{1}(N^{2})\right] = \left[\text{PSL}(2,\mathbb{Z}):\overline{\Gamma}_{0}(N^{2})\right] = \left[\text{PSL}(2,\mathbb{Z}):\overline{\Gamma}(N)\right] = \frac{N^{3}}{2}\prod\limits_{p\vert N}\left(1 - p^{-2}\right),\ N\geq 3
    \end{split}
\end{align}
where Here, $\phi$ is the Euler's totient or Euler's phi function defined as follows
\begin{align}\label{Euler_totient}
    \phi(N) = N\prod\limits_{p\vert N}\left(1 - \frac{1}{p}\right),\ p\in\mathbb{P},
\end{align}
where the product is over the distinct prime numbers dividing $N$. The index is multiplicative, i.e. suppose $A,B,C$ are groups such that $A\subset B\subset C$ and the indices $[A:B]$ and $[B:C]$ are finite, then we have
\begin{align}
    [A:C] = [A:B][B:C].
\end{align}
For example, at level $N=7$, we have $\left[\Gamma_{1}(7):\Gamma(7)\right] = 7$, $\left[\Gamma_{0}(7):\Gamma_{1}(7)\right] = \phi(7) = 6$, $\left[\text{SL}(2,\mathbb{Z}:\Gamma_{0}(7)\right] = 7\prod\limits_{p\vert 7}\left(1+p^{-1}\right) = 8$. Using the multiplicative nature of the index, we find
\begin{align}
    \begin{split}
        \left[\text{SL}(2,\mathbb{Z}):\Gamma_{0}(7)\right]\left[\Gamma_{0}(7):\Gamma_{1}(7)\right]\left[\Gamma_{1}(7):\Gamma(7)\right] = 336 = \left[\text{SL}(2,\mathbb{Z}):\Gamma(7)\right],
    \end{split}
\end{align}
which matches with the result obtained using definition \ref{index}. For the Hall divisor $\ell\vert N$, i.e. in addition to $\ell$ being a divisor, $\ell$ and $\tfrac{N}{\ell}$ are coprime, the Atkin-Lehner involution at $\ell$ is defined to any matrix that has the following form
\begin{align}
    A_{\ell} = \frac{1}{\sqrt{\ell}}\begin{pmatrix}
    \ell\cdot a & b\\ N c& \ell\cdot d
  \end{pmatrix}
\end{align}
for $a,b,c,d\in\mathbb{Z}$ such that $\text{det}\ A_{\ell} =1$. These operators define involutions on the space of weakly holomorphic modular forms of weight $k$ denoted by $\mathcal{M}_{k}^{!}(\Gamma_{0}(N))$. Products of these involutions correspond to cusps of the Hecke group $\Gamma_{0}(N)$ and slashing by these matrices yields expansions at those cusps. The Atkin-Lehner involutions can be chosen such that they square to the identity matrix $\mathbb{1}_{2}$ and hence, they act as involutions when considered as modular transformations on the upper half-plane $\mathbb{H}^{2}$. For the special case of $k = N$, this is called the Fricke involution,
\begin{align}\label{Fricke_involution}
   \begin{split}
       W_{N} = \frac{1}{\sqrt{N}}\begin{pmatrix}
  0 & -1\\ N & 0
   \end{pmatrix},\\
   W_{N}:\tau\mapsto -\frac{1}{N\tau}.
   \end{split}
\end{align}
Let $[\tau]$ denote the equivalence class of $\tau\in\mathbb{H}^{2*}$ under the action of the group $\Gamma$ on $\mathbb{H}^{2}$. The Fricke involution exchanges two cusp classes $[i\infty]$ and $[0]$ while fixing the fixed point. For a positive integer $N$, the Fricke group denoted by $\Gamma^{+}_{0}(N)$ is generated by the Hecke group $\Gamma_{0}(N)$ and the Fricke involution $W_{N}$, i.e. $\Gamma_{0}^{+}(N) = \langle\Gamma_{0}(N),W_{N}\rangle$ or
\begin{equation}\label{Fricke}
    \Gamma_{0}^{+}(N) \equiv \Gamma_{0}(N)\cup \Gamma_{0}(N)W_{N}.
\end{equation}
The group $\Gamma_{0}^{*}(N)$ on the other hand, is one that is generated by the Hecke group and the Atkin-Lehner involution, i.e.
$\Gamma_{0}^{*}(N) = \langle\Gamma_{0}(N),A_{N}\rangle$. It is important to note that, unlike the Hecke groups, the Fricke and Atkin-Lehner groups are not subgroups of $\text{SL}(2,\mathbb{Z})$. When $N \leq 5$ and when $N = p\in\mathbb{P}$, $\Gamma_{0}^{*}(N) = \Gamma_{0}^{+}(N)$. Let $M\in\mathbb{N}$ and let $h$ be the largest divisor of $24$ such that $M = Nh^{2}$. The normalizer of $\Gamma_{0}(m)$ is given by
\\
\begin{align}
    \mathcal{N}\left(\Gamma_{0}(m)\right) = \Gamma_{0}^{+}(Nh\vert h) = \begin{pmatrix} h & 0\\ 0 & 1\end{pmatrix}^{-1}\Gamma_{0}^{+}(N)\begin{pmatrix} h & 0\\ 0 & 1\end{pmatrix},
\end{align}
\\
where we notice that we recover the Fricke group when $h=1$. 
\begin{tcolorbox}[colback=blue!5!white,colframe=gray!75!black,title=]
  \begin{theorem}[\cite{Apostol}]\label{thm:Apostol_eta}
  For $\gamma = \left(\begin{smallmatrix}a & b\\ c & d\end{smallmatrix}\right)\in\text{SL}(2,\mathbb{Z})$ with $c>0$ and $\tau\in\mathbb{H}^{2}$,
    \begin{align}
      \begin{split}
          \eta(\gamma(\tau)) =& \eta\left(\frac{a\tau + b}{c\tau + d}\right) = \epsilon(a,b,c,d)\left(-i(c\tau + d)\right)^{\frac{1}{2}}\eta(\tau),\\
          \epsilon(a,b,c,d) =& \text{exp}\left[\pi i\left(\frac{a+d}{12c} + s(-d,c)\right)\right],\\
          s(a,b)=& \sum\limits_{\ell=1}^{b-1}\frac{\ell}{b}\left(\frac{a\ell}{b} - \left\lfloor\left.\frac{a\ell}{b}\right\rfloor\right. - \frac{1}{2}\right).
      \end{split}
  \end{align}
  \end{theorem}
\end{tcolorbox}
\noindent
It follows from this theorem that if $k\vert N$ and $\gamma = \left(\begin{smallmatrix}a & b\\ Nc & d\end{smallmatrix}\right)\in\Gamma_{0}(N)$ with $c>0$, then 
\begin{align}
    \begin{split}
        \eta(k\gamma(\tau)) =& \eta\left(\frac{ka\tau + kb}{Nc\tau + d}\right) = \eta\left(\begin{pmatrix}
        a & kb\\ Nc & d
        \end{pmatrix}(k\tau)\right)\\
        =& \epsilon\left(a,kb,\tfrac{cN}{k},d\right)\left(-i(Nc\tau + d)\right)^{\frac{1}{2}}\eta(k\tau).
    \end{split}
\end{align}
Under the $S$-transform, the Dedekind eta-function behaves as follows
\begin{align}\label{Dedekind_S_transform}
    \begin{split}
        \eta(S(\tau)) =& (-i\tau)^{\tfrac{1}{2}}\eta(\tau),\\
        \eta(N\cdot S(\tau)) =& \left(-\frac{i\tau}{N}\right)^{\tfrac{1}{2}}\eta\left(\frac{\tau}{N}\right).
    \end{split}
\end{align}

\subsection*{Signature and genus of congruence subgroups}
We consider a somewhat larger set of congruence subgroups described by
\begin{equation}
    H(p,q,r;\chi,\tau) = \left\{\left.\begin{pmatrix}
    1 + ap & bq\\ cr & 1 + dp
    \end{pmatrix} \in \Gamma\right\vert a,b,c,d\in\mathbb{Z},\ c\equiv \tau a\ (\text{mod}\ \chi)\right\},
\end{equation}
where $p$ divides $qr$ and $\chi$ divides $\text{gcd}\left(p,\tfrac{qr}{p}\right)$. These groups include $\Gamma(N) = H(N,N,N;1,1)$, the Hecke subgroup $\Gamma_{0}(N) = H(1,1,N;1,1)$, and $\Gamma_{1}(N) = H(N,1,N;1,1)$. A subgroup of $\Gamma$ is called a regular subgroup if it contains $I  = \mathbb{1}_{2}$, the identity element of $\Gamma$. If a subgroup is not regular, it is called irregular. In general, the subgroup $H(p,q,r;\chi,\tau)$ is irregular and the associated regular subgroup is denoted by $\pm H(p,q,r;\chi,\tau)$. Let $H(p,N;\chi) = H(p,N,1;\chi,1)$, for $p,N,\chi\in\mathbb{Z}$ such that $p\vert N$ and $\chi\vert \text{gcd}\left(p,\tfrac{N}{p}\right)$. 
\begin{tcolorbox}[colback=magenta!5!white,colframe=gray!75!black,title=]
  \begin{lemma}[\cite{Cummins}]
    For $p,q,r,\chi,\tau\in\mathbb{Z}_{+}$, such that $p\vert qr$and $\chi\vert\text{gcd}\left(p,\tfrac{qr}{p}\right)$. Let $g = \text{gcd}(\chi,\tau)$. Then the groups $H(p,q,r;\chi,\tau)$ and $H(p,gqr;\chi/g)$ have the same signatures. 
  \end{lemma}
\end{tcolorbox}
\noindent
Let $\Gamma'$ be a subgroup of finite index $\mu = \mu(\Gamma')$ of the modular group $\Gamma$ or in other words, $\Gamma'$ can be described as a finitely generated Fuchsian group of the first kind contained in $\Gamma$. Then, the index $\mu = [\text{SL}(2,\mathbb{Z}): \Gamma(N)]$ defined in \ref{index} is given by the following formula
\begin{equation}\label{index_euler_char}
    \mu = \frac{\text{Area}(\Gamma'\backslash\mathbb{H}^{2})}{\text{Area}(\Gamma\backslash\mathbb{H}^{2})} = 6\left(2g-2 + \sum\limits_{j=1}^{v}\left(1 - \frac{1}{m_{j}}\right)\right) = 6\chi(\Gamma'),
\end{equation}
where $\chi(\Gamma')$ is the (negative) Euler characteristic of $\Gamma'$. The group $\Gamma$ contains elliptic elements of orders $2$ and $3$ only. Hence, let us define the following
\begin{itemize}
    \item $\nu_{0} = \nu_{0}(\Gamma')$ be the degree of covering of Riemann surfaces $X(\Gamma')\to X(\Gamma)$ given by the index of the subgroup $\Gamma'$ in $\Gamma$, i.e.
    \begin{equation}
        \nu_{0}(\Gamma') = \mu(\Gamma'),
    \end{equation}
    \item $\nu_{2} = \nu_{2}(\Gamma')$ be the number of $\Gamma'$-inequivalent elliptic fixed points (in $\mathbb{H}^{2}$) of order $2$, 
    \item $\nu_{3} = \nu_{3}(\Gamma')$ be the number of $\Gamma'$-inequivalent elliptic fixed points (in $\mathbb{H}^{2}$) of order $3$,
    \item $\nu_{\infty} = \nu_{\infty}(\Gamma')$ be the number of $\Gamma'$-inequivalent parabolic fixed points (in $\mathbb{Q}\cup \{\infty\}$) given by
    \begin{equation}
      \nu_{\infty}(\Gamma') = \nu_{\infty}^{\text{reg}}(\Gamma') + \nu_{\infty}^{\text{irr}},  
    \end{equation}
    where $\nu_{\infty}^{\text{reg}}(\Gamma')$ and $\nu_{\infty}^{\text{irr}}(\Gamma')$ are the number of regular and irregular cusps of $\Gamma'$ respectively. 
    \item If the discriminant $D$ of the imaginary quadratic field $K = \mathbb{Q}\left(\sqrt{-D}\right)$ is $5$, $8$, and $12$ then $\nu_{5}$, $\nu_{4}$, and $\nu_{6}$ respectively occur.
    \item If the discriminant $D$ of the imaginary quadratic field $K = \mathbb{Q}\left(\sqrt{-D}\right)$ is different from $5$, $8$, and $12$ then only $\nu_{2}$ and $\nu_{3}$ exist.
\end{itemize}
\noindent
\begin{tcolorbox}[colback=blue!5!white,colframe=gray!75!black,title=]
\begin{theorem}[\cite{Shimura}]\label{thm:genus}
  The genus of the curve $X(\Gamma') = \Gamma'\backslash\mathbb{H}^{2*}$ is
  \begin{equation}
      g(X(\Gamma')) = 1 + \frac{\nu_{0}(\Gamma')}{12} - \frac{\nu_{2}(\Gamma')}{4} - \frac{\nu_{3}(\Gamma')}{3} - \frac{\nu_{\infty}(\Gamma')}{2}. 
  \end{equation}
\end{theorem}
\end{tcolorbox}
\noindent 
\begin{tcolorbox}[colback=blue!5!white,colframe=gray!75!black,title=]
\begin{theorem}[\cite{Kok}]\label{thm:genus_Fricke_index}
  Let $\Gamma'$ be an intermediate group of $\Gamma_{0}(M)\leq \mathcal{N}\left(\Gamma_{0}(M)\right) = \Gamma_{0}(Nh\vert h)$, where $M\in\mathbb{N}$ and $M = Nh^{2}$ with $h$ as the largest divisor of $24$. The genus of the curve $X(\Gamma') = \Gamma'\backslash\mathbb{H}^{2}$ reads 
  \begin{equation}\label{genus_Fricke_eqn}
      g(X(\Gamma')) = 1 + \chi(\Gamma_{0}^{+}(N))\frac{\left[\Gamma_{0}^{+}(Nh\vert h):\Gamma'\right]}{2} -
       \frac{5\nu_{6}(\Gamma')}{12} - \frac{3\nu_{4}(\Gamma')}{8} - \frac{2\nu_{3}(\Gamma')}{6} - \frac{\nu_{2}(\Gamma')}{4} - \frac{\nu_{\infty}(\Gamma')}{2},
  \end{equation}
  where $\chi_{0}(\Gamma_{0}^{+}(M)) = \tfrac{M}{6}\prod\limits_{p\vert M}\tfrac{p+1}{2p}$ is the Euler characteristic.
\end{theorem}
\end{tcolorbox}
\noindent 
Let us consider $H$ to be one of the groups of the set $H(p,q,r;\chi,\tau)$ or $\pm H(p,q,r;\chi,\tau)$. The signature of $H$ is given by $(\mu,\nu_{2},\nu_{3},\nu_{\infty}^{\text{reg}},\nu_{\infty}^{\text{irr}})$ and the signature of $\pm H$ is given by $(\mu, \nu_{2},\nu_{3},\nu_{\infty}^{\text{reg}} + \nu_{\infty}^{\text{irr}}, 0)$. 
\begin{tcolorbox}[colback=blue!5!white,colframe=gray!75!black,title=]
  \begin{theorem}[\cite{Cummins}]\label{thm:reg_irr_cusps}
    For $p,q,r,\chi,\tau\in\mathbb{Z}_{+}$ such that $q\vert N$ and $\chi\vert \text{gcd}\left(p,\tfrac{N}{p}\right)$. Let $c = c(p,N;\chi)$ be the number of orbits of group $H(p,N;\chi)$ acting on a set related to the cups of $\Gamma(N)$. then, the number of regular and irregular cusps of $H(p,N;\chi)$ is given by
    \begin{align}
        \left(\nu_{\infty}^{\text{reg}},\nu_{\infty}^{\text{irr}}\right) = \begin{cases}
        (c,0),\ \text{if}\ p=2,\ \text{and}\ \chi = 1,\\
        \ \ \ \ \ \ \ \ \text{or}\ p=1,\\
        \left(\frac{2c}{5},\frac{c}{5}\right),\ \text{if}\ p=2,\ \chi = 2,\ 2\vert\vert (N/p),\\
        \left(\frac{c}{4},\frac{c}{2}\right),\ \text{if}\ p=2,\ \chi = 2,\ 2^{k}\vert\vert (N/p),\ k\ \text{odd},\ k>1\\
        \left(\frac{c}{3},\frac{c}{3}\right),\ \text{if}\ p=2,\ \chi = 2,\ 2^{k}\vert\vert (N/p),\ k\ \text{even}\\
        \left(\frac{2c}{5},\frac{c}{5}\right),\ \text{if}\ p=4,\ \chi = 2,\ 2\nmid (N/p),\ (\text{so}\ \chi=1)\\
        \left(\frac{c}{2},0\right),\ \text{otherwise}
        \end{cases}
    \end{align}
  \end{theorem}
\end{tcolorbox}
\noindent
Here, we have used the notation $a\vert\vert b$ for $a,b,\in\mathbb{Z}$ that is to be read as $a$ exactly divides $b$ and $\text{gcd}(a,b)=1$.
\\
\begin{tcolorbox}[colback=blue!5!white,colframe=gray!75!black,title=]
  \begin{theorem}\label{thm:eliptic_points}
    The number of elliptic points of order $2$ for $\Gamma_{0}(N)$ is given by 
    \begin{align}
        \begin{split}
            \nu_{2}(\Gamma_{0}(N)) = \begin{cases} \prod\limits_{p\vert N}\left(1 + \left(\frac{-1}{p}\right)\right),\ &\text{if}\ 4\nmid N,\\
            0,\ &\text{if}\ 4\vert N,
        \end{cases}
        \end{split}
    \end{align}
    where $\left(\tfrac{-1}{p}\right) = \pm 1$ if $p = \pm 1(\text{mod}\ 4)$ and is $0$ if $p = 2$, and the number of elliptic points of order $3$ for $\Gamma_{0}(N)$ is given by 
    \begin{align}
         \begin{split}
            \nu_{3}(\Gamma_{0}(N)) = \begin{cases} \prod\limits_{p\vert N}\left(1 + \left(\frac{-3}{p}\right)\right),\ &\text{if}\ 9\nmid N,\\
            0,\ &\text{if}\ 9\vert N,
        \end{cases}
        \end{split}
    \end{align}
     where $\left(\tfrac{-3}{p}\right) = \pm 1$ if $p = \pm 1(\text{mod}\ 3)$ and is $0$ if $p = 3$.
  \end{theorem}
\end{tcolorbox}
\noindent
The cases when $H(p,q,r;\chi,\tau)$ contains the identity element $I = \mathbb{1}_{2}$ are
  \begin{enumerate}
      \item $p=2,\ \chi = 2,\ \tau$ even.
      \item $p = 2,\ \chi=1$.
      \item $p=1,\ \chi = 1$.
  \end{enumerate}
\noindent
For the case of Hecke subgroups $\Gamma_{0}(k)$ defined in \ref{Hecke_subgroup} with $k$ an odd prime, we have
\begin{align}
    \mu =& k+1,\ \ \nu_{\infty} = 2,\\
    \nu_{2} =& \begin{cases}
     2,\ \text{if}\ k\equiv 1(\text{mod}\ 4),\\ 0,\ \text{if}\ k\equiv 3(\text{mod}\ 4),
    \end{cases}\\
    \nu_{3} =& \begin{cases}
    1,\ \text{if}\ k = 3\\
    2,\ \text{if}\ k\equiv 1(\text{mod}\ 3),\\
    0,\ \text{if}\ k\equiv 2(\text{mod}\ 3).\\
    \end{cases}
\end{align}
The genus for $k=p$ prime reads
\begin{equation}
    g(\Gamma_{0}(p)) = \begin{cases}
    0,\ \text{if}\ p=2\ \text{and}\ 3,\\
    \frac{p-13}{12},\ \text{if}\ p\equiv 1(\text{mod}\ 12),\\
    \frac{p-n}{12},\ \text{if}\ p\equiv n(\text{mod}\ 12)\ \text{for}\ n=5\ \text{and}\ 7,\\
    \frac{p+1}{12},\ \text{if}\ p\equiv 11(\text{mod}\ 12).
    \end{cases}
\end{equation}
The (negative) Euler characteristic for $k$ prime is given by 
\begin{equation}
    \chi = \frac{k+1}{6}. 
\end{equation}
For the case of the Fricke group $\Gamma^{+}_{0}(p) = \Gamma^{*}_{0}(p)$ where $p$ is $1$ or a prime for which the genus of $\Gamma_{0}^{+}(p)$ is zero, i.e. $p\in\{1,2,3,5,7,11,13,17,19,23,29,31,41,47,59,71\}$.
\\
\begin{tcolorbox}[colback=magenta!5!white,colframe=gray!75!black,title=]
  \begin{lemma}[\cite{Choi}]\label{lemma:Fixed_points}
   The number of fixed points $\#$ of the Frick involution $W_{p}$ on $X_{0}(p)$ is given by 
   \begin{align}
       \begin{split}
           \# = \begin{cases}
           \tfrac{p-1}{6},\ &p\equiv 1\ (\text{mod}\ 12),\\
           \tfrac{p+7}{6},\ &p\equiv 5\ (\text{mod}\ 12),\\
           \tfrac{p+5}{6},\ &p\equiv 7\ (\text{mod}\ 12),\\
           \tfrac{p+13}{6},\ &p\equiv 11\ (\text{mod}\ 12),\\
           \end{cases}
       \end{split}
   \end{align}
   \end{lemma}
\end{tcolorbox}
\noindent
The genus of $\Gamma_{0}^{+}(p)$ reads
\begin{equation}\label{genus_Fricke}
    g(\Gamma^{+}_{0}(p)) = \frac{1}{4}\left(g(\Gamma_{0}(p)) + 2 - \#)\right).
\end{equation}

\subsection*{The dimension of spaces of forms of congruent subgroups}
We state useful theorems that provide us with a working formula to find the dimensions spaces of forms. Let $\mathcal{M}_{k}(\Gamma')$ and $\mathcal{S}_{k}(\Gamma')$ be the space of modular forms and the space of cusp forms of the congruent subgroup $\Gamma'$ respectively.
\section*{Case I: $\Gamma'\subseteq\Gamma$}
\begin{tcolorbox}[colback=blue!5!white,colframe=gray!75!black,title=]
\begin{theorem}[\cite{Shimura}]\label{thm:space_dim_G}
   If $k\geq 2$ is even, then the dimension of the space of cusp forms of $\Gamma'$ and the space of modular forms is
  \begin{align}
      \begin{split}
          \text{dim}\ \mathcal{S}_{k}(\Gamma') =& \delta_{2,k} + \frac{k-1}{12}\nu_{0}(\Gamma') + \left(\left\lfloor\left.\frac{k}{4}\right\rfloor\right. - \frac{k-1}{4}\right)\nu_{2}(\Gamma')\\
          +& \left(\left\lfloor\left.\frac{k}{3}\right\rfloor\right. - \frac{k-1}{3}\right)\nu_{3}(\Gamma') - \frac{\nu_{\infty}(\Gamma')}{2},\\
      \text{dim}\ \mathcal{M}_{k}(\Gamma') =& \text{dim}\ \mathcal{S}_{k}(\Gamma') + \nu_{\infty}(\Gamma')  - \delta_{2,k}.
      \end{split}
  \end{align}
  If $k\geq 3$ is odd and $-I\notin \Gamma'$, then 
   \begin{align}
      \begin{split}
          \text{dim}\ \mathcal{S}_{k}(\Gamma') =&  \frac{k-1}{12}\nu_{0}(\Gamma') + \left(\left\lfloor\left.\frac{k}{3}\right\rfloor\right. - \frac{k-1}{3}\right)\nu_{3}(\Gamma')  - \frac{\nu_{\infty}^{\text{reg}}(\Gamma')}{2},\\
      \text{dim}\ \mathcal{M}_{k}(\Gamma') =& \text{dim}\ \mathcal{S}_{k}(\Gamma') + \nu_{\infty}^{\text{reg}}(\Gamma'),
      \end{split}
  \end{align}
  and the spaces $\mathcal{M}_{k}(\Gamma')$ are trivial for $k$ odd when $-I\in \Gamma'$. 
\end{theorem}
\end{tcolorbox}
\noindent 

\section*{Case II: $\Gamma' = \Gamma_{H}(N)\subseteq\Gamma$}
\begin{tcolorbox}[colback=blue!5!white,colframe=gray!75!black,title=]
    \begin{theorem}[\cite{Cohen-I}]\label{thm:space_dim_G_0}
    Consider the congruence subgroups $\Gamma'\subset\Gamma$ defined by \begin{equation}
    \Gamma_{H}(N) = \left\{\left.\begin{pmatrix}
    a & b\\ c & d
    \end{pmatrix}\in\text{SL}(2,\mathbb{Z})\right\vert c\equiv 0(\text{mod}\ N),\ a,\ d\in H\right\}
    \end{equation}
    for $H$ a subgroup of the multiplicative group $(\mathbb{Z}/N\mathbb{Z})^{*}$. this includes two cases $\Gamma_{1}(N)$ and $\Gamma_{0}(N)$ corresponding to groups $H = \{1\}$ and $H = (\mathbb{Z}/N\mathbb{Z})^{*}$ respectively. Then, the dimension of the space of cusp forms and modular forms read 
    \begin{align}
        \begin{split}
            \text{dim}\ \mathcal{S}_{k}(\Gamma_{H}(N)) =& 2\left\lfloor\left. \frac{k}{3}\right\rfloor\right.-1 \ \text{if}\ k\geq 3,\\ \text{dim}\ \mathcal{M}_{k}(\Gamma_{H}(N)) =& 2\left\lfloor\left. \frac{k}{3}\right\rfloor\right.+1 \ \text{if}\ k\geq 1.
        \end{split}
    \end{align}
  \end{theorem}
\end{tcolorbox}
\noindent 

\section*{Case III: $\Gamma' = \Gamma^{+}_{0}(N)$}
\begin{tcolorbox}[colback=blue!5!white,colframe=gray!75!black,title=]
\begin{theorem}[\cite{Choi}]\label{thm:space_dim_G*0}
   For $k> 2$ with $k$ being an even integer, the dimension of the space of cusp forms of $\Gamma'$ is
  \begin{align}
      \begin{split}
          \text{dim}\ \mathcal{S}_{k}(\Gamma^{+}_{0}(2)) =&  \begin{cases}
          \left\lfloor\left.\frac{k}{8}\right\rfloor\right. - 1,\ k\equiv 2\ (\text{mod}\ 8)\\
          \left\lfloor\left.\frac{k}{8}\right\rfloor\right.,\ \ \ \ \ \ k\not\equiv 2\ (\text{mod}\ 8),\\
          \end{cases}\\
          \text{dim}\ \mathcal{S}_{k}(\Gamma^{+}_{0}(3)) =& \begin{cases}
          \left\lfloor\left.\frac{k}{6}\right\rfloor\right.-1,\ k\equiv 2,6\ (\text{mod}\ 12),\\
          \left\lfloor\left.\frac{k}{6}\right\rfloor\right.,\ \ \ \ \ \  k\not\equiv 2,6\ (\text{mod}\ 12),\\
          \end{cases}\\
          \text{dim}\ \mathcal{S}_{k}(\Gamma^{+}_{0}(p)) =& \begin{cases}
          \left(\tfrac{p+5}{6}\right)\left\lfloor\left.\frac{k}{4}\right\rfloor\right. + \left\lfloor\left.\frac{k}{3}\right\rfloor\right. - \frac{k}{2},\ p\equiv 1,7\ (\text{mod}\ 12),\\
          \left(\tfrac{p + 13}{6}\right)\left\lfloor\left.\frac{k}{4}\right\rfloor\right. - \frac{k}{2},\ \ \ \ p \equiv 5,11\ (\text{mod}\ 12),
          \end{cases}
      \end{split}
  \end{align}
  where the third expression is for $p>3$. The dimension of the space of modular forms is 
  \begin{align}
    \text{dim}\ \mathcal{M}_{k}(\Gamma^{+}_{0}(p)) = 1 + \text{dim}\ \mathcal{S}_{k}(\Gamma^{+}_{0}(p)),    
  \end{align}
  for $p\in\{1,2,3,5,7,11,13,17,19,23,29,31,41,47,59,71\}$.
\end{theorem}
\end{tcolorbox}
\noindent
We follow \cite{Junichi} for finding the basis decomposition of spaces of modular forms and valence formulae of various Fricke and Hecke groups.
\noindent

\section{Kissing numbers in d = 48 and d = 96}\label{appendix:C}
\subsection*{d = 48}
For a $c = 48$ CFT with a partition function involving the $J$-function, the following procedure helps us find the kissing number.
\begin{align}
    \begin{split}
        &\eta^{48}\left(J + 24 + a\right)\left(J + 24 + b\right)(\tau)\\
        =& 1 + (a+b)q + (2\cdot 196560 - 24 (a + b) + a b)q^{2}\\
        +& 12 (2795520 + 16402 (a + b) - 4 a b)q^{3}\\
    +& 8(4928987700 + 1506776 (a + b) + 135a b)q^{4} + \ldots
    \end{split} 
\end{align}
This can be rewritten as follows
\begin{align}\label{Theta_series_d=48_J}
    \begin{split}
        \eta^{48}(\tau)&\left(J(\tau) + 24 + a\right)\left(J(\tau) + 24 + b\right)\\
        =&1 + a_{1}q + a_{2}q^{2} + \left(52416000 + 195660a_{1} -48a_{2}\right)q^{3} + \ldots
    \end{split}
\end{align}
Setting $a_{1} = a_{2} = 0$, we find kissing number $\mathscr{K} = 52416000$ and lattice radius $\rho = \tfrac{\sqrt{3}}{2}$. We now repeat this procedure for a $c = 48$ CFT with partition function corresponding to groups $\Gamma_{0}^{+}(p)$ for $p = 5$. For $\Gamma_{0}^{+}(5)$, we have
\begin{align}
    \begin{split}
    &\eta^{48}\left(j_{5^{+}} + 8 + a\right)\left(j_{5^{+}} + 8+ b\right)(\tau)\\
    = 1 +& (a+b)q + (-380 - 24 (a + b) + a b)q^{2}\\
        +& 2(2120 + 31 (a + b) - 24 a b)q^{3} +  \ldots
    \end{split}
\end{align}
This can be rewritten as follows
\begin{align}
     \begin{split}
        &\eta^{48}\left(j_{5^{+}}+ 8+ a\right)\left(j_{5^{+}}+ 8+ b\right)(\tau)\\
        = 1 +& a_{1}q + a_{2}q^{2} + (14000 - 1090a_{1} - 48 a_{2} - a_{3})q^{3}\\
        +& (-143250 + 9328a_{1} + 120a_{2} - 20a_{3} - a_{4})q^{4}\\ 
        +& \ldots
    \end{split}
\end{align}
Setting $a_{1} = a_{2} = a_{3} = 0$, we find kissing number $\mathscr{K} = 14000$ and lattice radius $\rho = \tfrac{\sqrt{3}}{2}$.

\subsection*{d = 96}
For a $c = 96$ CFT with partition function involving the $J$-function, the following procedure helps us find the kissing number.
\begin{align}
    \begin{split}
        &\eta^{96}\left(J + 24 + a\right)\left(J + 24 + b\right)\left(J + 24 + c\right)\left(J + 24 + d\right)(\tau)\\
        =& 1 + (a+b+c+d)q + (4\cdot 196560 - 24 (a + b + c + d) + a b + a c + a d + b c + b d + c d)q^{2}\\
        +& 12 (2\cdot2795520 + 49161 (a + b + c + d) - 4 (a b + a c + b c + a d + b d + c d) + a b c + a b d + a c d + b c d)q^{3}\\
    +& (233407137600 + 36165568 (a + b + c + d) + 394200 (a b + a c + a d + b c + b d + c d)\\
    -& 72 (a b c + a b d + a c d + b c d) + a b c d)q^{4}\\
 +& (39581611130880 + 116042542350 (a + b + c + d) + 
 14661440 (a b + a c + a d + b c + b d + c d)\\
 +& 
 199044 (a b c + a b d + a c d + b c d) - 96 a b c d)q^{5} + \ldots
    \end{split} 
\end{align}
This can be rewritten as follows
\begin{align}
    \begin{split}
        \eta^{96}(\tau)&\left(J(\tau) + 24 + a\right)\left(J(\tau) + 24 + b\right)\left(J(\tau) + 24 + c\right)\left(J(\tau) + 24 + d\right)(\tau)\\
        =&1 + a_{1}q + a_{2}q^{2} + a_{3}q^{3} + a_{4}q^{4}\\
        +& \left(19028983034880 + 111347591658 a_{1} + 53273168 a_{2} + 16011 a_{3} - 96 a_{4}\right)q^{5} + \ldots
    \end{split}
\end{align}
Setting $a_{1} = a_{2} = a_{3} = a_{4} = 0$, we find kissing number $\mathscr{K} = 19028983034880$ and lattice radius $\rho = \tfrac{\sqrt{5}}{2}$. Repeating this procedure for a $c = 96$ CFT with partition function corresponding to the group $\Gamma_{0}^{+}(7)$, we get
\begin{align}
    \begin{split}
        &\eta^{96}\left(j_{7^{+}}+ 15+ a\right)\left(j_{7^{+}}+ 15+ b\right)\left(j_{7^{+}}+ 15+ c\right)\left(j_{7^{+}}+ 15+ d\right)(\tau)\\
        =& 1 + (a+b+c+d)q + (-1092 - 24 (a + b + c + d) + a b + a c + a d + b c + b d + c d)q^{2}\\
        +& (14224 - 567 (a + b + c + d) - 48 (a b +  a c + a d + b c + b d + c d) + a b c + a b d + a c d + b c d)q^{3}\\
        +& (359730 + 28852 (a + b + c + d) + 534 (a b +  a c + a d + b c + b d + c d)\\
        -& 72 (a b c + a b d + a c d + b c d) + a b c d)q^{4}\\
        +& (-11362176 - 299586 (a + b + c + d) + 18280 (a b +  a c + a d + b c + b d + c d) \\
        +& 2211 (a b c + a b d + a c d + b c d) - 96 a b c d)q^{5} + \ldots
    \end{split}
\end{align}
This can be rewritten as follows
\begin{align}
    \begin{split}
        &\eta^{96}\left(j_{7^{+}}+ 15+ a\right)\left(j_{7^{+}}+ 15+ b\right)\left(j_{7^{+}}+ 15+ c\right)\left(j_{7^{+}}+ 15+ d\right)(\tau)\\
        =& 1 + a_{1}q + a_{2}q^{2} + a_{3}q^{3} + a_{4}q^{4}\\
        +& (80426640 - 3941757a_{1}-156104a_{2}-4701a_{3}-96a_{4}-a_{5})q^{5} + \ldots
    \end{split}
\end{align}
Setting $a_{1} = \ldots = a_{4} = 0$, we find kissing number $\mathscr{K} = 80426640$ and lattice radius $\rho = \tfrac{\sqrt{5}}{2}$.

\end{document}